\documentclass[11pt, a4paper]{article}
\usepackage[top=12mm,bottom=12mm,left=20mm,right=20mm,head=12mm,includeheadfoot,a4paper]{geometry}
\pdfoutput=1

\usepackage[utf8]{inputenc}
\usepackage[english]{babel}
\usepackage[T1]{fontenc}
\usepackage{amsmath,amssymb,graphicx, latexsym, theorem, enumerate, subcaption, csquotes, tocloft, titlesec}
\usepackage[nottoc,notlot,notlof]{tocbibind}
\usepackage[font={small}]{caption}
\usepackage{booktabs, multirow, nicefrac, float, xspace}
\usepackage[dvipsnames]{xcolor}
\usepackage{hyperref}

\normalsize

\newcommand{\im}{\mathrm{Im}}
\newcommand{\re}{\mathrm{Re}}
\newcommand{\Fc}{ \widetilde{F} }

\newcommand{\deltac}{ \widetilde{\delta} }

\DeclareMathOperator\arctanh{arctanh}

\titlespacing*{\section}{0pt}{1.8\baselineskip}{\baselineskip}
\newcommand{\tmop}[1]{\ensuremath{\operatorname{#1}}}

\hypersetup{
    colorlinks=true,
    linkcolor=blue,
    filecolor=magenta,      
    urlcolor=cyan,
    citecolor=red
}

\usepackage[sorting=none, maxbibnames=6]{biblatex} 
\begin{document}

\begin{flushleft}
 \hfill \parbox[c]{40mm}{CERN-TH-2023-161}
\end{flushleft}

\vspace{2cm}

\thispagestyle{empty}

\begin{center}
\bf \Large{Reconstructing $S$-matrix Phases with Machine Learning}
\end{center}

\begin{center}
{\textsc {Aurélien Dersy$^{a,b}$, 
Matthew D. Schwartz$^{a,b}$, Alexander Zhiboedov$^c$}}
\end{center}

\begin{center}
{\it $^a$Department of Physics, Harvard University , \\
02138 Cambridge, MA, USA\\
$^b$ NSF Institute for Artificial Intelligence and Fundamental Interactions\\[5mm]
}
{\it $^c$CERN, Theoretical Physics Department, CH-1211 Geneva 23, Switzerland}
\end{center}

\begin{center}

\texttt{\small schwartz@g.harvard.edu, adersy@g.harvard.edu, alexander.zhiboedov@cern.ch}  \\

\end{center}

\vspace{2cm}

\begin{abstract} 
An important element of the $S$-matrix bootstrap program is the relationship between the modulus of an $S$-matrix element and its phase. Unitarity relates them by an integral equation. Even in the simplest case of elastic scattering, this integral equation cannot be solved analytically and numerical approaches are required. We apply modern machine learning techniques to studying the unitarity constraint. We find that for a given modulus, when a phase exists it can generally be reconstructed to good accuracy with machine learning. Moreover, the loss of the reconstruction algorithm provides a good proxy for whether a given modulus can be consistent with unitarity at all. In addition, we study the question of whether multiple phases can be consistent with a single modulus, finding novel phase-ambiguous solutions. In particular, we find a new phase-ambiguous solution which pushes the known limit on such solutions significantly beyond the previous bound. 
\end{abstract}

\newpage{}
{
  \hypersetup{linkcolor=black}
  \tableofcontents
}

\newpage{}

\section{Introduction}
A crucial ingredient in the $S$-matrix bootstrap program is the relation between the modulus of an amplitude $F(z)$ and its phase, as constrained by unitarity. An important question is, is it always possible to find a phase given the magnitude of a scattering amplitude, and, if so, is that phase unique? If an algorithm were known to find the phase, then it could be applied to physical data, from the differential cross section, to reconstruct the underlying quantum mechanical amplitude. It is generally believed that the amplitude is uniquely fixed, up to the trivial ambiguity, $F(z)\to -F(z)^\star$, provided one has access to the differential cross section across all energies and all angles \cite{Bessis1967, ALVAREZESTRADA1971196}.\footnote{The proof \cite{Bessis1967} assumes that the external particles are scalars, that the amplitude is symmetric under crossing, and finally that there are no bound states.}  In a more realistic setting such information is never known, so we can ask how much information about the phase can be deduced from scattering data in a limited range of energies, or even at fixed energy. Even at fixed energy, and even in the elastic scattering regime where only $2\to2$ scattering is possible, the problem of finding the amplitude from the differential cross section is a hard one~\cite{Martin1969, Martin1973, Chadan_1989}. For inelastic scattering with an infinite number of partial waves, there can be continuous families of phases with the same modulus~\cite{Atkinson:1972hr,Bart:1973cf,Atkinson:1973wt}; for inelastic scattering with $L$ partial waves there are at most $2^{L}$ phases for a given modulus~\cite{GERSTEN1969537}; for elastic scattering, the number of phase-ambiguous solutions is expected to be at most \emph{two} \cite{Martin:2020jlu}.
The question of how to constrain the phase given the cross section has been around since the 1960s, but not much progress has been made since the 1970s. Given the revitalization of the $S$-matrix bootstrap program \cite{Kruczenski:2022lot}, partly inspired by modern computational techniques, we propose to revisit some of the questions using modern tools such as machine learning. We focus here on a clear well-defined problem: given a differential cross-section of a scalar $2\to 2$ scattering process in the elastic region (energy below the first inelastic threshold), under what circumstances does an underlying complex amplitude producing it exists and under what circumstances is the amplitude unique?

We focus on the elastic scattering regime at fixed energy. Since energy is fixed, a $2 \to 2$ amplitude is a function only of the scattering angle $\theta$ and we use $z \equiv \cos\theta$ throughout. We write $F(z)$ for the amplitude, $B(z) \equiv |F(z)|$ for its modulus, and $\phi(z)$ for its phase. The partial wave decomposition of the amplitude is
\begin{equation}\label{eq:partialwaves}
    F (z) = B(z) e^{i \phi(z)} = \sum_{\ell=0}^\infty (2 \ell + 1) f_\ell P_\ell (z) ,
\end{equation} 
where $P_\ell (z)$ are the standard Legendre polynomials of spin $\ell$.
In this notation, unitarity requires $\text{Im} f_\ell = |f_\ell|^2$ for all $\ell$.\footnote{This unitarity relation holds for elastic scattering of non-identical scalar particles $AB \to AB$. The relationship between $F(z)$ and the standard amplitude $\langle p_3, p_4 | \hat T |p_2, p_1 \rangle = (2 \pi)^4 \delta^4(p_1 + p_2 - p_3 - p_4) T(s,t)$ at fixed $s$, see e.g. \cite{Correia:2020xtr,Martin:1969ina}, is $F(z) \equiv \frac{1}{16 \pi} \sqrt{\frac{s-4m^2}{s}} T\Big(s,-\frac{s-4 m^2} {2}(1-z) \Big)$, where in writing this formula we assumed for simplicity that all particles have equal mass $m$. For scattering of identical particles $AA \to AA$ only even spin partial waves appear in the sum \eqref{eq:partialwaves}, so that $F(z)=F(-z)$, and the relationship to the standard scattering amplitude becomes $F(z) \equiv \frac{1}{32 \pi} \sqrt{\frac{s-4m^2}{s}} T\Big(s,-\frac{s-4 m^2}{2}(1-z) \Big) $. } 
Writing the partial waves as $f_\ell = \sin \delta_\ell e^{i\delta_\ell}$, unitarity is equivalent to all the phase shifts $\delta_\ell$ being real. The differential cross section depends only on $|F(z)|^2 = B(z)^2$, so up to trivial kinematic factors the differential cross section and modulus are equivalent. 

Although the partial-wave decomposition is general, if there are an infinite number of partial waves it may not be so useful. One can instead phrase the unitarity condition as an integral equation for the modulus $B(z)$ and phase $\phi(z)$~\cite{newton1968,atkinson1970}:
\begin{equation} \label{eq:sinphieq}
       \sin \phi (z) = \int_{- 1}^1 d z_1  \int_0^{2 \pi} d
   \phi_1 \frac{B(z_1) B(z_2)} {4 \pi\, B(z)} \cos\big[\phi (z_1) -  \phi (z_2)\big]
\end{equation}
where
\begin{equation}
    z_2(z, z_1, \phi_1) \equiv zz_1 + \sqrt{1 - z^2} \sqrt{1-z_1^2}\cos \phi_1 \,.
\end{equation}

 By evaluating Eq.~\eqref{eq:sinphieq} at $z=1$ so that $z_2=z_1$ we immediately get a \emph{necessary} condition on $B(z)$ to be a valid modulus, namely
 \begin{equation} \label{eq:dual_bound}
\int_{- 1}^1 d z_1 \frac{B(z_1)^2} {2 B(1)} \leq 1    .
 \end{equation}
This ``dual'' bound already severely restricts the space of allowable $B(z)$. 

Given $B(z)$ it is in general very difficult to solve Eq.~\eqref{eq:sinphieq} to find $\phi(z)$. Indeed, there are two closely related, but unanswered, questions we can ask
\begin{enumerate}
  \item For which $B(z)$ is there a solution to Eq.~\eqref{eq:sinphieq}? That is, which elastic-scattering cross sections can conceivably be realized in a unitary quantum field theory? 
  \item For which $B(z)$ can there be more than one solution to Eq.~\eqref{eq:sinphieq}? That is, when is the phase unique?

\end{enumerate}
Both of these questions were studied some time ago and only partially answered, as we now review.  

The sharpest statements so far have been made by applying the contraction mapping principle to the unitarity equation, where the search for a phase solution can be recast as a problem of finding the mapping's associated fixed point \cite{newton1968, atkinson1970, JEBowcock_1975}.
The current bounds on both existence and uniqueness have been derived based on the integrated form of the kernel in  Eq.~\eqref{eq:sinphieq}
\begin{equation} \label{kernelkz}
    K(z) \equiv \int_{- 1}^1 d z_1  \int_0^{2 \pi} d
   \phi_1 \, \frac{B(z_1) B(z_2)} {4 \pi\, B(z)} \,.
\end{equation}
The maximum of this function was denoted by Martin as
\begin{equation}
    \sin \mu \equiv \max_{-1 \leq z \leq 1} K(z) \label{sinmudef} \,.
\end{equation}
To motivate this, we note that since $| \cos[\phi(z_1) - \phi(z_2)]  | \leq 1$ Eq.~\eqref{eq:sinphieq} implies
\begin{equation}
    | \sin \phi(z) | \le \sin \mu \,.
\end{equation}
If the phase $\phi(z)$ is constant then by Eq.~\eqref{eq:sinphieq} $B(z)$ must be constant as well and $\sin \phi = \sin\mu  = B$, so this bound is saturated. Furthermore, since $\phi$ must be real we have that for constant phases, $\sin\mu \le 1$ and $B \le 1$.
It has also been proven that as long as $\sin\mu \le 1$ for any given $B(z)$, a corresponding phase always exists. The proof treats the Eq.~(\ref{eq:sinphieq}) as a non-linear operation $\phi_{n+1} = O(\phi_n)$, and applies the Leray-Schauder principle to argue for the existence of a fixed point \cite{Martin1969}. We discuss this approach more in Section~\ref{sec:simplepolynomial}.
Conversely, there exist differential cross sections with $\sin \mu >1$ for which no phase exists, the simplest example being constant $B>1$. So one cannot hope to push this sufficient criterion for the existence of a phase further. If $\sin \mu >1$ there is no known test to determine whether or not a phase exists for a given $B(z)$ beyond \eqref{eq:dual_bound}.

Regarding the question of uniqueness, the contraction mapping principle was applied to demonstrate that any solution with $\sin \mu < \sqrt{\frac{\sqrt{5}-1} {2} } \approx 0.79$ is unique \cite{Martin1969, newton1968}, while further refinements \cite{Gangal1984} pushed the bound up to $\sin \mu < 0.86$. 
For polynomial amplitudes (finite number of partial waves) $\sin \mu \le 1$ is enough to ensure both existence and uniqueness \cite{Martin1969}. For polynomial amplitudes, it has also been shown that if the average modulus $\sigma_T = \frac{1}{2} \int_{-1}^1 dz B^2(z)$ satisfies $\sigma_T < 1.38$ then uniqueness is guaranteed.
For general amplitudes with an infinite number of partial waves, it has been conjectured~\cite{Martin1969, atkinson1970}  but not proven or disproven that uniqueness should still hold if $\sin \mu <1$.
In the elastic scattering region that we consider, any nontrivial (i.e. excluding $F(z) \to -F(z)^\star$) phase ambiguity is expected to be twofold at most, as has been proven for genuine entire functions (i.e not a polynomial) \cite{Itzykson1973}, see also \cite{Martin:2020jlu}. 

A modulus with two corresponding phases was found by Crichton in 1966 \cite{Crichton1966}. Crichton's solution has only the $L=2$ partial waves and $\sin \mu = 3.2$. Shortly after, the complete set of $L=2$ phase ambiguous solutions was characterized \cite{ATKINSON1973125}. The lowest value of $\sin\mu$ among these was $2.6$. Solutions with $L=3$ and $L=4$ have also been studied \cite{BERENDS1973507, Cornille1974}. These solutions are discussed in Section~\ref{sec:finiteL}.

Phase-ambiguous solutions have also been found with an infinite number of partial waves in \cite{Atkinsonambiguity}. The lowest published value of $\sin\mu$ among these, to our knowledge, is $\sin\mu \approx 2.15$. These results are reviewed in more detail in Section~\ref{sec:infiniteL}. Applying modern numerical methods we are able to find a phase-ambiguous solution with $\sin\mu \approx 1.67$.

In this paper, we revisit some of these old questions about phase determination in the elastic regime using modern numerical methods and machine learning. Recent advances in machine learning have given rise to a multitude of applications in physics, from jet tagging algorithms \cite{Butler2019}, to fast detector simulators \cite{calogan} or AI-driven symbolic regression \cite{Tegmark2020, kamienny2023deep} and give us the perfect tool for tackling hard numerical problems for which classical algorithms are challenging to design. Is it well known that neural networks are universal function approximators \cite{HORNIK1989359} and as such are ideal candidates for solving integro-differential equations. In particular Physics-informed neural networks have been shown to be able to resolve multi-dimensional differential equations \cite{RAISSI2019686}, where they act as a functional ansatz and have a loss function given by the differential equation of interest. The extension to integral equations usually involves a discretization scheme for the actual integral and has been studied and implemented in various libraries \cite{neuralpde, DeepXDE, YUAN2022111260, PANG2020109760}. In the following, we will explore how similar techniques can be applied to study and solve the unitarity integral equation Eq.~(\ref{eq:sinphieq}), demonstrating how to numerically recover the phase corresponding to a given input differential cross section. We will highlight the interesting duality between the convergence properties of the machine learning algorithm and the kernel function, making the link with bounds derived in the literature. Finally, we will deploy various neural networks and impose a repulsive loss in order to probe the uniqueness of the recovered solution.

We begin in Section~\ref{sec:nnsetup} by describing the machine learning setup and approach we take to establishing consistency between a modulus and a phase. The unitarity constraint is encoded in Eq.~\eqref{eq:sinphieq}. There are different ways to solve this equation. Given a known modulus $B(z)$, for example from experimental cross-section data or some $S$-matrix-bootstrap computation, one can then search for a phase $\phi(z)$ consistent with unitarity. To find moduli with phase ambiguities, one can alternatively search for 3 functions $B(z)$, $\phi_1(z)$ and $\phi_2(z)$ all consistent with unitarity. To guarantee that $\phi_1(z)$ and $\phi_2(z)$ are not trivially equivalent (e.g. by $\phi_1(z) = \pi - \phi_2(z)$), we will also need to add a repulsive loss to keep the solutions apart. Section~\ref{sec:singleamplitude} discusses how to use our machine learning set-up to find $\phi(z)$ given $B(z)$. Section~\ref{sec:finiteL} discusses the case of phase-ambiguous solutions for amplitudes with a finite number of partial waves and Section~\ref{sec:infiniteL} the infinite partial-wave case. By exploring this space through a combination of a machine learning search and a refinement using a classical algorithm capable of high precision, we find a large number of new phase-ambiguous solutions. The lowest $\sin\mu$ among these is $\sin\mu \approx 1.67$, a value significantly closer to $\sin\mu=1$ than the best previously known example $\sin\mu \approx 2.15$. A summary and conclusions are in Section~\ref{sec:conclusions}.

\section{Machine Learning implementation}\label{sec:nnsetup}
Finding a unitary amplitude for a given input differential cross section boils down to solving the Eq.~\eqref{eq:sinphieq}. Solving differential or integral equations with machine learning is a problem that has already been tackled efficiently in the literature, through the use of Physics-Informed Neural Networks (PINNs) \cite{RAISSI2019686}. In a PINN the idea is to use a neural network $u_\sigma (\vec{x})$ as a surrogate of the solution $u (\vec{x})$. Here $u_\sigma$ is a neural network that takes as input the data $\vec{x}$ and has parameters (weights and biases) described by $\sigma$. The precise architecture of $u_\sigma$ can be fine-tuned given the problem at hand but typical setups consider simple feed-forward neural networks. Having the neural network ansatz allows one to take derivatives efficiently with respect to the inputs, a desirable property for solving differential equations. Indeed the loss function for PINNs is usually taken to be the differential equation itself, evaluated at a set of collocation points. 

In our problem, we have a few notable particularities that depart from the typical PINN use case:
\begin{enumerate}
  \item We are solving an integral equation as opposed to a differential equation. In practice, this is done by approximating the integral, for instance with Gaussian quadrature or a trapezoidal rule, which will inevitably lead to some numerical errors.
  \item We will be interested in understanding whether the phase solution is unique. Probing this property could be done by training different neural networks that are initialized with different random seeds. Another approach, which is one that we will prefer, is to add a repulsion term in the loss function. This allows us to simultaneously train different neural networks, each corresponding to a distinct solution of the integral equation.
  \item We are not always guaranteed to find a solution for any given $B(z)$, so our networks are not always expected to converge to low loss values. This will lead us to study the loss landscape in more detail for simple input differential cross sections.
  \item We are interested in parsing through the space of differential cross sections to probe existence and uniqueness criteria. When explicitly looking for ambiguous solutions we will see that after parameterizing the amplitude in some simple way we are able to learn both the phase $\phi(z)$ and the modulus $B(z)$.
\end{enumerate}

\begin{figure}
    \centering
    \includegraphics[width=0.75\textwidth]{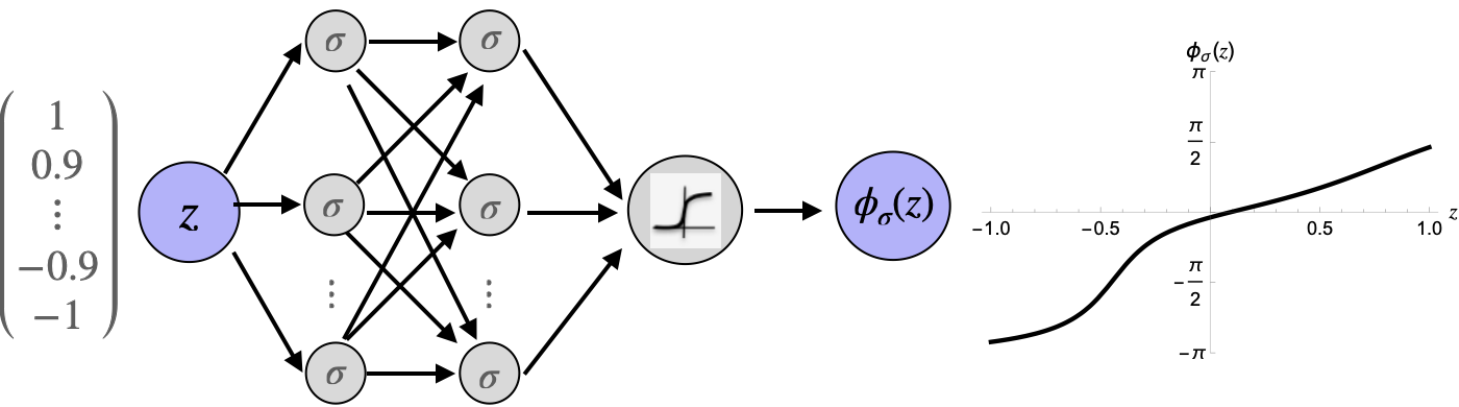}
    \caption{We utilize a neural network ansatz for parametrizing the phase solution. The feedforward neural network has a series of layers with learnable parameters $\sigma$ ending with a final \textit{Tanh} activation function, constraining the outputs to lie within the range $[-\pi, \pi]$. }
    \label{fig:phinet}
\end{figure}

Similarly to PINNs however we will parametrize the phase $\phi(z)$ by a neural network $\phi_\sigma$ and ask for it to solve the Eq.~\eqref{eq:sinphieq}. The parameterized phase $\phi_\sigma(z)$ shown in Fig.~(\ref{fig:phinet}) is a network that takes in a single input and depends on a set of neural network parameters $\sigma$. These parameters are to be updated and optimized to satisfy a given objective or loss function, typically given by the unitarity integral equation. Contrarily to PINNs, we will not have any specific boundary condition to satisfy, rather we will force the output of $\phi_\sigma(z)$ to lie within the range $[-\pi, \pi]$. This is done by adding a scaled sigmoid or tanh activation function at the end of the network. Since we are interested in solving a formal equation we are free to take any $z$ point as part of our training data, provided $z\in[-1,1]$.

 \subsection{Implementation details}
Following the discussion of the previous section we parametrize $\phi(z)$ by a network with a simple feed-forward architecture, which we implement with \textit{PyTorch} \cite{pytorch}. We will restrict ourselves to small architectures, typically 4 layers with 64 nodes each using Rectified Linear Unit (ReLU) activation functions. The outputs are constrained in the $[-\pi,\pi]$ range by adding a scaled hyperbolic tangent function after the final layer\footnote{If we are in the $\sin \mu <1$ regime, where existence is guaranteed, we can further restrict the range to $[-\pi/2,\pi/2]$ in order to eliminate the trivial ambiguity relating $\phi(z)\rightarrow \pi - \phi(z)$.}.  The loss function is taken to be the Mean Squared Error (MSE) of the integral equation, averaged over a set of $N_c$ randomly sampled collocation points, namely:
 \begin{equation}\label{eq:lossfunc}
   \mathcal{L}_E = \mathbb{E}_z \left|\left|B (z) \sin \phi (z) -  \frac{1}{4 \pi} \int_{- 1}^1 d z_1  \int_0^{2 \pi} d
   \phi_1 B (z_1) B (z_2) \cos\left(\phi (z_1) - \phi (z_2)\right)\right|\right|^2  \,.
 \end{equation}
 In practice, the expectation value $\mathbb{E}_z$ appearing in the loss is estimated by averaging over the set of collocation points $\{z_c\}$ as $\mathbb{E}_z \sim N_c^{-1} \sum_{z \in \{z_c\}}$. The two-dimensional integral is also estimated, discretizing the integrand over a grid in $(z_1,\phi_1)$ space. For training, we will also consider a scaled version of this loss defined as 
\begin{equation}\label{eq:lossfunc_scaled}
    \mathcal{L}_E^S = \mathbb{E}_z \left|\left|\sin \phi (z) -  \frac{1}{4 \pi B (z) } \int_{- 1}^1 d z_1  \int_0^{2 \pi} d
   \phi_1 B (z_1) B (z_2) \cos\left(\phi (z_1) - \phi (z_2)\right)\right|\right|^2   
\end{equation}
which has the desirable feature of having terms of order 1. When learning $B(z)$ this loss will also discourage the network from learning arbitrarily small moduli. One could be tempted to further normalize and divide the loss function by $\sin \phi(z)$, but this leads to numerical instabilities if the expected phase value nears 0.
 
During training the expectation value in the loss function is approximated by averaging over a batch of 64  randomly sampled $\{z_c\}$ collocation points. Random sampling ensures that the entire angle range is properly resolved and not overfitted. The network parameters are updated at the end of each epoch, defined here by the complete processing of a single batch. The parameter update is done via the Adam optimizer \cite{Adam2014} where we set the coefficients $\beta_1 = 0.9$ and  $\beta_2 = 0.999$ to their default values. These coefficients correspond to the exponential decay rates of respectively the first and second moment estimates of the gradient. In order to calculate the loss function and the relevant two-dimensional integrals we have to resort to a numerical approximation. We use the trapezoidal rule\footnote{The trapezoidal rule is implemented in \textit{PyTorch} directly which ensures that we will end up with a final loss that is fully differentiable and supports backpropagation.}, where at each fixed $z$ value we pick out $25\times 25$ reference points, linearly spaced out in the $z_1$ and $\phi_1$ directions, giving us an evaluation grid for approximating the two-dimensional integral. In general the trapezoidal rule for a function $f(x)$ will give a numerical error scaling as $K N^{-2}$, where $N$ is the number of points picked along a single direction and $|f''(x)|<K$. In our problem of interest, with our choice of points,  we expect the trapezoidal rule to give errors in the range of $10^{-4}-10^{-6}$ at different $z$ values. This implies that as $\mathcal{L}_E \sim 10^{-8}$ the loss becomes of the order of the numerical precision that we operate at. Our default model choice and hyperparameters are summarized in Table~\ref{tab:hyperparameters} and any deviation from those in the numerical experiments will be mentioned explicitly. Different choices of hyperparameters have been considered but those listed ended up giving the best performance after a brief optimization search. 

 \begin{table}[htp!]
    \centering
   \caption{Model architecture and hyperparameters used for the parameterization of the phase $\phi_\sigma(z)$ with a single neural network.}
   \label{tab:hyperparameters}
    \begin{small}
\begin{tabular}{l l c }
    \toprule
Parameter class &Parameter type& Value  \\
\midrule
 \multirow{3}{*}{Architecture} & Activation & ReLU\\
 & Layers & [64,64,64,64] \\
 & Final layer& $\pi \tanh(z)$\\
\midrule
\multirow{5}{*}{Optimizer} & Optimizer name & Adam\\
& $\beta_1$ & 0.9\\
& $\beta_2$ & 0.999\\
 & Learning rate & $3\cdot10^{-3}$\\
 & Scheduler & MultiplicativeLR with 0.999 decay\\
\midrule
\multirow{3}{*}{Training loss}
& Batch size & 64\\
& Integral approximator & Trapezoidal rule\\
& Integral sampling points & 25 $\times$ 25\\
 \bottomrule
\end{tabular}
\end{small}
   \end{table}

\section{Single phase determination} \label{sec:singleamplitude}
We start our analysis by probing the question of existence of $\phi(z)$ given $B(z)$. That is, we will be interested in training a single neural network to recover a phase $\phi(z)$, which is a solution to Eq.~\eqref{eq:sinphieq}, assuming that the modulus $B(z)$ is a known function. We will start by verifying that the network can be trained to recover solutions in the regime where $\sin \mu <1$, where we have guarantees on the existence of the function. Focusing on simple polynomial differential cross sections, we will illustrate that the loss landscape is sensitive to the value of $\sin \mu$ and that the existence bounds are respected. We will also demonstrate that our method is able to recover solutions when $\sin \mu >1$, taking examples where the amplitudes are parameterized by an either finite or infinite partial wave decomposition.

\subsection{Warmup: simple examples}
\label{sec:simplepolynomial}
\begin{figure}[t]
     \centering
         \includegraphics[width=0.4\textwidth]{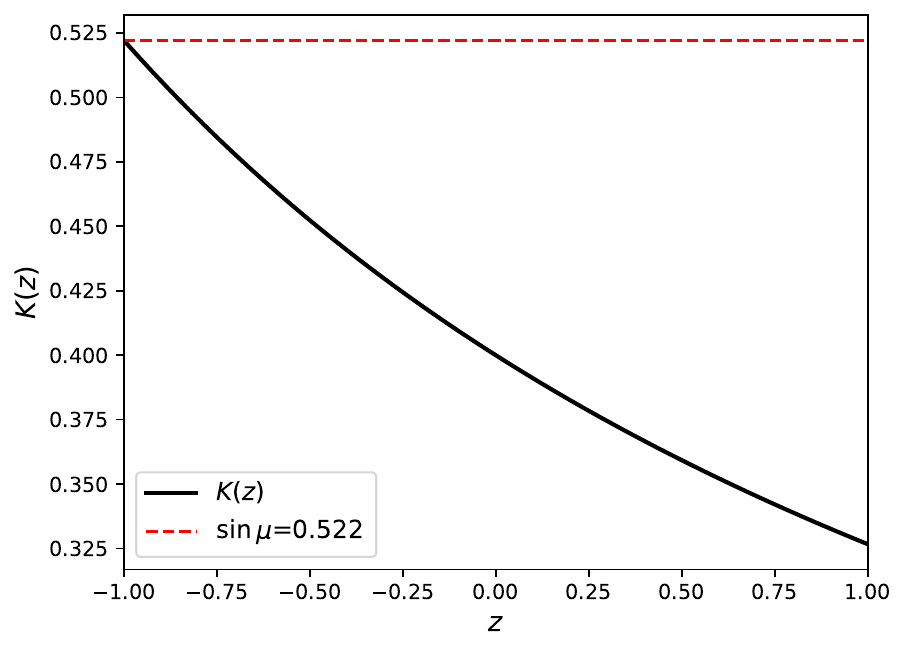}
         \hspace{5mm}
        \includegraphics[width=0.4\textwidth]
         {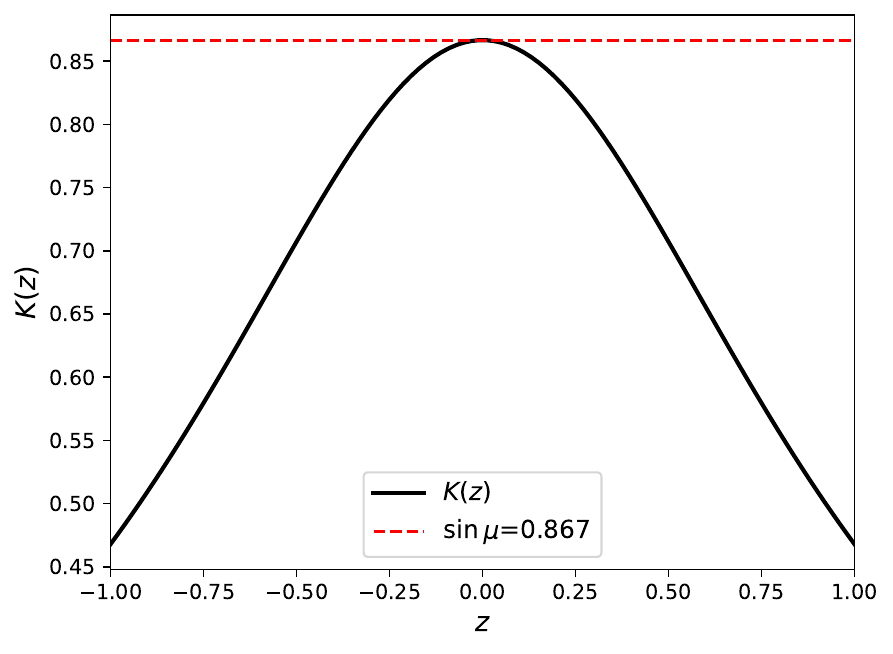}
         \includegraphics[width=0.4\textwidth]{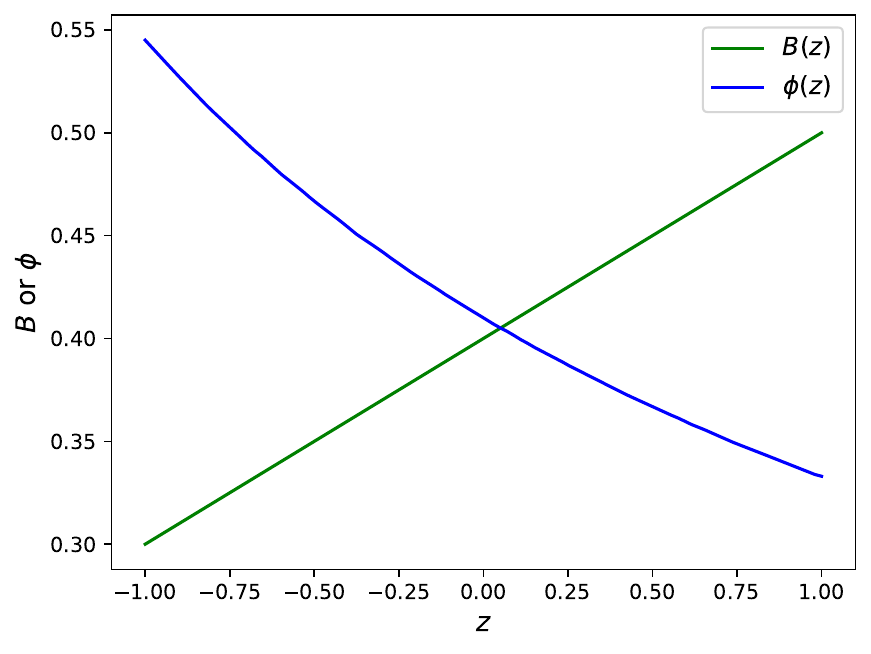}
         \hspace{5mm}
        \includegraphics[width=0.4\textwidth]
        {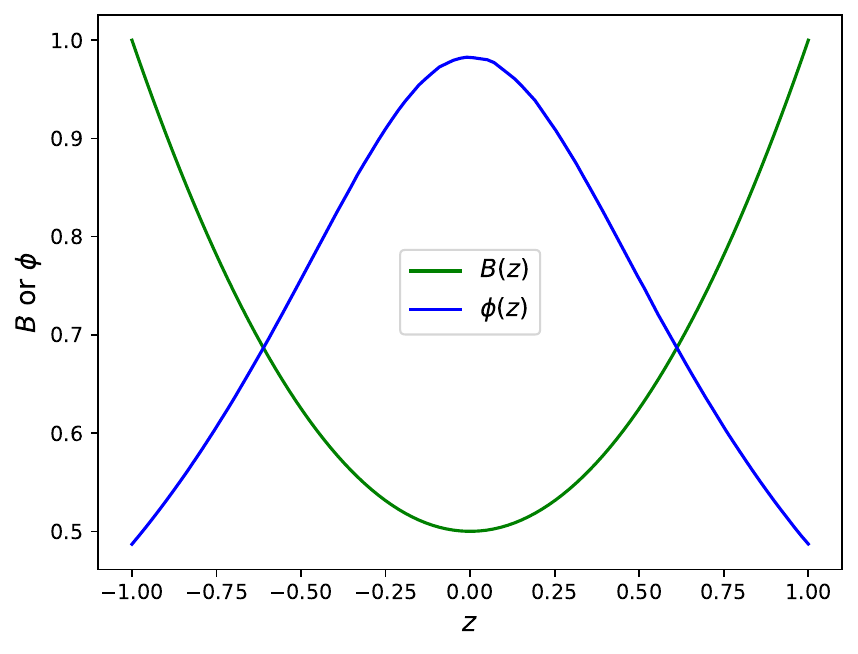}
     \caption{We warm up by determining the phase $\phi(z)$ for a linear modulus  $B(z)=\frac{z+4}{10}$ (left) and a quadratic modulus $B(z)=\frac{z^2+1}{2}$ (right). Top panels show the integrated kernel $K(z)$ and its maximum $\sin\mu$. Bottom panels show $B(z)$ and the phase $\phi(z)$ found with machine learning.
     \label{fig:kernelslinquad}
     }
\end{figure}

To get started, we first consider simple polynomial forms for the modulus. We consider a linear function $B(z)=(z+4)/10$ and a quadratic function $B(z) = (z^2+1)/2$. Both moduli are positive across the $z$ range and have $\sin \mu$ values that are respectively $\sin \mu_1 = \frac{47}{90}\approx 0.522$ and $\sin \mu_2 = \frac{13}{15} \approx 0.867$, guarantying the existence of a solution. Their integrated kernels $K(z)$ (cf. Eq.~\eqref{kernelkz}) whose maximum gives $\sin \mu$ are shown in Fig~\ref{fig:kernelslinquad}. We implement different neural networks following the setup described in Section~\ref{sec:nnsetup} and let them run for 5000 epochs. The  final performance is evaluated on a test set of 100 linearly spaced out $z$ points. We show at the bottom of Fig.~\ref{fig:kernelslinquad} the predicted phases for both cases considered. The final evaluation losses are both of the order of $\mathcal{L}_E^S \sim 10^{-8}$, thus at the order of the numerical precision which is supported by the numerical integration scheme. For moduli satisfying $\sin \mu <1 $ such as these, the numerical fixed point iteration \cite{atkinson1970} applies and we have verified with a classical algorithm that the solutions agree with the ones found by our framework.

In Fig.~\ref{fig:kernelslinquad}, we can observe that for both the linear and quadratic cases, the phases $\phi(z)$ look a lot like the integrated kernel $K(z)$. This is straightforward to understand. When $\phi(z) \ll 1$, one can expand the unitarity equation Eq.~\eqref{eq:sinphieq} using $\sin \phi(z) \approx \phi(z)$ and $\cos[\phi(z_1)-\phi(z_2)]\approx 1$ to see that $\phi(z) = K(z)$ to first order in $\phi(z)$. Indeed, one can then expand to second order in $\phi(z)$ giving
\begin{equation} 
        \phi (z) = K(z) +  \int_{- 1}^1 d z_1  \int_0^{2 \pi} d
   \phi_1 \frac{B(z_1) B(z_2)} {4 \pi\, B(z)} \frac{1}{2}[K (z_1) -  K(z_2)\big]^2 + \cdots
\end{equation}
and so on. This is actually just a version of the fixed-point iteration scheme starting with $\phi(z) = 0$. One can in principle use this procedure even if $\phi(z)$ is not small. There the iterative scheme takes $\phi_{n+1} = \Phi(\phi_n)$ where 
\begin{equation}
    \Phi(\phi_n) = \arcsin\left(\frac{1}{4\pi}\int_{-1}^1 d z_1 \int_0^{2\pi} d\phi_1 \frac{B(z_1)B(z_2)}{B(z)} \cos[\phi(z_1)-\phi(z_2)]\right)
\end{equation}
and aims at finding the fixed point $\phi^\star=\Phi(\phi^\star)$. However, if $\sin\mu > 1$ then $K(z)>1$ for some $z$ and $\sin \phi(z) = K(z)$ has no solution for real phases $\phi(z)$.  This is one reason we only expect phase ambiguities for $\sin\mu>1$. Generally, we find that for $\sin\mu<1$ the iteration tends to converge fairly quickly (although we cannot prove it is independent of the initial condition for the iteration), but when $\sin\mu>1$ it often does not converge at all.

The unitarity equation also implies the bound $\sin \phi(z) \leq K(z)$, which is respected in our results, giving an important cross-check. For these simple polynomial amplitudes, we can see from Fig.~\ref{fig:kernelslinquad} that $K(z)\approx B(z)^{-1}$. This is once again expected in the regime where the integral appearing in Eq.~\eqref{kernelkz} is slowly varying. In particular, for the linear modulus, we have
\begin{equation}
K(z) B(z)= \int_0^{2\pi} d\phi_1 \int_{-1}^1 \frac{B(z_1)B(z_2)}{4\pi} = \frac{z+48}{300}
\end{equation}
Since $-1<z<1$ this function is essentially constant and $K(z) \approx \frac{c}{B(z)}$ for some $c$ results.

As a side note, when $\phi(z)$ is a solution then so is $\pi - \phi(z)$. These solutions are said to be trivially related and either could have been expected. Changing the initialization seed of the networks is a way to recover such alternative solutions.

\subsection{Scanning the loss landscape}
Having validated our method on two simple polynomial examples, we can now look into the performance of our implementation on families of $B(z)$. We consider two families: a linear one, $B(z)=az+b$ and a quadratic one $B(z)= c z^2 + d$. For each value of $a$ and $b$ or $c$ and $d$ we can search for a phase. Although the network cannot tell us for sure whether unitarity is exactly satisfied, the loss of the neural network provides a good proxy for satisfaction. We thus explore the loss landscape and compare it to other indicators of whether unitarity can be satisfied.

\subsubsection{Linear functions} \label{sec:linearamplitudes}
We consider the family $B(z) = az+b$ with $a$ and $b$ real and $b>|a|$, which ensures positivity of $B(z)$ for all $z$ values. Although $B(z)$ is a polynomial, the amplitude $F(z) = B(z) e^{i\phi(z)}$ will generally not be. Indeed, this parameterization for $B(z)$ is not compatible with any unitary polynomial $F(z)$ (see Appendix~\ref{sec:appendixfinite}). As such, any numerical solution for $\phi(z)$ has to be understood as possessing an infinite partial wave decomposition.

\begin{figure}[t]
    \centering
    \includegraphics{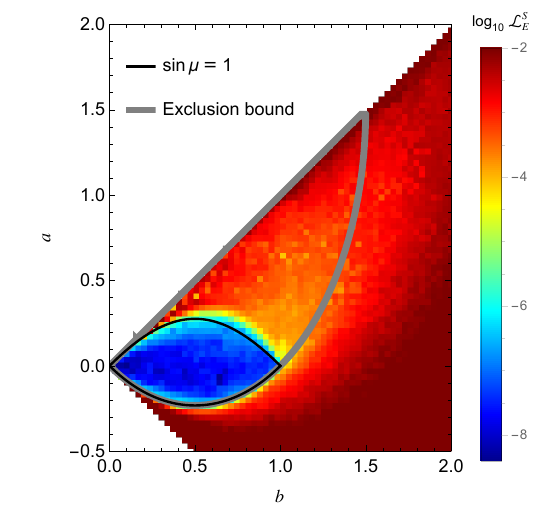}
    \caption{Scaled loss landscape for the two-parameter family of moduli $B(z)=az+b$. For each $a$ and $b$ we find a phase $\phi(z)$. Red regions indicate that no solution is likely. The black curve is $\sin\mu=1$ and delimits the region within which we are guaranteed the existence of a solution. The grey curve delimits the bound of Eq.~(\ref{eq:dual_bound}) and solutions cannot be found outside of its enclosed region.}
    \label{fig:linearfulllandscape}
\end{figure}

\begin{figure}[ht]
     \centering
     \begin{subfigure}[b]{0.45\textwidth}
         \centering
         \includegraphics[width=\textwidth]{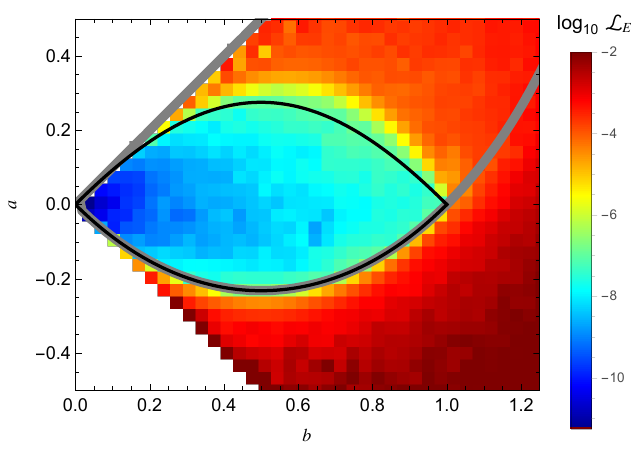}
         \caption{Base log loss landscape}
         \label{fig:landscapelinear}
     \end{subfigure}
          \hspace{5mm}
          \begin{subfigure}[b]{0.45\textwidth}
         \centering
         \includegraphics[width=\textwidth]{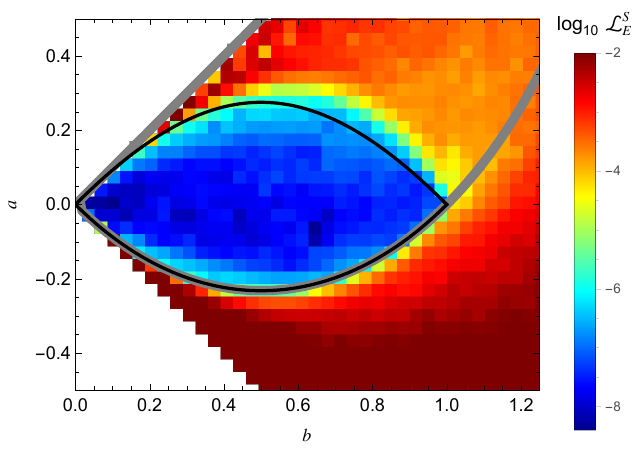}
         \caption{Scaled log loss landscape}
         \label{fig:linlandscapebase}
     \end{subfigure}
     \centering
     \begin{subfigure}[b]{0.45\textwidth}
         \centering
         \includegraphics[width=\textwidth]{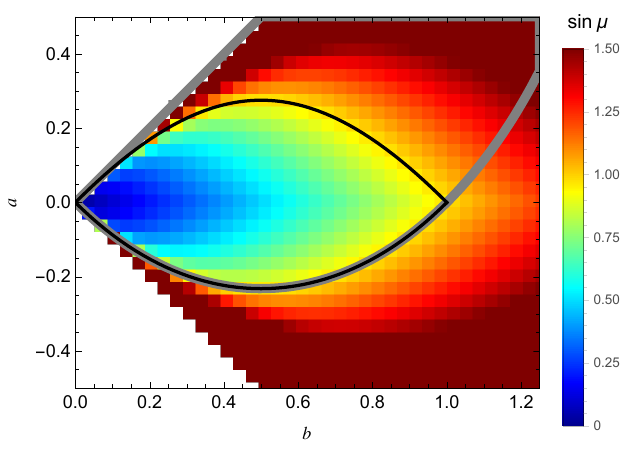}
         \caption{$\sin \mu$ landscape}
         \label{fig:sinmulinlandscape}
     \end{subfigure}
          \hspace{5mm}
          \begin{subfigure}[b]{0.45\textwidth}
         \centering
         \includegraphics[width=\textwidth]{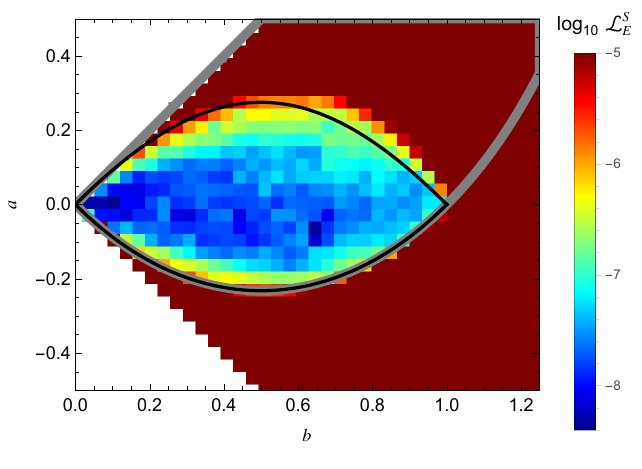}
         \caption{Cut on the scaled log loss landscape}
         \label{fig:landscapelinearzoom}
     \end{subfigure}     
     \caption{Zoom on the loss landscapes for the two-parameter family of moduli $B(z)=az+b$. Black and grey curves follow Fig.~\ref{fig:linearfulllandscape}. Top panels show the base loss and scaled loss. Panel (c) shows a heat map of $\sin\mu$ values over the family $B(z)=az+b$ (no phase is determined or needed). Right shows the scaled loss landscape with a hard cut of $\mathcal{L}_E^S \sim 10^{-5}$. This loss boundary agrees very well with the $\sin\mu=1$ boundary. }
      \label{fig:linearlosslandscapes}
\end{figure}

We conduct a scan over this family of $B(z)$ by taking a $75 \times 60$ grid over different $a,b$ values, with $a\in[-0.5, 2.0]$ and $b \in [0,2.0]$. For every parameter pair, we train a new neural network using the scaled loss function of Eq.~\eqref{eq:lossfunc_scaled} for 2000 epochs. We then evaluate the base and scaled losses,  Eqs.~\eqref{eq:lossfunc} and \eqref{eq:lossfunc_scaled}, on the resulting solutions. In Fig.~\ref{fig:linearfulllandscape} we show the complete scaled loss landscape, along with the $\sin \mu = 1$ and  $\int_{-1}^1 B(z_1)^2 = 2B(1)$ contours. Those are contours for respectively guaranteeing and excluding solutions. A zoomed-in perspective on the regions of low losses is shown in Fig.~\ref{fig:linearlosslandscapes}. $\sin\mu<1$ seems to give a good indication that a solution exists or not. This is non-trivial --  the ML algorithm knows nothing about $\sin\mu$ and there could equally well have been an entirely different functional of $B(z)$ which characterized the existence of a solution. 
The correspondence of $\sin\mu=1$ with the boundary of the allowed region is further explored in the bottom panels. There we also
show the values of $\sin\mu$ across this linear $B(z)$ family. 

On the bottom right we show that the boundaries of $\sin \mu \sim 1$ and $\mathcal{L}_E^S \sim 10^{-5}$ almost perfectly overlap. As the loss crosses this threshold it rapidly grows by many orders of magnitude, up to $10^{-4} - 10^{-2}$, indicative of a regime where the network is unable to find a solution outside of $\sin \mu > 1$ for linear moduli. In particular, the bottom boundary of $\sin \mu =1 $ overlaps with the dual exclusion bound, across which the network has $\mathcal{L}_E^S > 10^{-5}$ and does not find any solutions. However, near the top $\sin \mu =1$ boundary, we have a very thin region of potential solutions with $\sin \mu > 1$. All of these solutions fall within the grey region and are not forbidden by the dual exclusion bound. Additional dual bounds constraints are explored in Appendix~\ref{sec:dual_bound} but do not go beyond the simple exclusion bound we have considered.

\begin{figure}[t]
    \centering
    \includegraphics{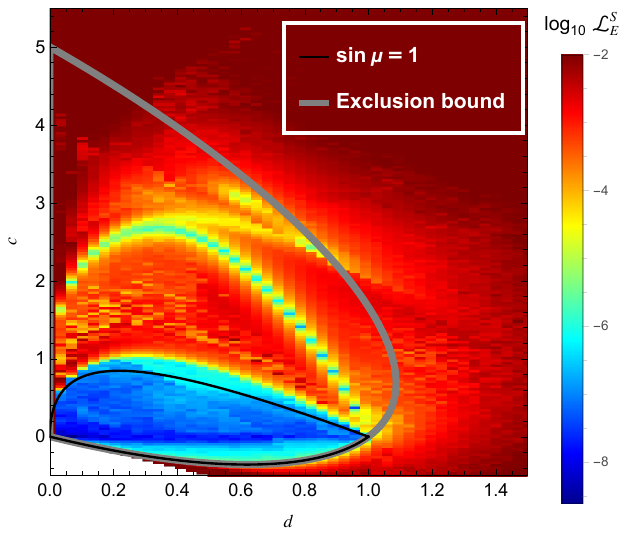}
    \caption{Scaled loss landscape for the two-parameter family of moduli $B(z)=c z^2+d$. For each $c$ and $d$ we find a phase $\phi(z)$. Red regions indicate that no solution is likely. The black curve is $\sin\mu=1$ and delimits the region within which we are guaranteed the existence of a solution. The grey curve delimits the bound of Eq.~(\ref{eq:dual_bound}) and solutions cannot be found outside of its enclosed region.}
    \label{fig:quadraticfulllandscape}
\end{figure}

\begin{figure}[t]
     \centering
     \begin{subfigure}[b]{0.42\textwidth}
         \centering
         \includegraphics[width=\textwidth]{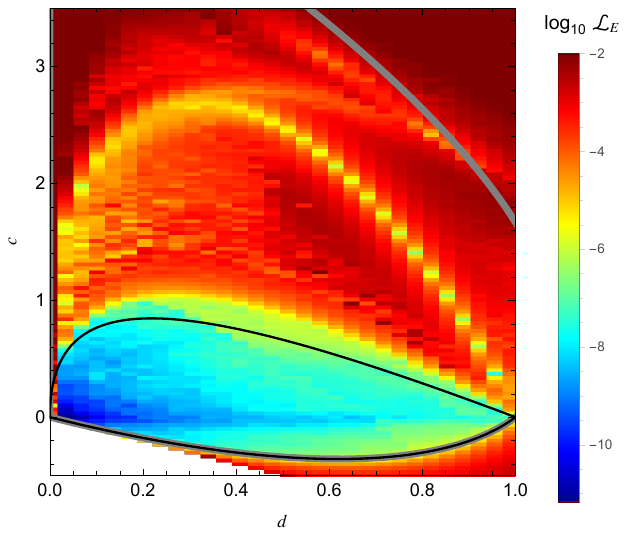}
         \caption{Base log loss landscape}
         \label{fig:landscapequad}
     \end{subfigure}         
     \hspace{5mm}
          \begin{subfigure}[b]{0.42\textwidth}
         \centering
         \includegraphics[width=\textwidth]{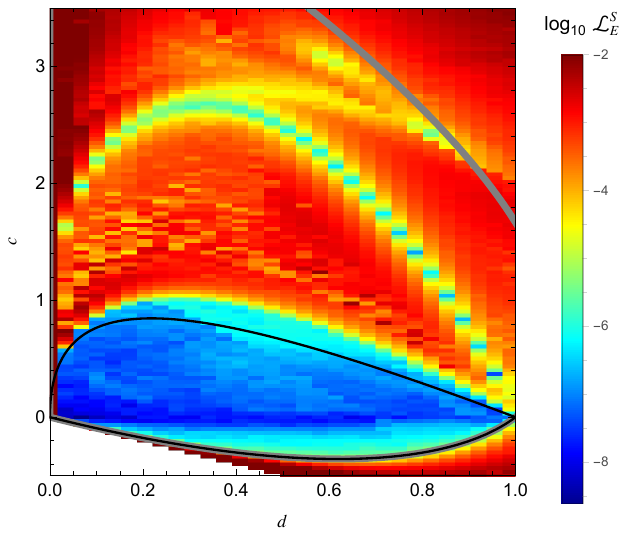}
         \caption{Scaled log loss landscape}
         \label{fig:quadlandscapebase}
     \end{subfigure}
     \\
          \centering
     \begin{subfigure}[b]{0.4\textwidth}
         \centering
         \includegraphics[width=\textwidth]{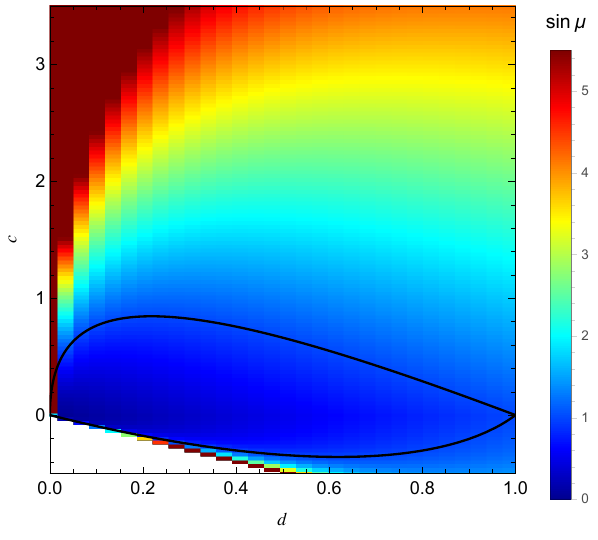}
         \caption{$\sin \mu$ landscape}
         \label{fig:sinmuquadlandscape}
     \end{subfigure}
          \hspace{5mm}
          \begin{subfigure}[b]{0.42\textwidth}
         \centering
         \includegraphics[width=\textwidth]{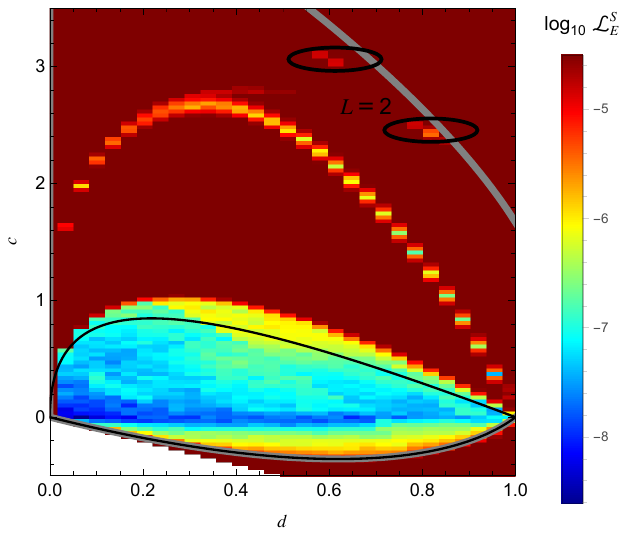}
         \caption{Cut on the scaled log loss landscape}
         \label{fig:landscapequadzoom}
     \end{subfigure}
     \caption{Loss landscapes for the two-parameter family of moduli $B(z)=c z^2+d$.  Black and grey curves follow Fig.~\ref{fig:linearfulllandscape}. For each $c$ and $d$ we find a phase $\phi(z)$. Red regions indicate that no solution is likely.
    Top panels show the base loss and scaled loss.  Panel (c) shows a heat map of $\sin\mu$ values over the family $B(z)=cz^2+d$. Right shows the scaled loss landscape with a hard cut of $\mathcal{L}_E^S \sim 10^{-4.5}$, along with the $L=2$ finite partial wave solutions circled in black. The 1D curve corresponds to finite $L>2$ solutions. 
    }      \label{fig:quadlosslandscapes}
\end{figure}

\subsubsection{Quadratic functions} \label{sec:scansquad}
Next, we consider the family of symmetric quadratic moduli with $B(z) = c + d z^2$. To keep $B(z)$ positive we restrict to $c>|d|$. We proceed in a similar fashion as in the previous section, constructing a grid of $45 \times 180$ points for $c\in [0, 1.5]$ and $d\in[-0.5,5.5]$, where we train a new neural network at each point for 2000 epochs.

In Figs.~\ref{fig:quadraticfulllandscape}-\ref{fig:quadlosslandscapes} we repeat the same plots as for the linear function: the loss landscapes for the base and scaled losses along with the $\sin\mu$ values and the $\mathcal{L}_E^S$ cut.
Within the $\sin \mu < 1$ region, where a solution is guaranteed, the loss is generally small and $\mathcal{L}_E^S < 10^{-5}$ is always satisfied. This is a strong sign that our network is able to properly find the expected solutions in this region. We notice however that the network still finds approximate numerical solutions with $\mathcal{L}_E^S \sim 10^{-5}$ up to $\sin \mu \sim 1.1 -1.2$.

In Fig.~\ref{fig:quadlosslandscapes} we can see two regions of low loss away from $\sin\mu<1$: two small islands circled in panel $d$ and an additional one-dimensional curve along which the loss is around $\mathcal{L}_E^S \sim 10^{-6} - 10^{-5}$. Both regions are significantly outside of the $\sin \mu = 1$ boundary. 

As shown in Appendix \ref{sec:appendixfinite} the two islands correspond to the genuine finite partial wave solutions of order $L=2$, which have quadratic differential cross sections. Although genuine solutions, their associated loss is around $\mathcal{L}_E^S \sim 10^{-5}$ as their corresponding $\sin \mu$ values are quite high, around $2.95$ and $3.67$ respectively.  Indeed, as $\sin \mu$ grows the networks become harder to train as can be understood from the structure of the loss function of Eq.~(\ref{eq:lossfunc_scaled}). For $\sin \mu >1$ the cosine term in the integral needs to precisely modulate the kernel function in order to have a value that can be matched with $|\sin \phi(z)| < 1$. Resolving these high $\sin \mu$ solutions to better accuracy will require using a higher number of training epochs and some additional fine-tuning of the neural network parameters. Alternatively, provided one knows that a finite partial wave solution is expected, it is also possible to first parametrize the unitarity amplitude as in Eq.(\ref{eq:partialwaves}). One can then fit the phase shifts $\delta_\ell$ by ensuring that modulus $|F(z)|$ matches the $B(z)$ given as input. In that context, one can go back and forth between the machine learning scans and the classical fitting algorithm in order to fully characterize the low loss landscape.

 Similarly, as described in Appendix \ref{sec:appendixfinite}, the 1D curve of $\sin\mu>1$ solutions is associated with finite partial wave solutions of order $L \neq 2$. Their corresponding moduli are numerically well approximated by a quadratic $B(z)$ in the $z \in [-1,1]$ region. Thus, even though their moduli are not quadratic per see, these spurious solutions show up in our scans and are associated with low loss values.

\subsection{Extremal amplitudes}
Up until now, we have considered only toy amplitudes where the modulus is a low order polynomial. Here we demonstrate that the same technique can also be applied when using differential cross sections obtained in the context of the nonperturbative $S$-matrix bootstrap program, see e.g. \cite{Paulos2019, Chen2022, EliasMiro:2022xaa}. 
One class of amplitudes of interest are  ``extremal'' amplitudes that maximize or minimize the value of the amplitude at the crossing-symmetric point $\lambda = \frac{1} {32 \pi} T(\frac{4m^2}{3}, \frac{4m^2}{3})$, which intuitively measures the strength of the interaction between particles. By combining the so-called primal and dual methods, see \cite{Guerrieri:2021tak}, the maximal value of the coupling was obtained to be
\begin{equation}
2.66 \leq \max \lambda \leq 2.73    ,
\end{equation}
whereas for the minimal coupling \cite{EliasMiro:2022xaa}, the current bound is
\begin{equation}
-8.02 \leq \min \lambda \leq -7.0 .
\end{equation}
It was also found that when maximizing/minimizing various couplings, while elastic unitarity was not imposed it effectively emerged with a good precision in the extremization process. This structure was further explored in \cite{Tourkine:2023xtu}. In the context of the present work, it is therefore interesting to consider the differential cross-sections produced by the extremal amplitudes and check that elastic unitarity can indeed be satisfied by finding the appropriate $\phi(z)$.

To analyze this case in more detail, we take the numerical results from~\cite{Chen2022} and use them to compute $B(z)$ that we then use as input. The functions $B(z)$ computed in this way are more physical than our toy functions in that they  come from amplitudes that satisfy analyticity, unitarity and crossing at all energies. Due to the fact that they describe the scattering of identical particles, they obey $B(z)=B(-z)$. This symmetry is also respected by the expected phase, and we force our networks to output a symmetric result by considering $[\phi_\sigma(z) + \phi_\sigma(-z)]/2$ as the final output. In order to expedite the numerical evaluations of the input $B(z)$ functions, we will fit them by symmetric polynomials of finite order.

\begin{figure}[hbt!]
     \centering
     \begin{subfigure}[b]{0.45\textwidth}
         \centering
         \includegraphics[width=\textwidth]{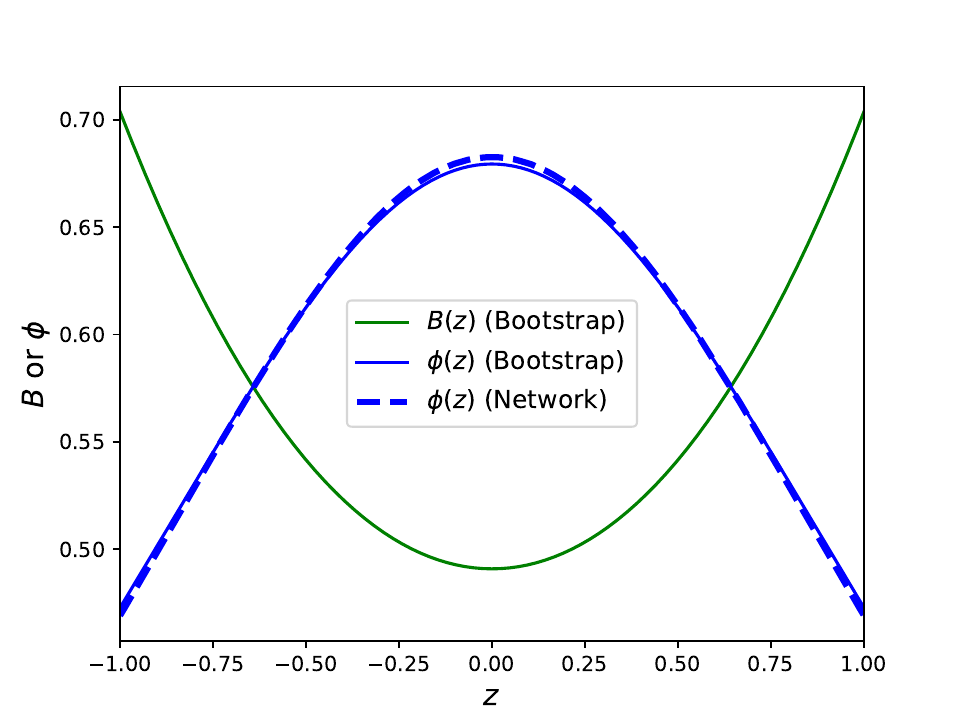}
         \caption{Amplitudes at $s=4+1/2 m^2$}
         \label{fig:s4halfamp}
     \end{subfigure}
  \hspace{5mm}
          \begin{subfigure}[b]{0.45\textwidth}
         \centering
         \includegraphics[width=\textwidth]{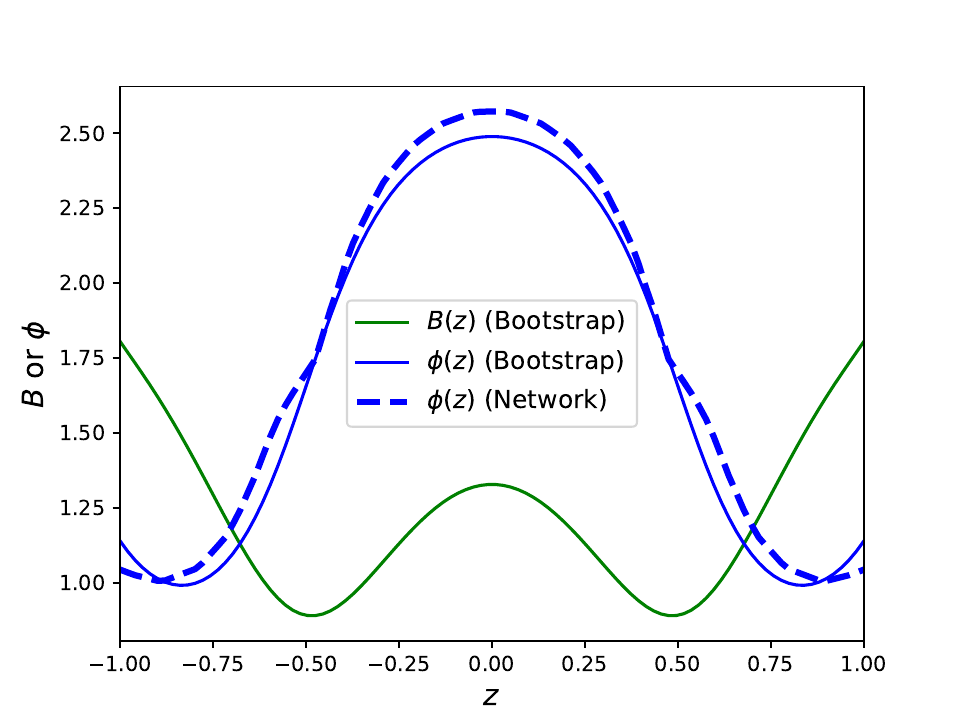}
         \caption{Amplitudes at $s=8 m^2$}
         \label{fig:s8amp}
     \end{subfigure}
     \\
     \begin{subfigure}[b]{0.45\textwidth}
         \centering
         \includegraphics[width=\textwidth]{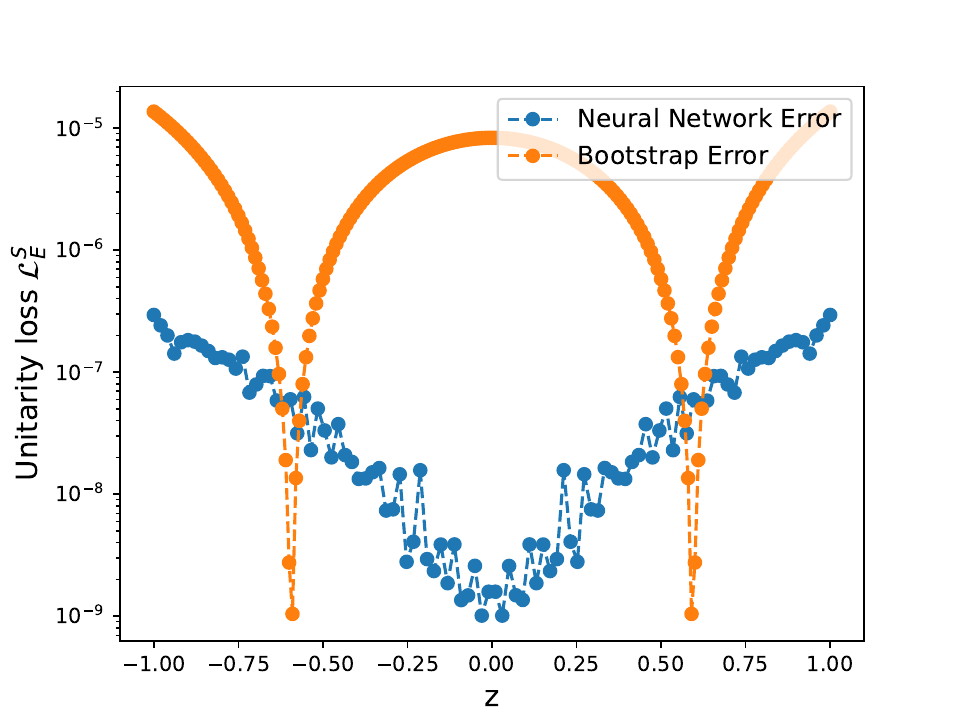}
         \caption{Unitarity loss at $s=4+1/2 m^2$}
         \label{fig:s4halfunitarity}
     \end{subfigure}
  \hspace{5mm}
          \begin{subfigure}[b]{0.45\textwidth}
         \centering
         \includegraphics[width=\textwidth]{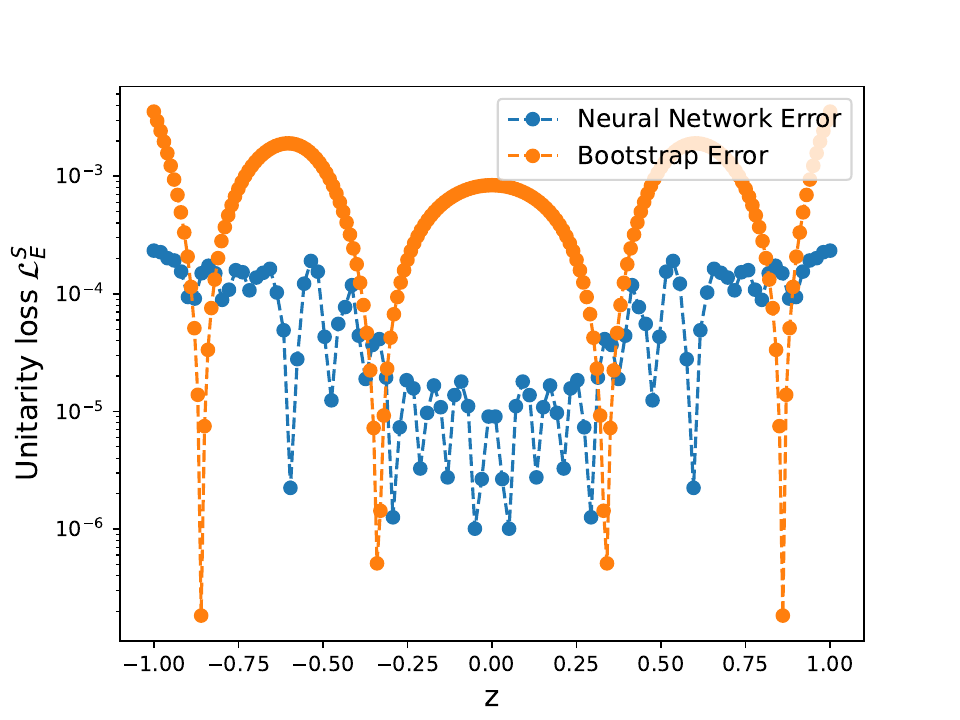}
         \caption{Unitarity loss at $s=8 m^2$}
         \label{fig:s8unitarity}
     \end{subfigure}
     \caption{(Top pannels) $B(z)$ moduli and associated $\phi(z)$  phases coming from the bootstrap program of \cite{Chen2022} for the amplitude with $\lambda \approx - 6.48$ and $N_{\text{max}}=26$. We extract $B(z)$ at different $s$ energies and use a neural network to predict the associated phase. (Bottom panels) Evaluation of the unitarity loss of Eq.~(\ref{eq:lossfunc_scaled}) on the predicted amplitudes coming from the bootstrap program and the trained neural networks.}\label{fig:boostrap_min}
\end{figure}

We find that more interesting results arise from the amplitudes that are closer to minimizing the coupling. Using the primal bootstrap results of \cite{Chen2022} for $N_{\text{max}}=26$, we get $\lambda \approx - 6.48$.  We choose energies to be $s= 4.5 m^2$ and $s=8 m^2$ and extract the appropriate $B(z)$ functions. Those are fitted by symmetric polynomials of order 14 and 40 respectively, which are then fed as inputs to our networks. Training is done over 5000 epochs, minimizing the scaled unitarity loss and using the default architecture and hyperparameters. The input $B(z)$ and associated phases are plotted on the top panels of Fig.~\ref{fig:boostrap_min}, where we also include the phase predictions from the bootstrap amplitudes. At $s=4.5 m^2$ the modulus has $\sin \mu=0.64$, while at $s=8 m^2$ the modulus is associated with $\sin \mu=1.64$.

To verify the accuracy of our results we also plot the evaluation of the unitarity loss of Eq.~\eqref{eq:lossfunc_scaled} on the bottom panels of Fig~\ref{fig:boostrap_min}. This is done for both the trained networks and the bootstrap predictions and allows us to verify to which extent unitarity is broken in both cases. We can immediately notice that at $s=4.5m^2$ the bootstrap and the neural network phases agree with one another and that in both cases unitarity is well respected. The unitarity loss is lower for the phase coming from the neural network as it has been specifically trained to minimize this quantity, while in the case of the bootstrap program elastic unitarity is only an emergent property. At $s=8m^2$ the difference between the neural network and bootstrap phases is more pronounced. Unitarity is not respected with very good accuracy, in particular for the bootstrap phase. At the $z=\pm 1$ edges the loss reaches the order of $10^{-2}$ and indicates that numerically the emergent elastic unitarity property does not hold accurately.\footnote{This is further confirmed by numerically calculating $K(z)$ of Eq.~(\ref{kernelkz}). For unitarity to hold we must have $\sin \phi(z) < K(z)$, which is respected by the neural network prediction. However, for the bootstrap phase, this is broken near the $z=\pm 1$ edge of the angle range.}

\noindent

\section{Phase ambiguities: finite partial waves \label{sec:finiteL}}
So far we considered the existence question: given a modulus $B(z)$ when does there exist a phase $\phi(z)$ so that the amplitude $F(z) = B(z) e^{i \phi(z)}$ is unitary? A related question is: when are there two possible phases for the same $B(z)$? These phases should not be related by the redefinition, $\phi(z) \rightarrow \pi - \phi(z)$  (such redefinition is called a trivial ambiguity). The value of $\sin\mu$ is often considered as an indicator of whether ambiguous solutions may exist. 
The best bound in the literature~\cite{Gangal1984} guarantees uniqueness for $\sin \mu < 0.89$, and it has been conjectured~\cite{Martin1969} that uniqueness should hold up to $\sin \mu =1$. In practice, however, most known examples have $\sin \mu$ values much higher (above 2.0) and require a dedicated construction.  In the following sections, we will see how machine learning can be used in conjunction with classical algorithms in order to study these ambiguous solutions. 
In this section, we focus on finite partial wave phase ambiguities and consider the infinite partial wave case in Section~\ref{sec:infiniteL}.

In the finite $L$ case, we both review known results and then discuss how machine learning can help. For finite $L$, the first phase-ambiguous solution was found by Crichton in 1966 and has $L=2$, so the amplitude is quadratic in $z$. As we will see, for finite $L$ there are an infinite number of phase-ambiguous solutions which decompose into 1d curves in the space of phase-shifts. The low dimensionality of the solution space makes the machine learning approach challenging. In the infinite $L$ case, the solution space is higher-dimensional and easier to explore with gradient descent.

\subsection{Classical solutions} \label{sec:classicalfiniteL}
We first review what is known about the finite $L$ phase-ambiguous solutions classically, i.e. without machine learning.
When there are a finite number of partial waves in an amplitude, the question
of whether there are multiple phases for the same amplitude reduces to whether
a finite set of equations can be solved simultaneously. For finite $L$ we write the
amplitude as
\begin{equation}
  F (z) = \sum_{\ell = 0}^L (2 \ell + 1) 
  e^{i \delta_{\ell}} \sin (\delta_{\ell})
  P_{\ell} (z)
\end{equation}
where $P_{\ell} (z)$ are Legendre polynomials and
$\delta_{\ell} \in \mathbb{R}$ are the phase shifts. 
This
parameterization in terms of real phases guarantees that the amplitude is
unitary. We are looking for another amplitude
\begin{equation}
  \Fc(z) = \sum_{\ell = 0}^L (2 \ell + 1) 
  e^{i \deltac_{\ell}} \sin (\deltac_{\ell})
  P_{\ell} (z)
\end{equation}
with the same norm as $F(z)$. We are interested in non-trivial ambiguities.

To find non-trivial ambiguities we need two sets of phases $\delta_{\ell}$ and
$\deltac_{\ell}$ not all equal (and not all opposite) for which $B(z)=| F(z) |^2=|\Fc(z)|^2$ is
the same. Since $P_{\ell} (z)$ is a polynomial of degree $\ell$, $| F(z)|^2$ is
a polynomial of degree $2L$. So setting the coefficients of $z^j$ from $|F(z)|^2$  equal to those of $|\Fc(z)|^2$ gives
$2 L + 1$ equations for the $2 L+2$ real phase shifts $\delta_{\ell}$ and
$\deltac_{\ell}$. This generically leads to a 1-dimensional solution space.
Indeed, the finite $L$ solutions for every $L$ correspond to a set of 1D
curves in $2 L + 1$ dimensions.

The term associated with $z^{2 L}$ in $|F(z) |^2$ is $(2 L + 1)^2 | P_L (z) |^2
\sin^2 \delta_L$. For this to be the same with $\delta_{\ell}$ and
$\deltac_{\ell}$ requires $\delta_L = \deltac_L$ so that the highest phase
shift for the two solutions must be equal. There may or may not be one of the
1D curves which has a given value of $\delta_L$. However, if we find a point on
one of these curves with a given $\delta_L$ we can then move along the curve
unambiguously to determine all the other connected solutions.

For example, with $L = 1$ equating the expression for $|F(z)|^2$ using the two sets of phase shifts gives 
\begin{align}
  | F (z) |^2 &= 
  \sin^2 \delta_0 - 6 z \cos (\delta_0 - \delta_1) \sin \delta_0
  \sin \delta_1 + 9 z^2 \sin \delta_1^2\\
  &=
  \sin^2 \deltac_0 - 6 z \cos (\deltac_0 - \deltac_1) \sin \deltac_0
  \sin \deltac_1 + 9 z^2 \sin {\deltac_1}^2
\end{align}
Matching the coefficients of $z^0$, $z^1$ and $z^2$ gives $2L+1=3$ equations for the 4 phase shifts $\delta_0,\delta_1,\deltac_0$ and $\deltac_1$.
The $z^2$ term forces $\delta_1=\deltac_1$ and the $z^0$ forces $\delta_0 = \deltac_0$ so that
there are no nontrivial solutions with $L = 1$.

The $L = 2$ case is already fairly complicated. A non-trivial solution with $L=2$ was found by Crichton~\cite{Crichton1966}:
\begin{equation}
    \text{Set 1 : } \left\{\begin{array}{ccc}
        \delta_0 &=& -\frac{ 7}{54}\pi \\[0.5em]
        \delta_1 &=& -\frac{869}{3600} \pi \\[0.5em]
        \delta_2 &=& \frac{1}{9} \pi
    \end{array}\right. \qquad \text{and} \qquad 
    \text{Set 2 : } \left\{\begin{array}{ccc}
        {\deltac}_0 &=& \frac{659}{1200} \pi \\[0.5em]
        {\deltac}_1 &=& -\frac{59}{400}\pi \\[0.5em]
        {\deltac}_2 &=& \frac{1}{9} \pi 
    \end{array}\right.
\end{equation}
that give rise to two amplitudes $F(z)$ and $\Fc(z)$ which share the same differential cross section.

\begin{figure}[t]
    \centering
    \hfill
        \includegraphics[width=0.45\textwidth]{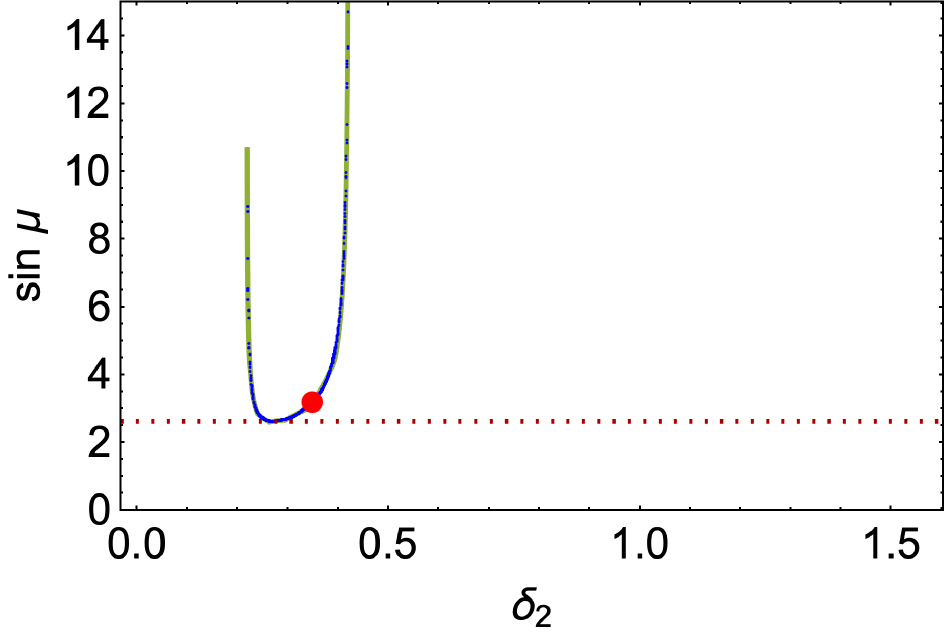}
    \hfill
        \includegraphics[width=0.45\textwidth]{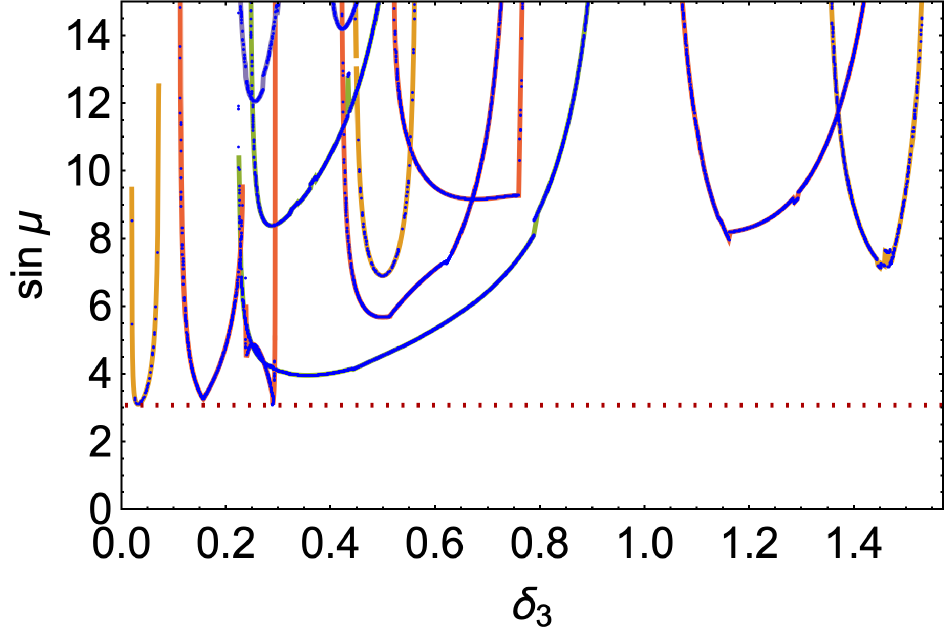}
    \hfill
        \\
    \hfill
        \includegraphics[width=0.45\textwidth]{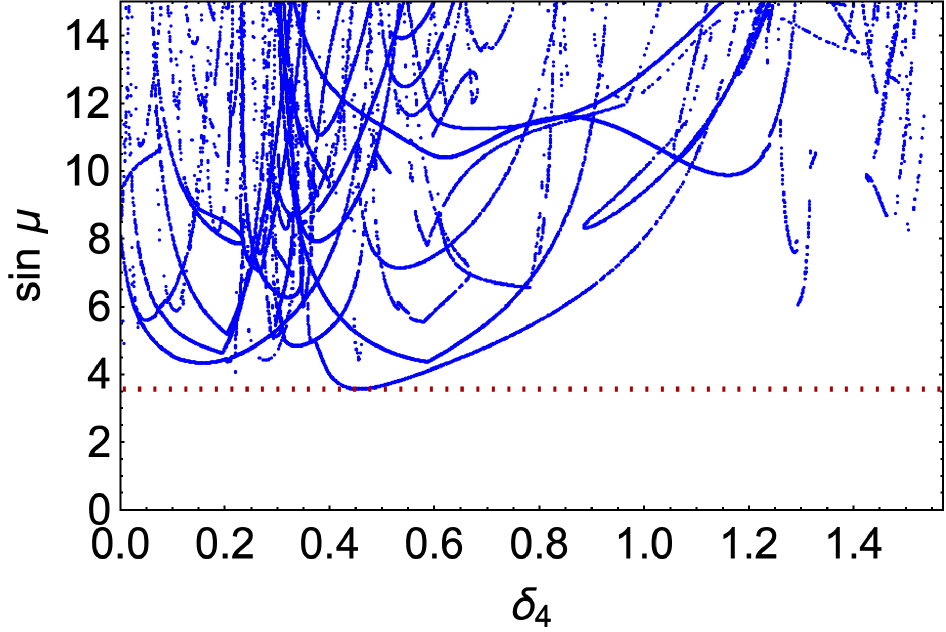}
    \hfill
        \includegraphics[width=0.45\textwidth]{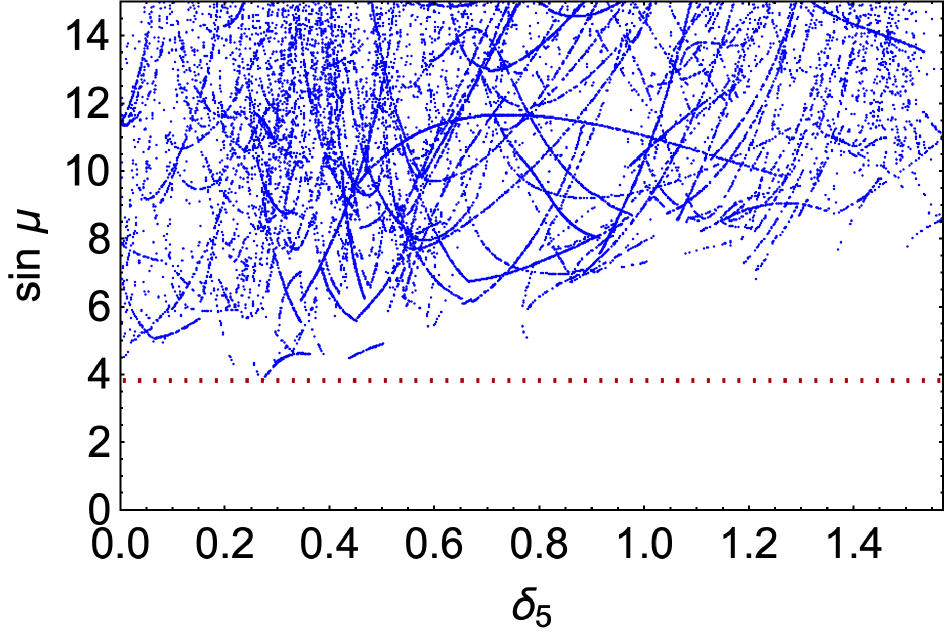}
    \hfill
     \caption{Finite partial-wave amplitudes with phase ambiguities separate into non-intersecting 1D curves in the $L+1$ dimensional space of phase shifts. For $L=2$ there is only one curve. These plots show a projection of these curves into the $\sin\mu,\delta_L$ plane for $L=2,3,4,5$. The point on the $L=2$ plot is Crichton's original ambiguity with $\delta_2 = \frac{\pi}{9}$. The solid curves are analytic solutions when known. The blue dots are a random scan. The red line gives the minimum $\sin\mu$ in each case.
     \label{fig:L234}
     }
\end{figure} 
Shortly after Crichton's paper, Atkinson, Johnson, Mehta and de Roo found the complete set of $L=2$ solutions~\cite{ATKINSON1973125}. These form a 1D curve in the space of phase shifts, as expected. The value of $\sin\mu$ for these solutions is shown in Fig~\ref{fig:L234} with Crichton's point indicated. Crichton's solution has $\sin\mu=3.2$. For $L=3$ the complete space of solutions is also known. For $L=4$ only a handful of solutions are known. 

Rather than attempting to improve these analytic results, we simply take a brute-force approach to finding solutions. To do so we first pick a random $\delta_L = \deltac_L$. Then we search for a solution to the remaining $2 L$ equations and
$2 L$ unknows close to random seed points for the other $\delta_{\ell}$ and
$\deltac_{\ell}$. Once the equations are solved, we then confirm that the
solutions are not trivially related. By sampling enough points one can see the emergence of a set of curves (one can also move along the curves to find connected solutions if desired). 

Results for $L=2,3,4$ and $5$ are shown in Fig.~\ref{fig:L234}. 
Interestingly we observe that the minimum value of $\sin \mu$ with $\delta_L \ne 0$
does not seem to decrease with $L$.\footnote{We have tentatively explored up to $L=10$. For $L>4$ there are so many curves that an enormous number of samples would be required to resolve them.}
For $L=2$, the lowest value has $\sin\mu \approx 2.63$. For $L=3$ it is $\sin\mu \approx 3.41$. The lowest value for $L=3$ is when $\delta_3=0$ which reduces it to $L=2$. Such points do not show up in our search since they must have  $\delta_L=0$ {\it{exactly}} which will never occur in a random scan.  For $L=4$ and $L=5$, the lowest values we found with $\delta_L \ne 0$ are $\sin\mu \approx 3.58$ and $\sin\mu \approx 3.83$ respectively. Based on these observations, we do not believe that going to higher finite $L$ will reveal ambiguous solutions with $\sin\mu$ values smaller than those from $L=2$.

\subsection{Machine learning with repulsive loss}\label{sec:repulsiveloss}
One might hope to use machine learning to find lower values of $\sin \mu$ for finite $L$. This becomes difficult because the solution space is one-dimensional. Thus, if one starts on a particular curve and does gradient descent in $\sin \mu$, one will only find the local minimum of that curve and never be able to jump to other curves. Fortunately, this is only a problem for finite $L$. Ambiguous solutions with infinite $L$ fill higher dimensional regions which are easier to explore. The finite $L$ case is still useful for exploring the machine learning approach as we have some exact phase ambiguous solutions, such as the Crichton one. So we will use finite $L$ as a testbed for constructing a neural network capable of finding phase ambiguities. The lessons learned from these examples can then be applied to the more promising infinite $L$ case.

\begin{figure}[t]
     \centering
     \begin{subfigure}[b]{0.45\textwidth}
         \centering
         \includegraphics[width=\textwidth]{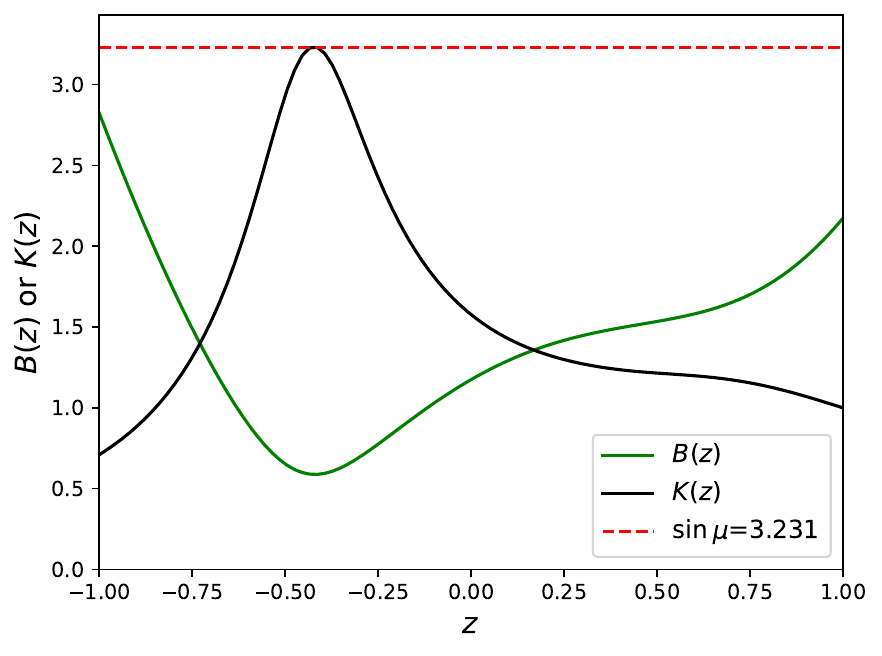}
         \caption{Crichton's modulus and integrated kernel}
         \label{fig:crichtonmod}
     \end{subfigure}
          \hfill
          \begin{subfigure}[b]{0.45\textwidth}
         \centering
         \includegraphics[width=\textwidth]{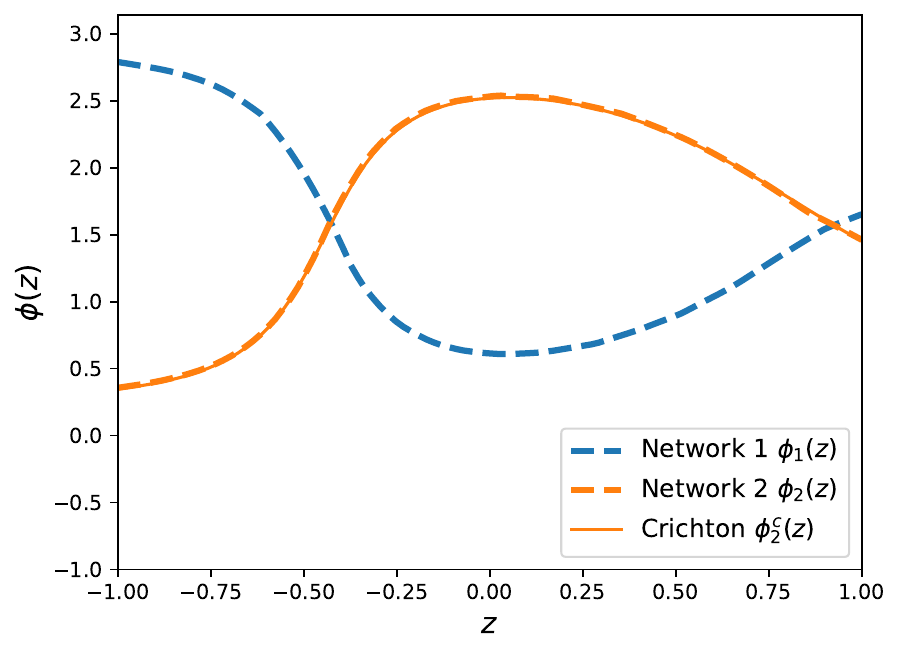}
         \caption{Redundant neural network phase solutions}
         \label{fig:crichtonredphases}
     \end{subfigure}
     \caption{Crichton ambiguity: naively training using two independently initialized neural networks. On the right panel, we notice that the two phases are trivially ambiguous.}
\end{figure}
We first attempt to recover both of Crichton's $L=2$ solutions. To begin, we simply try to find the phase multiple times and hope to get different answers based on different initialization seeds. We fix $B(z)=|F(z)|$ from Crichton's solution, as shown in Fig~\ref{fig:crichtonmod}. We first simply define two independent neural networks ($\phi_1(z)$ $\phi_2(z)$) and train them according to the principles of Section~\ref{sec:singleamplitude} with different random initializations. 
We let the networks run for 5000 epochs using the loss of Eq.~\eqref{eq:lossfunc}. We do find two phases this way, as displayed in Fig.~\ref{fig:crichtonredphases} alongside the theoretical solution coming from using the second set of phase shifts ${\deltac}$. 
Unfortunately, the phases are trivially related: $\phi_1(z) = \pi - \phi_2(z)$. As we randomly initialize new neural networks we can end up recovering either solution (or the second Crichton solution). What is apparent from this simple experiment is that we need a way to avoid trivial ambiguities. 

In order to study the uniqueness property associated with a given differential cross section we devise a methodology for consistently recovering different solutions with our neural networks. The setup is almost identical to the one used in Section~\ref{sec:singleamplitude}, where we start by independently initializing various neural networks that aim to solve the unitarity equation, minimizing the losses of either Eq.~(\ref{eq:lossfunc}) or Eq.~(\ref{eq:lossfunc_scaled}). The main deviation from this simple setup follows previous work in the literature~\cite{DiGiovanni2020FindingMS} and consists in introducing a new repulsive term in the loss function. The role of this term will be to push apart the various neural network solutions and ensure that they do not overlap. 

To introduce the repulsion term we must first define a measure for the closeness of two solutions $\phi_1(z)$ and $\phi_2(z)$. This measure should account for the periodicity properties of the phases to avoid boundary effects. One choice is to first define
\begin{equation}
    d(\phi_1, \phi_2) =  \mathbb{E}_z \left|\left|\big[\cos\phi_1(z) - \cos\phi_2(z) \big]^2 + \big[\sin\phi_1(z) - \sin\phi_2(z) \big]^2   \right|\right| 
\end{equation}
as a distance between two phase solutions. 

For  the repulsive loss itself, we will consider two  alternatives. The first one, the \textit{kick repulsion}, follows~\cite{DiGiovanni2020FindingMS} and consists in introducing the pairwise loss 
\begin{equation}\label{eq:lossrepulsivekick}
    \mathcal{L}_{R1}^{(1,2)} =\big[ d(\phi_1, \phi_2)\big]^{-p} + \big[ d(\phi_1, \pi  - \phi_2)\big]^{-p}
\end{equation}
where $p$ is some hyperparameter to be fixed. The first term in this loss ensures that we do not get exactly the same solutions, while the second term pushes us away from the trivial ambiguity. Since this repulsive term is always non-null, it will only be included in the loss function at intermediate epochs. More precisely, we let the networks first train for $e_i$ epochs, then activate the repulsion and then turn it off after a total of $e_f$ epochs. In that way, this repulsive term in the loss function acts like a kick that pushes the two solutions apart but doesn't prevent the network from achieving arbitrary low loss even when the solutions are similar.

At a given epoch $t$ the full loss function for $N$ networks now reads
\begin{equation}\label{eq:lossrepulsivekicktot}
\mathcal{L}_K = 
\begin{cases}
\sum\limits_i^N \mathcal{L}_E^{(i)}  + \lambda_R \sum\limits_{i<j}^N \mathcal{L}_{R1}^{(i,j)} & \text{for }   e_i<t<e_f  \\
\sum\limits_i^N \mathcal{L}_E^{(i)}   & \text{for } t<e_i \quad \text{or } t>e_f 
\end{cases}
\end{equation}
where $\lambda_R$ is another hyperparameter defining the relative repulsion strength and $\mathcal{L}_E$ is the unitarity loss. At the end of the training run, we can then evaluate the repulsion term to check if the two solutions found are indeed not identical or correspond to the trivial ambiguity.

\begin{table}[t]
    \centering
    \begin{small}
\begin{tabular}{l l c }
    \toprule
Repulsive loss type &Parameter& Parameter description  \\
\midrule
 \multirow{3}{*}{Kick repulsion} & $p$ & Strength of the inverse power law repulsion\\
 &$e_i$, $e_f$ & Start and end epochs where the repulsion is active\\
 & $\lambda_R$ & Repulsion loss to unitarity loss relative strength \\
\midrule
\multirow{4}{*}{Decaying repulsion} & $c_0$ & Initial repulsive factor\\
& $s_f$ & Total scaling of the repulsive factor\\
& $e_f$ & Epoch timescale where the repulsive factor grows\\
 & $\lambda_R$ & Repulsion loss to unitarity loss relative strength \\
  \bottomrule
\end{tabular}
\end{small}
    \caption{Hyperparameters to be tuned for the different repulsive losses considered}
   \label{tab:parametersrepulsion}
  \end{table}
  
The second  loss that we considered is a \textit{decaying repulsion}. It consists of an interaction term that is not singular even when the solutions overlap. This is done by adding the loss term
\begin{equation}\label{eq:decayingrepulsion}
    \mathcal{L}_{R2}^{(1,2)} = 2 - \tanh[c(t) d(\phi_1, \phi_2)] - \tanh[c(t) d(\phi_1, \pi - \phi_2)]
\end{equation}
where $c(t)$ is some hyperparameter. In order to precisely fit the different solutions we want the repulsion to be inconsequential as the training nears the end. This is achieved by making the parameter $c(t)$ epoch dependent, increasing throughout the training. In particular, we will take 
\begin{equation}
    c(t) = \Delta_1 \tanh\left(\frac{t}{a}-b\right) + \Delta_2
\end{equation}
where the parameters are chosen such that $c(t)$ starts at $c(0)=c_0$ and reaches 99$\%$ of its maximal final value $c_f$ after $e_f$ epochs, with $c(e_f) =  0.99 c_0 s_f$ where $s_f$ is the total scale factor\footnote{Additionally we ask for $c'(t)$ to be maximum at $e_f/2$. These constraints fix the parameters of $c(t)$ to be $a=e_f/(2 \arctanh{[(100-99s_f)/(100-101 s_f)]})$, $b=e_f/(2a)$, $\Delta_1 = c_0(100+99s_f)/200$ and $\Delta_2 = c_0(-100+101s_f)/200$.}. At a given epoch $t$ the full loss function for $N$ networks now reads 
\begin{equation}\label{eq:decayingrepulsiontot}
   \mathcal{L}_D = \sum_i^N \mathcal{L}_E^{(i)}  + \lambda_R \sum_{i<j}^N \mathcal{L}_{R2}^{(i,j)} 
\end{equation}
where the repulsive loss term is always included\footnote{In practice when the repulsive loss starts to be smaller than a tenth of the unitarity loss we will discard it to facilitate training. This check is to be performed throughout training and the repulsive loss can be reactivated as soon as we breach that threshold.}. We summarize the hyperparameters for the two types of repulsive losses in Table~\ref{tab:parametersrepulsion}.

\begin{figure}[t]
     \centering
     \begin{subfigure}[b]{0.45\textwidth}
         \centering
         \includegraphics[width=\textwidth]{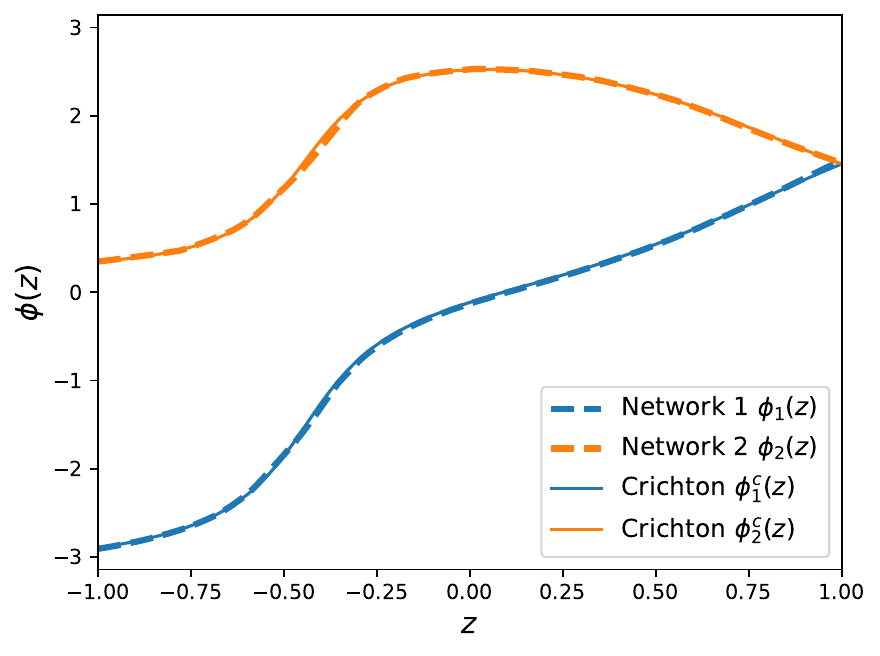}
         \caption{Learned neural network phases}
         \label{fig:crichtonambiguity}
     \end{subfigure}
          \hfill
          \begin{subfigure}[b]{0.45\textwidth}
         \centering
         \includegraphics[width=\textwidth]{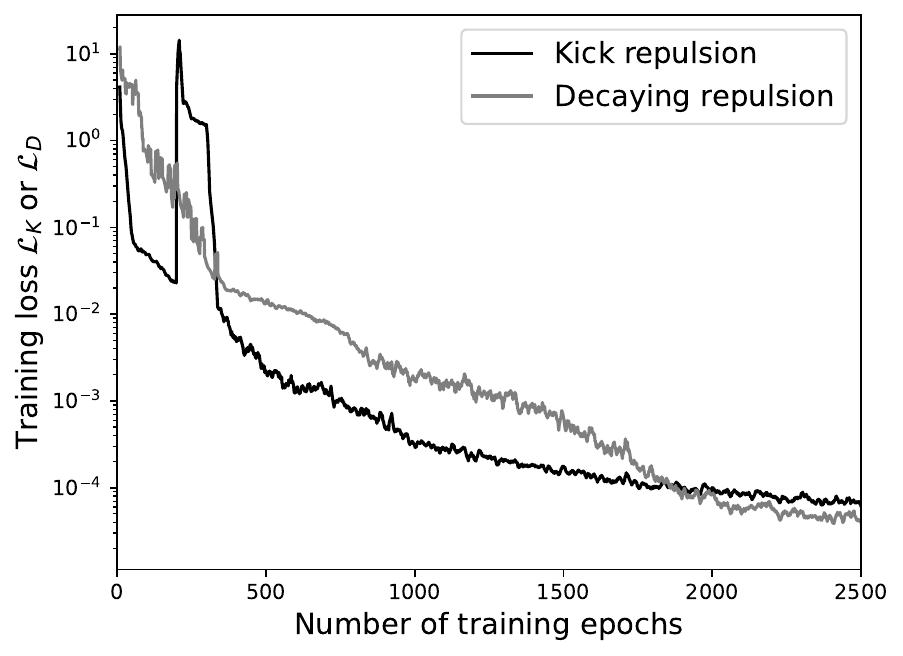}
         \caption{Loss functions for the different repulsion types}
         \label{fig:crichtonlosses}
     \end{subfigure}
     \caption{Recovering the Crichton ambiguity with the addition of a repulsive loss. On the left panel, we show the two distinct ambiguous phases recovered by our networks. On the right panel, we compare the evolution of the two repulsive losses introduced in Eq.~(\ref{eq:lossrepulsivekicktot}) and Eq.~(\ref{eq:decayingrepulsiontot}).}
\end{figure}

Adding the repulsive loss $\mathcal{L}_{R1}$, we train a neural network for 10000 epochs using the unitarity loss of Eq.~(\ref{eq:lossfunc}). After a quick search of the hyperparameter space, we use $p=2$, $e_i=200$, $e_f=300$ and $\lambda_R=1.0$, although other values could be considered with additional fine-tuning. The phases recovered by the networks are shown in Fig.~\ref{fig:crichtonambiguity} and are plotted against Crichton's prediction of the finite partial wave parameterization. The final losses for each phase are around $\mathcal{L}_E^S \sim 10^{-5}$, in good agreement with the expected answer. We note that for the solid blue curve in Fig.~\ref{fig:crichtonambiguity} we represent the phase coming from using Crichton's phase shifts $-\delta_l$, which corresponds to representing the trivial ambiguity associated with Crichton's first solution.

We also train another neural network with the  $\mathcal{L}_{R2}$ repulsive loss using $c_0=2, s_f=16, e_f=1000$ and $\lambda_R=50.0$. The resulting phases also have a final loss around $\mathcal{L}_E^S \sim 10^{-5}$. We compare the loss functions using the kick and decaying repulsion in Fig.~\ref{fig:crichtonlosses}, where we use a moving average window of 10 epochs to smooth the plot. Whereas the decaying repulsion has a smoothly decreasing loss we can indeed verify that the kick repulsion sharply boosts the loss across a given window.

\section{Infinite partial wave ambiguities} \label{sec:infiniteL}
Looking for amplitudes with ambiguous phases that have an infinite number of partial waves has also been explored classically. We first review some results in this direction, then apply machine learning to the infinite $L$ case.

\subsection{Classical solutions} \label{sec:infiniteLclassical}
One approach to  finding phase ambiguities with infinite $L$ is based on partially factorizing the amplitude and conjugating some of its zeros. The following discussion follows~\cite{Atkinsonambiguity}.

Any amplitude can be written (non-uniquely) as
\begin{equation}
\label{Fzform}
  F (z) = g (z) \prod_{\ell = 1}^L \frac{z - z_{\ell}}{1 - z_{\ell}}
\end{equation}
for some $L$ and some $g (z)$. When $g (z) = 1$ the amplitude is a polynomial
with a finite number of phase shifts. For $L=0$, this is just $F(z)=g(z)$ an arbitrary function.
This form is still useful, since if we conjugate any number of $z_{\ell}$
the amplitude will have the
same norm. In particular, the amplitude
\begin{equation}
  \Fc(z) = g (z) \prod_{\ell = 1}^L \frac{z - z_\ell^{\star}}{1 -
  z_\ell^{\star}}
\end{equation}
has $|F(z)|=|\Fc(z)|$. Constructing amplitudes in this way guarantees
they have the same norm but does not guarantee unitarity. 
And moreover, even if $F(z)$ is unitary, $\Fc(z)$ generally will not be.

In order to make progress, one needs to restrict the
class of functions searched.  Ref.~\cite{Atkinsonambiguity} focused exclusively on the simplest case where two amplitudes differ by a single zero. Doing a
partial wave decomposition of $g (z)$ we can write
\begin{align}
  F(z) &= \frac{z - z_1}{1 - z_1} \sum_{\ell = 0}^{\infty} (2 \ell + 1) \left(
  \frac{\gamma_{\ell} - 1}{2 i} \right) P_{\ell} (z) \nonumber\\
    \Fc(z) &= \frac{z - z_1^\star}{1 - z_1^\star} \sum_{\ell = 0}^{\infty} (2 \ell + 1) \left(
  \frac{\gamma_{\ell} - 1}{2 i} \right) P_{\ell} (z) =
  \frac{z - z_1^\star}{z-z_1}\frac{1 - z_1}{1 - z_1^\star} F(z)
  \label{onezero}
\end{align}
This is similar to a normal partial wave decomposition
\begin{equation}
  F(z) = \sum_{\ell = 0}^{\infty} (2 \ell + 1) \left( \frac{S_{\ell} - 1}{2
  i} \right) P_{\ell} (z)
\end{equation}
where $S_{\ell} = e^{2 i \delta_{\ell}}$ so that the unitarity condition
$\delta_{\ell} \in \mathbb{R}$ is equivalent to $| S_{\ell} | = 1$. Because of
the $\frac{z - z_1}{1 - z_1}$ prefactor the unitarity condition is not 
$|\gamma_{\ell} | = 1$ but rather $| S_{\ell} | = 1$. Solving for the
$S_{\ell}$ in terms of the $\gamma_{\ell}$ gives
\begin{equation}
  S_{\ell} = \frac{1}{1 - z_1} \left[ \frac{(\ell + 1) \gamma_{\ell + 1} +
  \ell \gamma_{\ell - 1}}{2 \ell + 1} - z_1 \gamma_{\ell} \right]
  \label{Sldef}
\end{equation}
The condition that $| S_{\ell} | = 1$ then gives a recursion relation among the
$\gamma_{\ell}$. Writing
\begin{equation}
    \gamma_{\ell} = 1 - \epsilon_{\ell}
\end{equation}
this relation
can be written in descending form:
\begin{equation}
  \epsilon_{\ell - 1} = \frac{2 \ell + 1}{\ell} \epsilon_{\ell} \tmop{Re}
  (z_1) - \frac{\ell + 1}{\ell} \epsilon_{\ell + 1} + \frac{2 \ell + 1}{\ell}
  (1 - \tmop{Re} z_1) \left[ 1 \pm \sqrt{1 + \frac{(\tmop{Im}
  z_1)^2}{(\tmop{Re} z_1 - 1)^2} \epsilon_{\ell} (2 - \epsilon_{\ell})}
  \right]
  \label{descend}
\end{equation}
or equivalently in ascending form:
\begin{equation}
  \epsilon_{\ell + 1} = - \frac{\ell}{\ell + 1} \epsilon_{\ell - 1} + \frac{2
  \ell + 1}{\ell + 1} \epsilon_{\ell} \tmop{Re} (z_1) + \frac{2 \ell + 1}{\ell
  + 1} (1 - \tmop{Re} z_1) \left[ 1 \pm \sqrt{1 + \frac{(\tmop{Im}
  z_1)^2}{(\tmop{Re} z_1 - 1)^2} \epsilon_{\ell} (2 - \epsilon_{\ell})}
  \right]
\end{equation}
Note the sign ambiguity: there are generally two solutions at each step
leading to $2^n$ sign choices. Generally, only one of them will give finite
amplitudes, with $\varepsilon_{\ell} \rightarrow 0$ as $\ell \rightarrow
\infty$. There is nevertheless no clear criterion for deciding which sign
to choose at each recursion step.

As discussed by \cite{Atkinsonambiguity} an additional necessary condition for the unitarity of
$F(z)$ following from $| S_{\ell} | = 1$ and finiteness is that
\begin{equation}
  \tmop{Im} (\gamma_{\ell}^{\star} \gamma_{\ell - 1}) = 0
\end{equation}
This means that each pair of successive $\gamma_{\ell}$ must have the same
phase. The phase can only change if $\gamma_L = 0$ for some $L$, in which case
the $\gamma_{L - 1}$ and $\gamma_{L + 1}$ can have different phases. So
overall the series of $\gamma_{\ell}$ comprises sequences with the same phase separated by zeros. Moreover,
for the amplitude to be finite $\gamma_{\ell} \rightarrow 1$ as $\ell
\rightarrow \infty$, which means that there has to be some $L$ beyond which all
of the $\gamma_{\ell}$ are real. Since some $\gamma_{\ell}$ has to be complex
(or else $F(z)$ and $\Fc(z)$ are complex conjugates), we conclude that
$\gamma_{L - 1} = 0$ for some $L$ and $\gamma_{\ell} \in \mathbb{R}$ for $\ell
\geqslant L$. Atkinson et al. call such solutions {\it{class $L$}}.

Class $2$ amplitudes have $\gamma_0 \in \mathbb{C}$, $\gamma_1 = 0$ and
$\gamma_{\ell} \in \mathbb{R}$ for $\ell > 1$. 
One way to find such solutions for
a given $z_1$ is by guessing a $\gamma_2 \in \mathbb{R}$ and recursing upwards. Given $z_1$ and $\gamma_2$ we can solve for $\gamma_0$ using the downward recursion relations. This gives
\begin{equation}
  | \gamma_0 | = \left| \frac{1 - z_1}{z_1} \right|, \quad \tmop{Re}
  (\gamma_0) = \frac{1}{4 \gamma_2} \left( 9 | 1 - z_1 |^2 - \left| \frac{1 -
  z_1}{z_1} \right|^2 - 4 \gamma_2^2 \right)
\end{equation}
From here, one can iterate upwards from $\gamma_2$ demanding that $
\gamma_{\ell}$ be real for $\ell > 2$. It is only possible for
$\gamma_{\ell}$ to be real if the upward-iterating discriminant is real, which
implies
\begin{equation}
  1 - \sqrt{\frac{(z_1 - 1) (z_1^{\star} - 1)}{(\tmop{Im} z_1)^2}} \le
  \varepsilon_{\ell} \le 1 + \sqrt{\frac{(z_1 - 1) (z_1^{\star} - 1)}{(\tmop{Im}
  z_1)^2}}
\end{equation}
This must hold for all $\ell$ so, in particular, it gives an allowed range for
$\varepsilon_2 = 1 - \gamma_2$. For a given value of
$\gamma_2$ and $z_1$ as one iterates upwards one may find that some higher $\gamma$ is imaginary. If this happens for all choices of signs in the recursion relation then that value of $\gamma_2$ is disallowed. This approach is called
the ascending iteration. Although Ref.~\cite{Atkinsonambiguity} found the ascending iteration inefficient, we find it can actually work quite well.

An alternative search procedure is the descending iteration. There one starts
at large $\ell$ where $| \varepsilon_{\ell} | \ll 1$ and one can linearize the
recursion relation. Solving the linearized version exactly gives
$\varepsilon_{\ell} = C Q_{\ell} \left( \tmop{Re} z_1 + \frac{\tmop{Im}^2
z_1}{\tmop{Re} z_1 - 1} \right)$ with $Q_{\ell}(x)$ the Legendre polynomial of
the second kind and $C$ a constant. One can then take this form for
$\varepsilon_L$ and $\varepsilon_{L + 1}$ for some large $L$ and iterate
downwards. Only for some value of $C$ will $\gamma_1 = 0$. Thus one can search
for $\gamma_1 = 0$ via the shooting method, as one might search for
eigenfunctions of a differential equation.

We find that  both the ascending and descending solutions are computationally extremely intense, sometimes requiring 60+ digits of precision to converge to a trustworthy solution. The choice of which sign
to pick on each step of the recursion is problematic as the search tree grows
exponentially. The descending solution is inferior in the sense that it can never produce an exact solution since the asymptotic form is assumed to be reached at some finite $L$;  the ascending solution can give an exact answer if the seed at low $\gamma_\ell$ is exact. In addition, we find that in some cases, the asymptotic behaviour at large $\ell$ is approached very slowly: even at $\ell=100$, the partial wave coefficients are not exponentially small. In particular, this happens for solutions that are strongly peaked at $z=\pm 1$ requiring large numbers of modes in their partial wave decomposition. Such solutions happen to be the ones which we found with low values of $\sin\mu$ (see Fig.\ref{fig:bestpoint} below).

\begin{figure}
    \centering
    \includegraphics[width=0.44\textwidth]{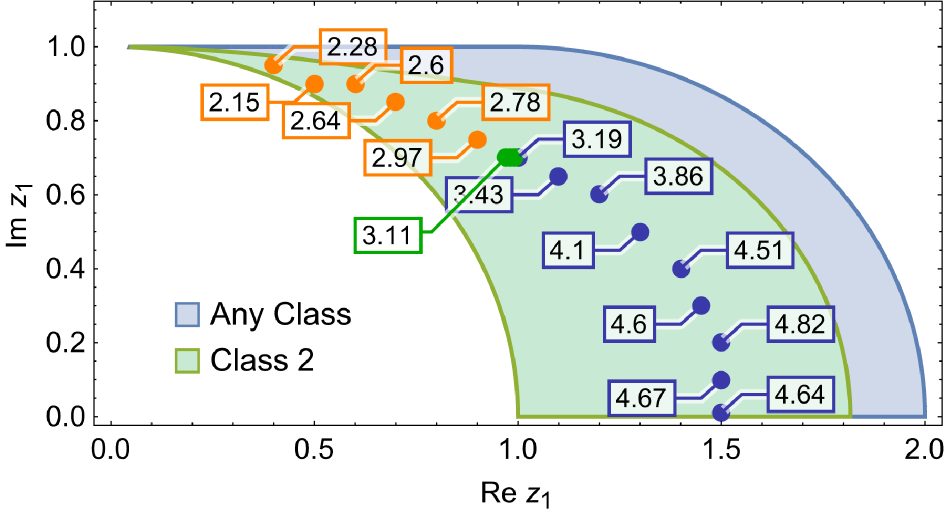}
    \includegraphics[width=0.46\textwidth]{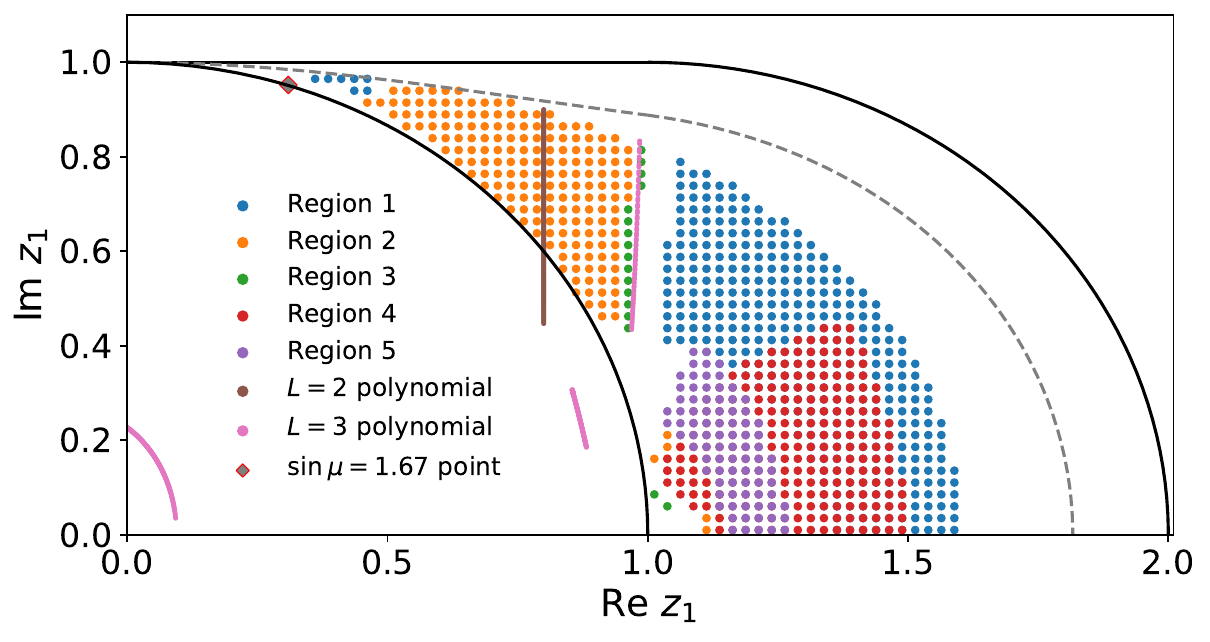}
    \caption{The left panel shows the region in $z_1$ space where non-polynomial amplitude pairs with a single conjugated zero are possibly allowed. The smaller green region is the possibly-allowed region for class 2 amplitudes where $\gamma_1=0$. Points are solutions found in~\cite{Atkinsonambiguity} labelled by their values of $\sin\mu$. Different colours correspond to different choices of signs in the descending iteration. The right panel shows solutions we found using the descending iterative scheme of Ref.~\cite{Atkinsonambiguity}. Different colours are different regions corresponding to different sign choices at various steps in the iteration. Class 2, $L=2,3$ partial waves polynomial amplitudes are also represented.} 
    \label{fig:regions}
\end{figure}
In Ref.~\cite{Atkinsonambiguity}, Atkinson et al. focused exclusively on parameterizations of the form of Eq.~\eqref{onezero} with one zero conjugated. They showed that non-polynomial solutions are only possible if $|z_1|>1$, $|\im\,z_1| < 1$ and if in the region where $|\re\,z_1|>1$ the additional constraint $\sqrt{|z_1|^2 - 2 |\re\,z_1| + 1}\le 1$ holds. 
The majority of the examples were class 2 (which has $\gamma_1 = 0$) for which one can impose the stronger constraints $|1-z_1| \ge \frac{3}{2}\left|\frac{1-z_1}{z_1}\right| (|z_1|-\frac{1}{3})|\im\,z_1|$ and $|1-z_1| \ge \frac{3}{2}\left|\frac{1-z_1}{z_1}\right| (|z_1|-\frac{1}{3})\sqrt{|z_1|^2 - 2 |\re\,z_1| + 1}$ in the $|\re\,z_1|>1$  region. These regions and the explicit points Atkinson et al. found are shown in Fig.~\ref{fig:regions}. The lowest value of $\sin\mu$ listed was $\sin\mu = 2.15$.

A challenge with the iterative approach is that there is no clear prescription for which sign to choose at each step in the iteration. When using the descending iteration Ref~\cite{Atkinsonambiguity} defined Region I to have all positive signs in the recursion relation in Eq.~\eqref{descend}. Region II is defined by choosing $-$ for $\ell=2$ and $+$ for all other $\ell$ in Eq.~\eqref{descend}. Region III has a minus sign for $\ell=2,3$ only, while region IV has minus signs at $\ell=2,3,4$. We also define a new region V which has a minus sign for $\ell=3$ only. The points belonging to these respective regions are coloured differently in Fig.~\ref{fig:regions}. We extend this study by searching for solutions in a 80$\times$40 grid in $z_1$ space. Our results are shown on the right side of 
 Fig.~\ref{fig:regions}. Although not clear from the figure, we find that the regions overlap: some points have solutions for the same $z_1$ but different sign choices in the iteration. Note that such pairs of solutions have different moduli so there are still only at most two phases for a given modulus (it is only $z_1$ which the solutions share).

Although Atkinson and collaborators found a number of phase-ambiguities at infinite $L$, they could only consider a small class of functions. They restricted to a single zero conjugated, as in Eq.~\eqref{onezero}, and moreover looked exclusively at class 2 amplitudes where $\gamma_1=0$ in their numerical work. Even with these assumptions, searching through the different regions is tedious, and trying to generalize to other classes or multiple conjugated zeros would be a herculean task.
We next explore how machine learning can search more efficiently for solutions.

\subsection{Machine learning complex functions}\label{sec:LMinfiniteL}
For the machine learning approach, we start with the class of functions in Eq.~\eqref{onezero} with a single conjugated zero. However, we do not need to do a partial wave decomposition. So we write simply
\begin{equation}\label{eq:mlF1}
    F(z) = \frac{z-z_1}{1-z_1} f(z), \quad\quad 
    \Fc(z) = \frac{z-z_1^\star}{1-z_1^\star} f(z)
\end{equation}
where $f(z)$ is the function to be learned and $z_1$ will be treated as a fixed input. The function $f(z)$ is parametrized by a neural network, closely following the implementation of Section~\ref{sec:nnsetup}, however for this application we need $f(z)$ to be a complex function instead of a real phase. In order to avoid complex numbers we have two obvious choices: have the network learn the modulus and phase of $f(z)$ or have it learn the real and imaginary parts of $f(z)$.
We tried both approaches but found that learning the real and imaginary parts was the most promising so we restrict to that choice in the following discussion.

For our neural network implementation, we modify the neural network depicted in Fig.~\ref{fig:phinet} by removing the final $\textit{Tanh}$ layer and having two final outputs instead of one. Removing the $\textit{Tanh}$ layer is justified since we are now predicting the real and the imaginary part of the amplitude directly and thus do not need to constrain them  within a given finite range. The two amplitudes are then reconstructed following the Eq.~\eqref{eq:mlF1} and training is done using the unitarity loss of Eq.~\eqref{eq:lossfunc_scaled} and the decaying repulsion of Eq.~\eqref{eq:decayingrepulsion}.

Although the iterative algorithm described in Section~\ref{sec:infiniteLclassical} and the machine learning implementation try to find similar solutions, the approaches differ in several key points:
\begin{enumerate}
   
    \item The iterative algorithm requires the choice of a particular coefficient $\gamma_L$ that is to be set to zero, yielding a \textit{class $L$} amplitude. In the machine learning implementation, the same neural network is used to recover any type of solution. For the same $z_1$ there can be phase-ambiguous solutions of different classes.
    Whereas these different class solutions can be individually picked out by the classical algorithm, the machine learning implementation will naturally tend to yield the one that is easiest to train, typically the one that has the lowest $\sin \mu$ value.
    \item The classical algorithm requires the choice of signs for the discriminant at each step in the iteration. So for $L$ steps, there are $2^L$ choices, leading to $2^L$ ``regions''. Most of these choices will not yield solutions, but there is no clear way to restrict the search. In the machine learning approach, the regions play no role: the algorithm will find solutions in all regions automatically.
    \item In the iterative algorithm unitarity is automatically enforced by the partial wave decomposition. However, the iteration may yield divergent results. A shooting method is used to find a sensible boundary condition for the iteration. In the machine learning implementation unitarity is enforced by minimizing a loss function. In practice one has to impose a cutoff on the loss $\mathcal{L}_E$ in order to assess whether the learned amplitudes do indeed respect unitarity.
    \item In the iterative algorithm, it is easy to see if the iterative solution yields a trivial ambiguity since the complex coefficients are known. In the machine learning approach, a repulsive loss has to be used to avoid trivial ambiguities. This makes the resolution of ambiguous solutions that are naturally close to one another more difficult. 
    \item Finding a solution with the classical algorithm can require high numerical precision. For example, we found that to confirm solutions close to the boundary of allowed $z_1$ values (where $\sin\mu$ is minimized) one can require 60 digits of precision or more. On the machine learning side, the precision is limited by the numerical integration scheme required when calculating the unitarity constraint of Eq.~(\ref{eq:lossfunc}). One generally cannot expect to have more than 5-10 digits of precision at best. However, to explore the space of solutions, high numerical precision is not required, as we could see in previous examples in Section~\ref{sec:singleamplitude} or Section~\ref{sec:repulsiveloss}.   
\end{enumerate}
In summary, the ML algorithm has the advantage of not needing a bunch of discrete choices and special cases to search. Thus it has the potential to search for a much broader class of solutions than the classical algorithm. On the other hand, its numerical precision is limited: you can never know if it actually finds a solution or not. An optimal approach may be to combine the two approaches: exploring the landscape of solutions with machine learning and then using a classical algorithm to refine particular solutions we find.

\subsection{Resolving the \texorpdfstring{$z_1$}{z1} landscape with ML}
\label{sec:z1landscape}
\begin{figure}[t]
     \centering
     \hfill
         \includegraphics[width=0.4\textwidth]{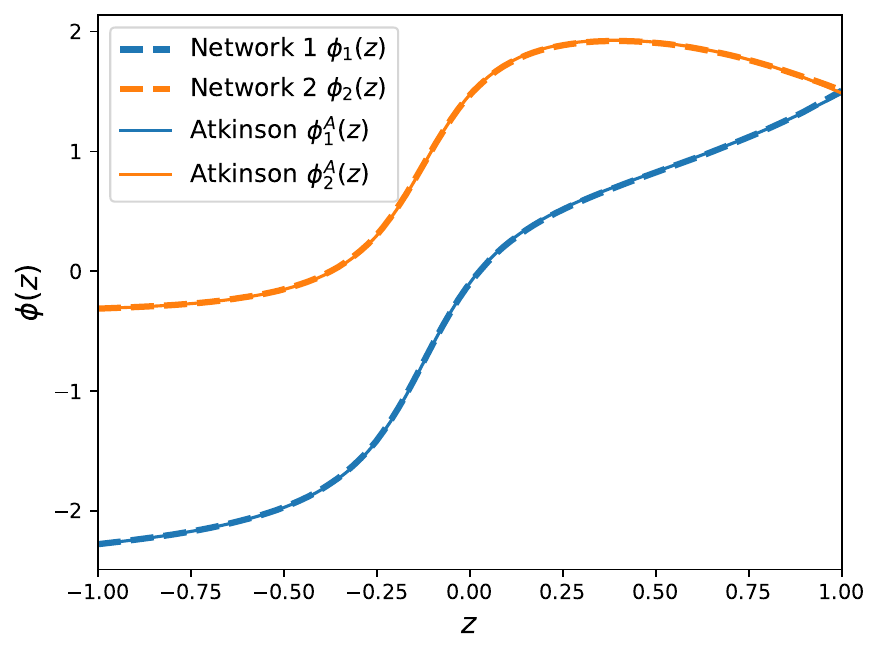}
     \hfill
         \includegraphics[width=0.4\textwidth]{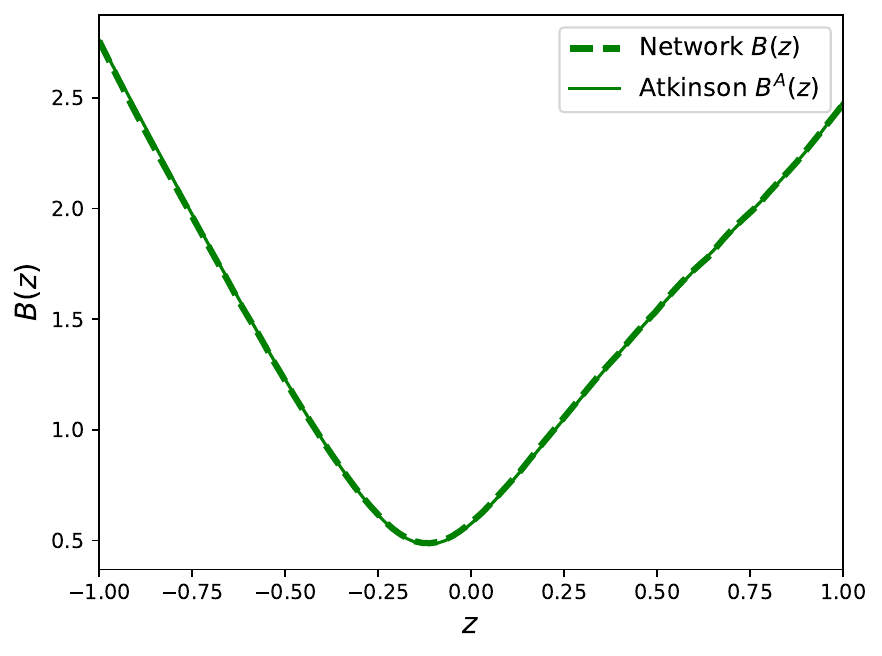}
     \hfill
     \caption{Phase-ambiguous solutions rediscovered with machine learning compared to previous results from~\cite{Atkinsonambiguity}. This solution has the form of Eq.~\eqref{eq:mlF1} with $z_1=\frac{6}{5} + \frac{3}{5} i$.
}
              \label{fig:atkinsonz1exmod}
\end{figure}

As a warm-up, we take one of the points and solutions found in~\cite{Atkinsonambiguity} which belongs to region I: $z_1= \frac{6}{5} + \frac{3}{5} i$. With this $z_1$ value, we train a network for 5000 epochs with the unitarity loss of Eq.~\eqref{eq:lossfunc_scaled} and using the decaying repulsion with  $c_0 =2$, $s_f = 16$, $e_f = 1000$ and $\lambda_R=2.0$. We extract the phases corresponding to the amplitudes of Eq.~\eqref{eq:mlF1} along with their respective moduli (which are identical by construction). We show these in Fig.~\ref{fig:atkinsonz1exmod}, along with the solutions recovered by the iterative algorithm. The final loss values are around $\mathcal{L}_E^S\sim 10^{-6}$ for both solutions, and we observe good agreement with the answer derived from the classical approach. 

We note that the amplitudes and phases for this solution are similar to the ones that we obtained when solving for the Crichton ambiguity in Section~\ref{sec:classicalfiniteL}. This observation will hold for a major part of the $z_1$ plane (where solutions are expected to be found) and prompts us to implement a better initialization for our neural network. 
When searching for another solution along the $z_1$ plane, it will be advantageous to initialize the network with a solution from a neighbouring point. The main training run will then be done over a smaller number of epochs and will not consider any repulsion term for the loss function. Since the phase solutions will be seeded at initialization as being properly distinct, we do not expect further training to modify them drastically and, instead, we will recover the two genuine ambiguous solutions.\footnote{One can view this property as starting the training run near the correct minimum of the loss landscape, as opposed to near a spurious minimum corresponding to a trivial ambiguity. In practice, after training, we will explicitly verify the nature of the solutions, for example by computing the value of the repulsion loss at evaluation.}

Next, we explore the space of $z_1$ values with phase ambiguities. To do so, we create a grid of $80 \times 40$ points in the complex $z_1$ plane with $\text{Re}\,z_1 \in [0,2]$ and $\text{Im}\,z_1 \in [0,1]$. This region is motivated by the known bound on the allowed range of $z_1$ (see Fig.~\ref{fig:regions}), but as we will see, that region will be rediscovered independently by the network.
At each point, we train for 500 epochs where the new neural networks are initialized with the trained networks of a nearest neighbor.\footnote{We start the procedure at the point $z_1= \frac{6}{5} + \frac{3}{5} i $, which we associate with the trained networks shown in Fig.~\ref{fig:atkinsonz1exmod}. A point is only trained if one of its neighbours has been previously resolved.}

\begin{figure}[t]
\centering
    \includegraphics[width = 0.48\textwidth]{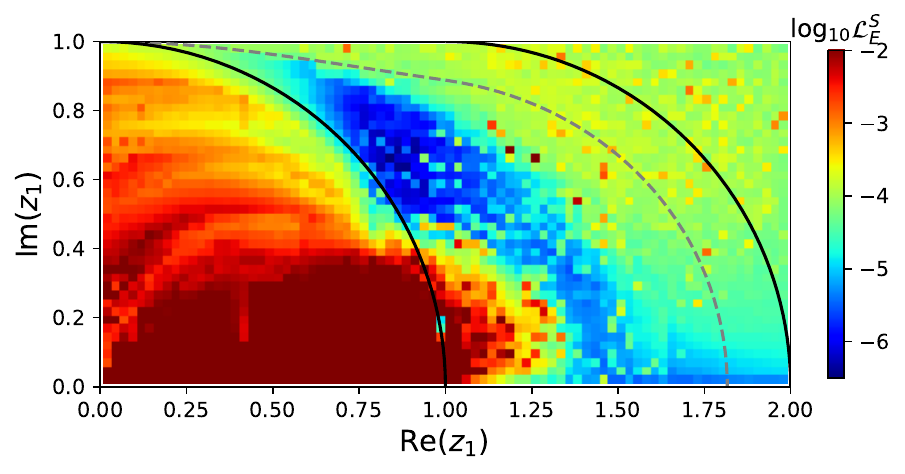}
\includegraphics[width = 0.48\textwidth]{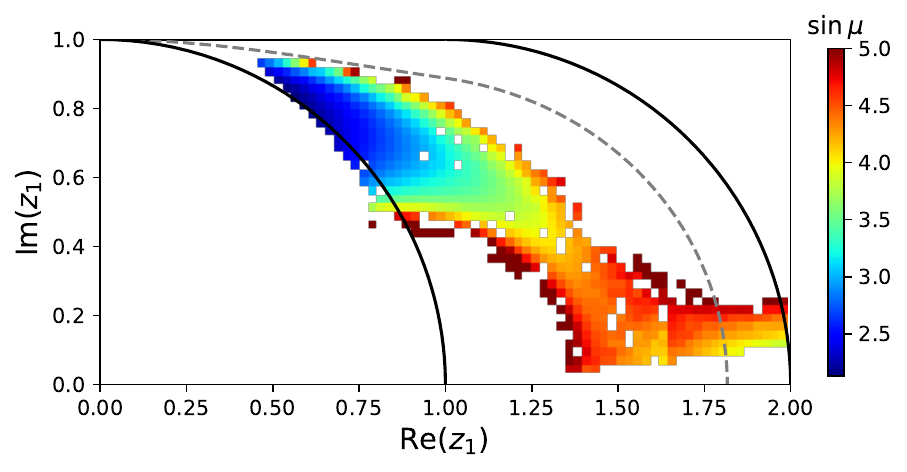}
    \caption{Loss landscape and $\sin\mu$ values in the search for phase ambiguities in amplitude pairs with a single conjugated root. The solid black lines delimit the region reachable with the descending algorithm of~\cite{Atkinsonambiguity}. 
     Class 2 amplitudes (those with $\gamma_1=0$) are only allowed between the grey dashed line and the lower black line $|z_1| = 1$. The low-loss points outside of the lower black curve ($|z_1|=1$) correspond to finite-$L$ solutions.
     Right panel shows the $\sin\mu$ landscape for low losses $\mathcal{L}_E^S < 10^{-4.5}$ and non degenerate solutions $\mathcal{L}_{R2} < 0.99$.
     }
    \label{fig:ininteLz1ML}
\end{figure}

The loss landscape from our scan is displayed on the left in Fig.~\ref{fig:ininteLz1ML}. The black curves here are an analytical bound on possible solutions: all possible infinite-$L$ single-conjugated-root solutions must lie within those curves. That does not mean that the entire region is allowed. The bounds also do not apply to the finite-$L$ polynomial solutions which can (and do) have $|z_1|<1$, as discussed in Appendix~\ref{sec:appendixfiniteatkinson}. From the figure, we can clearly see a region of low loss within the allowed region. In addition we recover a small domain with $|z_1|<1$ that has low loss, which is associated with polynomial solutions. We also note that not all of the allowed region has low loss. 
Indeed, for $\text{Re}(z_1) \in [1, 1.2]$ and $\text{Im}(z_1) \in [0, 0.3]$ the networks all have high evaluation losses with $\mathcal{L}_E^S > 10^{-3}$, indicating a failure to recover ambiguous solutions in that area. However, the iterative classical algorithm shows that some of the points should correspond to genuine solutions (see regions IV, V in Fig.~\ref{fig:regions}). Upon inspection we see that the solutions in this region are  highly oscillatory -- they do not resemble the functions obtained in the region I. Since these solutions are sufficiently dissimilar they cannot hope to be resolved by neural networks that have been initialized following another class of solutions and would require independent training and optimization trials in order to be recovered. This is certainly possible. But these solutions also have $\sin \mu$ values that are much higher than in the other domains of the $z_1$ plane. Because of the higher $\sin\mu$ values, this region is of no particular interest and we have not pursued its exploration further.

One feature distinguishing the machine learning loss landscape from the solution space of the classical algorithm is its continuity. The loss landscape does not suffer from the sharp boundaries that the classical algorithm experiences and is smooth across the domain in the $z_1$ plane that it resolves. Deforming $z_1$ slightly will result in another solution with a similar phase and differential cross section and the machine learning algorithm can do that interpolation easily. This is to be contrasted with the landscape emerging from the descending algorithm which struggles at the boundaries of the different regions, requiring enormous precision and fine-tuning to find solutions there. This is seen most clearly in the region near $\re\,z_1 = 1$ where regions I, II and III can possibly overlap (see Fig.~\ref{fig:regions}). The machine learning algorithm has no trouble in this region whereas the classical descending algorithm requires either an increasing number of partial waves or a broader search for its shooting method.

On the right of Fig.~\ref{fig:ininteLz1ML} we show the $\sin \mu$ values for the points in our scans where we only retain points that have $\mathcal{L}_E^S < 10^{-4.5}$. Additionally, we discard points corresponding to identical or trivial solutions. These points are characterized by having either of the terms in the loss of Eq.~\eqref{eq:decayingrepulsion} above 0.99. We note that possible phase-ambiguous solutions which happen to be very similar, such as those with $\re\,z_1 > 1.5$ and small $\im\,z_1$, are discarded by this second cut.

\begin{figure}[t]
\centering
    \includegraphics[width = 0.44\textwidth]{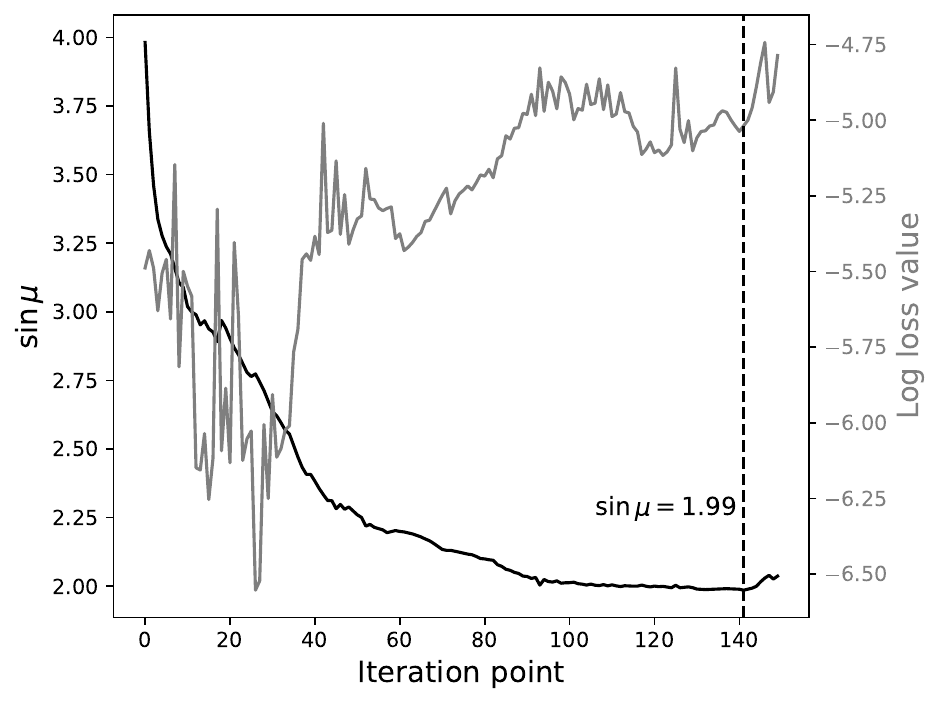}
\includegraphics[width = 0.42\textwidth]{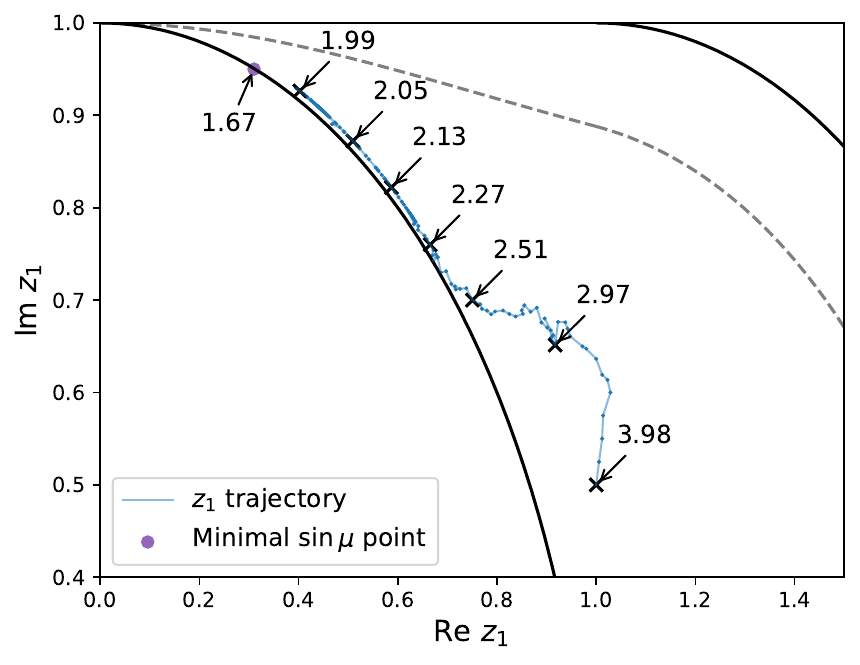}
    \caption{Gradient descent in the $z_1$ plane for minimizing $\sin \mu$. The left panel shows the value of $\sin \mu$ along the trajectory, along with the scaled loss $\mathcal{L}_E^S$ at each point. The right panel displays the trajectory in the $z_1$ plane. The point with the minimal $\sin \mu$ is found with the classical ascending algorithm by extending the gradient descent trajectory.} 
    \label{fig:z1gradientdescent}
\end{figure}

From this study, we see that the smallest $\sin\mu$ values tend to be close to the $|z_1|=1$ curve.
The lowest value of $\sin \mu$ that we found with this initial scan is $\sin \mu =2.13$ at $z_1=0.56+0.84i$ and is located in the region where the loss of the network starts to near $\mathcal{L}_E^S \sim 10^{-4.5}$. To get a lower value we can use the fact that the $\sin \mu$ landscape is continuous, allowing one to do a constrained gradient descent on it. The constraint that we have to respect here is one where the descent does not take us into regions of high loss. To implement the gradient descent we numerically estimate $\nabla_{z_1} \sin \mu (z_1)$ by using a central difference. This is done by training four networks at $z_1 \pm h$ and $z_1 \pm i h$ for 500 epochs with $h=10^{-3}$ and calculating the respective $\sin \mu$ value at those points. In order to accelerate convergence each network is initialized with the solved network at the $z_1$ point considered. We then take a step in the $z_1$ plane following\footnote{If $|z_1^{\text{new}}|<1$ we project out of the unit circle by considering $ 1.01\,z_1^{\text{new}}/|z_1^{\text{new}}|$ instead. This allows us to remain in regions of relatively low loss where we can trust our gradient descent.} 
\begin{equation}\label{eq:gradientdescent}
     z_1^{\text{new}} = \begin{pmatrix}
            \re\,z_1^{\text{new}} \\
            \im\,z_1^{\text{new}}
         \end{pmatrix} =  \begin{pmatrix}
            \re\,z_1^{\text{old}} \\
            \im\,z_1^{\text{old}}
         \end{pmatrix}  - \frac{\lambda_r}{2 h} \begin{pmatrix}
            \sin \mu (z_1^{\text{old}} + h) - \sin \mu (z_1^{\text{old}} - h) \\
            \sin \mu (z_1^{\text{old}} + i h) - \sin \mu (z_1^{\text{old}} - i h)
         \end{pmatrix} \,.
\end{equation}
A new network is then trained at $z_1^{\text{new}}$ and the process is iterated. In Fig.~\ref{fig:z1gradientdescent} we show such a gradient descent trajectory of 150 points where we used the learning rate $\lambda_r = 0.005$. The minimal value of $\sin \mu$ along the trajectory is 1.99, noticeably lower than in the initial scan. This is due to the fact that the region with the low $\sin \mu$ values is better resolved. The gradient descent takes small steps in the problematic region and the networks are trained to lower loss values, remaining under $\mathcal{L}_E^S \sim 10^{-5}$ as can be verified in the left panel of Fig.~\ref{fig:z1gradientdescent}.

\begin{figure}[t]
    \centering
    \hfill
    \includegraphics[width=0.4\textwidth]{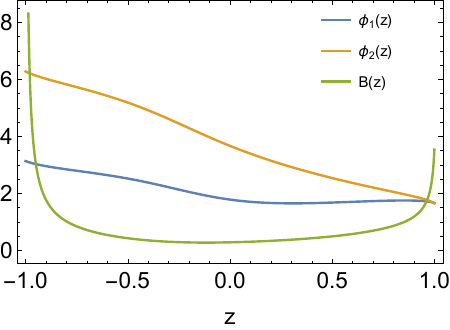}
    \hfill
    \includegraphics[width=0.42\textwidth]{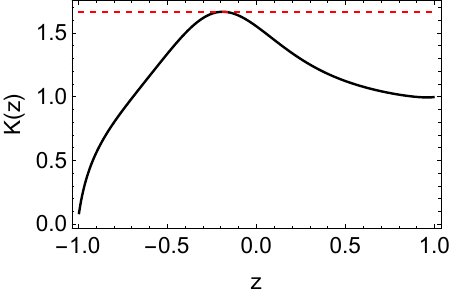}
    \hfill
    \caption{Left panel shows the modulus and unitary phases for the amplitude with $\sin\mu \approx 1.67$. Notice that it is peaked both in the forward, $z=1$, and backwards, $z=-1$, directions with $B(1) \approx 3.53$ and $B(-1) \approx 44.2$. For $z=1$ both phases approach $\frac{\pi}{2}$, whereas for $z=-1$ they approach $\pi$ and $2 \pi$ correspondingly. Curiously, the same features are exhibited by the simple toy model \eqref{Fguess}. Right panel shows the integrated kernel whose maximum is $\sin\mu$. This is the lowest known value of $\sin\mu$ for which a phase ambiguity exists.
    }
    \label{fig:bestpoint}
\end{figure}

Following this procedure, we can then use the classical ascending algorithm to further refine the lowest $\sin \mu$ point in a systematically improvable way. 
Following the gradient descent curve we are led to a point at $z_1 = 1.001 e^{0.4 \pi i} = 0.31 + 0.95 i$. When implementing the ascending algorithm we choose the Region II solution. The amplitude and phases for this point are shown in Fig.~\ref{fig:bestpoint}. It has $\sin\mu \approx 1.67$. This is the lowest known $\sin\mu$ value with phase ambiguous solutions. The phase shifts for this solution are given in Table~\ref{tab:phaseshifts}. At large $\ell$ the phase shifts in this solution oscillate and decay exponentially $\delta_{\ell} \sim (-1)^{\ell} \ell^a e^{- b \ell}$ with $b \approx 0.076$ and $a \approx 0.45$.

Looking at this solution it appears that $B (z)$ is peaked near $z =
\pm 1$ and the phases are roughly linear going between $\frac{\pi}{2}$ at $z =
1$ to either $\pi$ or $2 \pi$ at $z = - 1$. By exploring this structure, we can consider a toy amplitude of the form
\begin{equation}
  F (z) = a e^{i \phi_a} \delta (z - 1) + b e^{i \phi_b} \delta (z + 1) \label{Fguess}
\end{equation}
with $a, b, \phi_a$ and $\phi_b$ real numbers. Unitarity then implies
\begin{equation}
  \sin \phi_a = \frac{a^2 + b^2}{2 a}, \quad \sin \phi_b = a \cos (\phi_a -
  \phi_b), \quad \sin \mu = \max \left( \frac{a^2 + b^2}{2 a}, a \right)
\end{equation}
Recall that the expected lower bound on $\sin \mu$ for which there are
ambiguous solutions is $\sin \mu = 1$. If $a < 1$ then the condition $\sin \mu
= 1$ automatically leads to the value $\phi_a = \frac{\pi}{2}$ and then the
second constraint in Eq.~\eqref{Fguess}  becomes $\sin \phi_b = a \sin \phi_b$. Since we assumed
$a < 1$ we must have $\phi_b = \pi, 2 \pi$.
Remarkably, these are exactly the $z=\pm 1$ endpoints of the phases we found, see Fig.~\ref{fig:bestpoint}.
That is, we find two solutions
\begin{equation}
  F_{\pm} (z) = a i \delta (z - 1) \pm \sqrt{2 a - a^2} \delta (z + 1)
\end{equation}
with $a < 1$. Unfortunately, these two solutions are related by $F_+ = -
F_-^{\star}$ which is a trivial ambiguity. 
Based on this feature, it is interesting to contemplate a scenario in which any family of phase-ambiguous solutions will approach the trivial ambiguity as
$\sin \mu \rightarrow 1$. It would be very interesting to explore this further.

\begin{table}[ht]
{
\hfill
    \begin{tabular}{l|l}
\hline
$\delta_{0}$ & 	-0.7435785523	\\
$\delta_{1}$ &   0.3847784634	\\
\hline \\
$\deltac_{0}$ & -0.1982113969 \\
$\deltac_{1}$ &  0.5583871796	\\
\hline\\
$\delta_{2}$ & 	-0.314656996\\
$\delta_{3}$ & 	0.20231948\\
$\delta_{4}$ & 	-0.16811695\\
$\delta_{5}$ & 	0.14018563\\
\end{tabular}
\hfill
 \begin{tabular}{l|l}
\hline
$\delta_{6}$ & 	-0.12099793\\
$\delta_{7}$ & 	0.10455316\\
$\delta_{8}$ & 	-0.091873270\\
$\delta_{9}$ & 	0.080754325\\
$\delta_{10}$ & -0.071709663\\
$\delta_{11}$ & 0.063677489\\
$\delta_{12}$ & 	-0.056937764\\
$\delta_{13}$ & 	0.050906239\\
$\delta_{14}$ & 	-0.045741950\\
$\delta_{15}$ & 	0.041096977\\
\end{tabular}
\hfill
 \begin{tabular}{l|l}
\hline
$\delta_{16}$ & 	-0.037063302\\
$\delta_{17}$ & 	0.033422548\\
$\delta_{18}$ & 	-0.030227979\\
$\delta_{19}$ & 	0.027337227\\
$\delta_{20}$ & 	-0.024780662\\
$\delta_{21}$ & 	0.022462722\\
$\delta_{22}$ & 	-0.020400075\\
$\delta_{23}$ & 	0.018527050\\
$\delta_{24}$ & 	-0.016852096\\
$\delta_{25}$ & 	0.015329178
    \end{tabular}
\hfill
    }
    \caption{Here we present first 25 phase shifts for the $\sin\mu \approx 1.67$ solution with two possible phases. Only the first two phase shifts differ between the two amplitudes. To find the results  we ran the ascending algorithm with  $\sim 100-200$ modes and checked that the significant figures quoted above are convergent, in that they are not sensitive to how many modes are included.}
    \label{tab:phaseshifts}
\end{table}

\subsection{Extensions beyond one zero}
So far we have concentrated our efforts on finding solutions where a single zero $z_1$ is complex conjugated. This is only a small subset of all possible ambiguous solutions. In general, one could consider probing ambiguous solutions constructed from complex conjugating multiple different zeros. With the classical approach, one could try to construct an iterative algorithm in the style of~\cite{Atkinsonambiguity}. With more zeros, there is a larger space to search and many more discrete choices to make. While progress is possible, it seems extraordinarily tedious to search this way. 
The complexity of the classical approach is to be contrasted with the flexibility of the machine learning implementation, where one simply needs to modify the parametrization of Eq.~(\ref{eq:mlF1}) by taking out the appropriate number of zeros $z_1, \ldots, z_n$. Training can then proceed in the exact same way with no additional conceptual work. 

As a warm-up with multiple zeros, we show that a known ambiguous $L=3$ solution can be reproduced with the same ML construction. Ref. \cite{BERENDS1973507} considered cases where one root $z_3$ is held fixed and the other two roots $z_1$ and $z_2$ are conjugated:
 
\begin{align}
F(z) &= a \frac{(z-z_1)(z-z_2)(z-z_3)}{(1-z_1)(1-z_2)(1-z_3)}    \\
\Fc(z) &= a \frac{(z-z_1^\star)(z-z_2^\star)(z-z_3)}{(1-z_1^\star)(1-z_2^\star)(1-z_3)}    
\end{align}
For the machine learning setup, we parameterize
\begin{align}
F(z) &=  \frac{(z-z_1)(z-z_2)}{(1-z_1)(1-z_2)}  f(z)   \\
\Fc(z) &=  \frac{(z-z_1^\star)(z-z_2^\star)}{(1-z_1^\star)(1-z_2^\star)}    f(z)
\end{align}
we can then take $z_1$ and $z_2$ as inputs and learn a single complex function $f(z)$ as we did in the case of one zero.

\begin{figure}[t]
     \centering
     \begin{subfigure}[b]{0.45\textwidth}
         \centering
         \includegraphics[width=\textwidth]{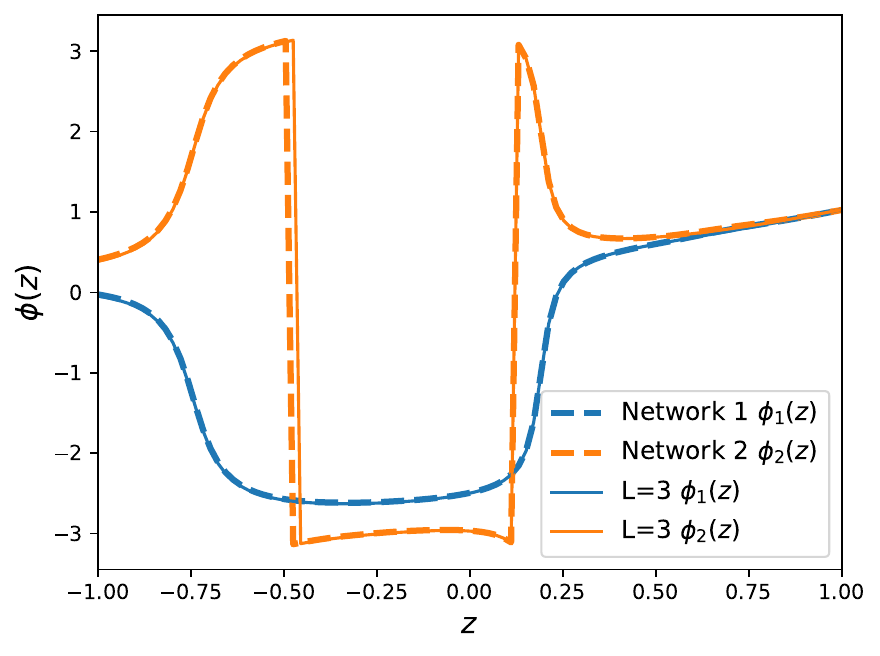}
     \end{subfigure}
          \hspace{5mm}
          \begin{subfigure}[b]{0.44\textwidth}
         \centering
         \includegraphics[width=\textwidth]{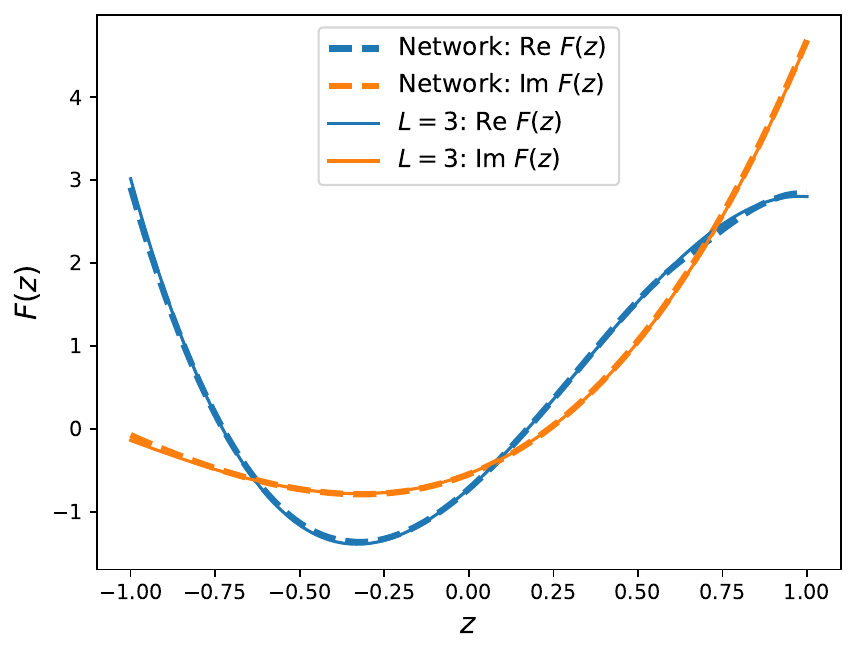}
     \end{subfigure}
     \centering
     \begin{subfigure}[b]{0.45\textwidth}
         \centering
         \includegraphics[width=\textwidth]{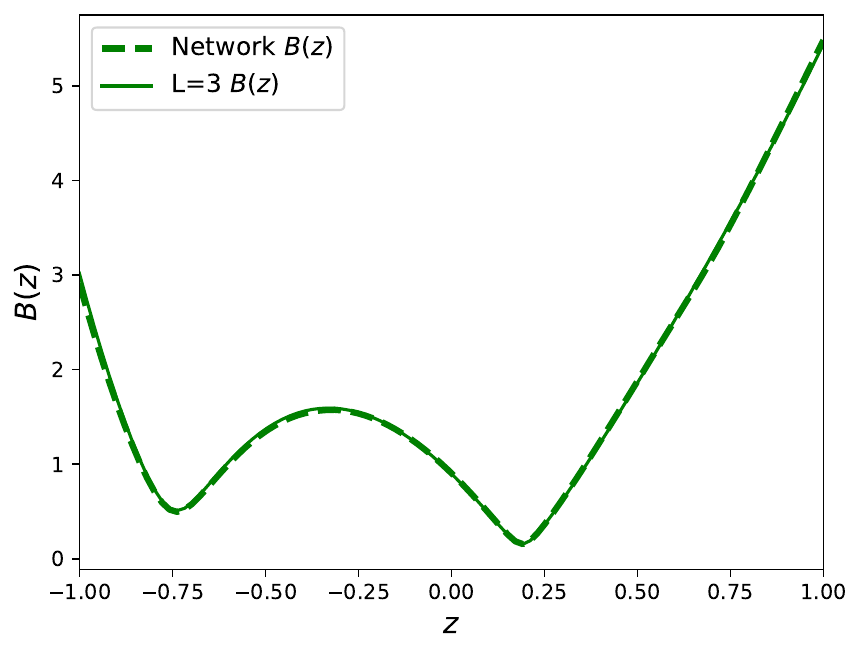}
     \end{subfigure}
          \hspace{5mm}
          \begin{subfigure}[b]{0.46\textwidth}
         \centering
         \includegraphics[width=\textwidth]{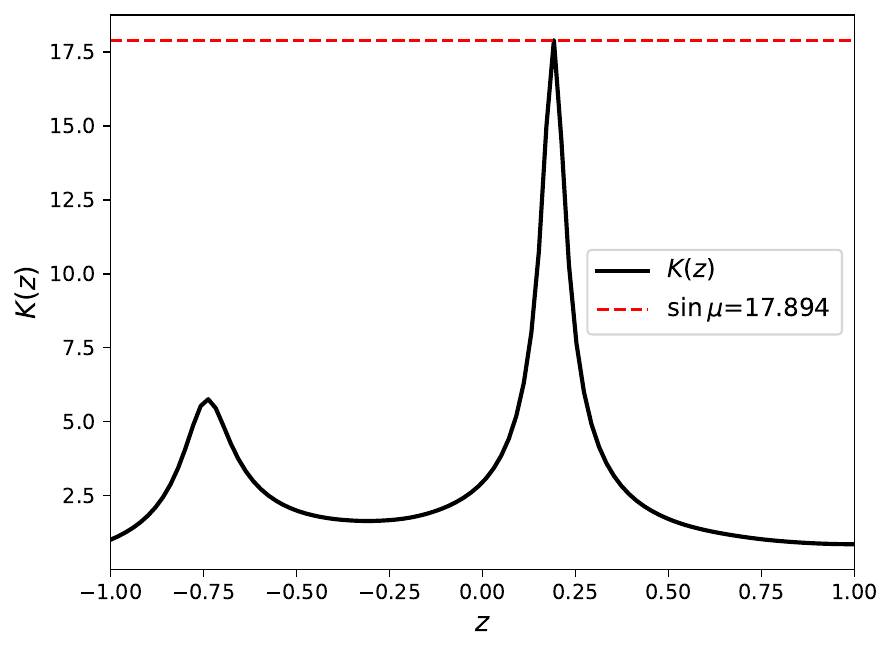}
     \end{subfigure}     
 \caption{Resolving the ambiguous solutions associated with a finite $L=3$ partial wave differential cross section. The two solutions differ by complex conjugation of two of their zeros. We display the phase outputs of the neural network compared to the exact solutions (top left panel) and the prediction for the real and imaginary parts of the first solution's amplitude  $F(z)$ (top right panel). The bottom panels show the learned modulus $B(z)$ and the corresponding kernel $K(z)$.}
     \label{fig:L3ML}
\end{figure}

To be concrete, one example solution found in~\cite{BERENDS1973507} had $a \approx 2.80+4.67 i$, $z_1\approx -0.74 - 0.06 i$, $z_2 \approx 0.19 + 0.03i$ and $z_3 \approx 1.32+0.95 i$. We take only $z_1$ and $z_2$ from this known solution and then train to find $f(z)$. We train the network for 5000 epochs using the decaying repulsion with the parameters of Section~\ref{sec:z1landscape}. To speed up training we also seed the neural network for $f(z)$ by the one that has been trained on the $z_1 =\frac{6}{5}+\frac{3}{5}i$ point. The result is shown in Fig.~\ref{fig:L3ML}. We find that the phases and moduli of the two solutions agree  with those found in~\cite{BERENDS1973507}. The agreement is robust, with an evaluation loss around $\mathcal{L}_E^S \sim 10^{-5}$ indicating that we indeed recovered the expected solution. We note that although the phases in the left panel of Fig.~\ref{fig:L3ML} appear to be discontinuous, the functions predicted by the network, $\text{Re} f(z)$ and $\text{Im} f(z)$, are themselves continuous. This translates into the real and imaginary parts of the amplitude $F(z)$ being continuous, as displayed on the right panel of Fig.~\ref{fig:L3ML}. The modulus and integrated kernel $K(z)$ are also shown. The $\sin\mu$ value comes from the maximum of $K(z)$, which gives $\sin \mu \approx 18$ for this example.

\begin{figure}[t]
\centering
    \includegraphics[width = 0.4\textwidth] {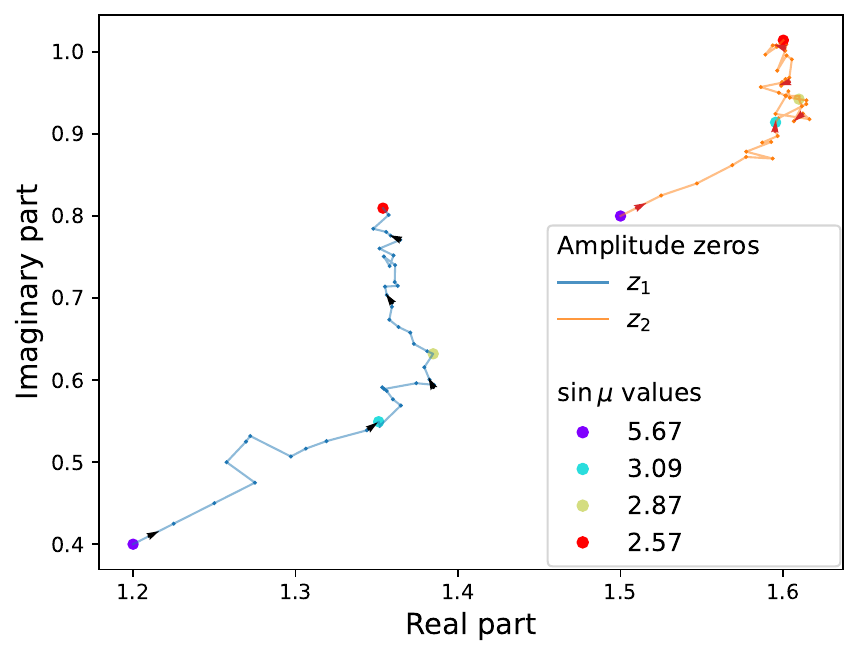}
\includegraphics[width = 0.42\textwidth]
{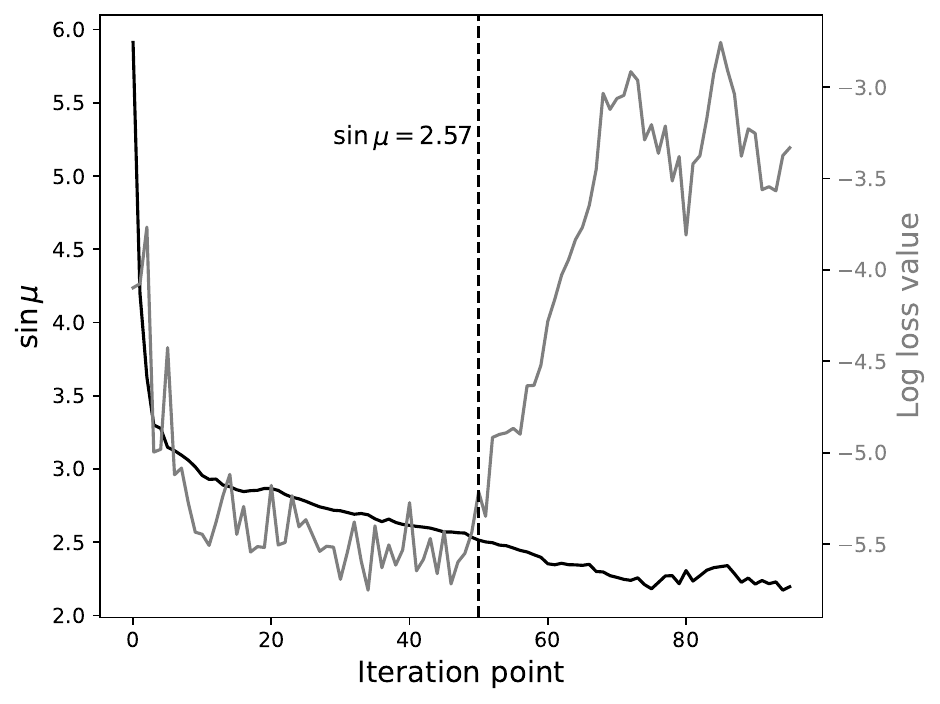}
\caption{Gradient descent in the $z_1, z_2$ space for minimizing $\sin \mu$. The left panel displays the trajectories of both the $z_1$ and $z_2$ roots for the points along the descent where $\mathcal{L}_E^S$ remains small. The right panel shows the value of $\sin \mu$ along the trajectory, along with the scaled loss $\mathcal{L}_E^S$ at each point. The gradient descent trajectory cannot be trusted once the evaluation loss blows up. }
    \label{fig:z1z2gradientdescent}
\end{figure}

Following the procedure outlined in Section~\ref{sec:z1landscape} one can also proceed to do gradient descent in the $z_1, z_2$ space in order to minimize $\sin \mu$. An example of a gradient descent trajectory is displayed in Fig.~\ref{fig:z1z2gradientdescent} where both roots are simultaneously updated at each point of the descent. The main difficulty we encounter is ensuring that the gradient descent follows a trajectory of low loss. As can be observed in the right panel of Fig.~\ref{fig:z1z2gradientdescent}, the evaluation loss associated with our trained networks starts to blow up at a given point along the trajectory, after which we cannot trust that viable ambiguous solutions are being recovered. The last value that can be trusted yields an ambiguous solution with $\sin \mu \approx 2.57$. In order to reach a trustworthy lower value of $\sin \mu$ one could envision modifying the gradient descent update of Eq.~\eqref{eq:gradientdescent} by forbidding updates that increase $\mathcal{L}_E^S$ substantially. While finding low $\sin\mu$ phase-ambiguous solutions in this way is conceivable, our initial assessment suggests that the amount of oversight required for this approach outweighs its probability for success, so have not pursued this direction further.

\clearpage
\section{Conclusions} \label{sec:conclusions}
In this paper, we have explored the problem of determining the phase of an amplitude from its modulus using modern machine learning. In the elastic scattering regime, the modulus and phase of an amplitude are constrained by a non-linear integral equation which enforces unitarity. Although the equation is difficult to solve analytically or with traditional numerical methods, it is easily solved with machine learning. Given a modulus $B(z)$ with $z$ the cosine of the scattering angle, a phase $\phi(z)$ for the amplitude can be parameterized as a neural network, then determined from unitarity through gradient descent. Using this technique we were able to reproduce known results for finite partial wave amplitudes, infinite partial wave amplitudes, and amplitudes determined by other $S$-matrix bootstrap principles. A few obvious extensions of our work include focusing on the scattering of identical particles, considering elastic scattering of spinning and/or flavored particles which would lead to a coupled system of unitarity equations, as well as doing the computation in the general number of spacetime dimensions $d$. 

More generally, it would be very interesting to apply machine learning to explore the full amplitude in both energy and angle, as in the classic analysis of pion scattering in \cite{Ananthanarayan:2000ht}, or more recent explorations of the space of nonperturbative amplitudes starting from \cite{Paulos2019}. This would require imposing in addition to unitarity the constraints of analyticity and crossing. In fact, methods very similar to the ones used to analyse elastic scattering at fixed energy were developed for the full amplitude by Atkinson, see e.g. \cite{Atkinson:1968hza,Atkinson:1968exe,Atkinson:1969eh,Atkinson:1970pe}, and were recently successfully implemented numerically \cite{Tourkine:2023xtu}. They are based on iterations of unitarity and are expected to converge only for a small subset of admissible amplitudes. More powerful gradient-descent type methods to construct the full amplitude have not been developed yet and it is to be seen if machine learning could be useful to tackle this problem.

Coming back to the present paper, there are two important open questions in $S$-matrix theory which we have shown machine learning approaches can address. The first is whether a phase $\phi(z)$ exists at all for a given differential cross section, or equivalently, a given modulus $B(z)$ of the scattering amplitude. This problem is solvable in a straightforward manner with machine learning. Code to find $\phi(z)$ from $B(z)$ is available here \url{https://github.com/aureliendersy/S-Matrix-Bootstrap}. It has been proposed that a functional $\sin\mu$ which involves a non-linear integral over $B(z)$ (see Eq.~\eqref{sinmudef}) is a good criterion for whether a solution exists. It has been shown that for $\sin\mu<1$ a solution always exists. If $\sin\mu>1$ there are three possibilities 1) no phase may exist 2) a unique phase may exist 3) two non-trivially related phases may exist.  Examples are known in all three cases. No clear criterion is known however to determine which case applies for a given $B(z)$ in general. Only options 2) and 3) are possible for $\sin\mu<1$ but no criterion is known to decide which. The analytical bound is that uniqueness must hold if $\sin\mu<0.86$ or the average modulus $\frac{1}{2}\int_{-1}^1 dz B^2(z)$ is less than 1.38. In the literature, ambiguous solutions are known with at best $\sin\mu \approx 2.15$. Using machine learning, we have found $B(z)$ with ambiguous phases with $\sin\mu \approx 1.67$. This is the first improvement on this bound in 50 years.

The machine learning approach offers several distinct advantages over classical approaches. The framework we have developed for solving the unitarity integral equation is general and recovering a phase can be attempted for any input modulus. This is to be contrasted with classical fixed point iteration schemes, which only converge if $\sin \mu < 1$. This straightforwardness has enabled us in Section~\ref{sec:simplepolynomial} to extensively explore various polynomial moduli and determine which ones are consistent with unitarity. We confirmed the existence bounds set by $\sin \mu < 1$ and identified the region in moduli space where $\sin \mu > 1$ solutions could be expected. Extending the setup to probe the uniqueness of the solution space was equally conceptually simple and only required the addition of a repulsive term to the loss function. Classical approaches  have instead focused only on analytically solvable cases, such as finite partial waves of low order, or on specific parametrizations for the amplitudes. One such parametrization proposed in~\cite{Atkinsonambiguity} was reviewed in Section~\ref{sec:infiniteLclassical} and contrasted with a machine-learning solution of the same problem. Whereas the classical algorithm required multiple discrete choices of parameters, carving out separate solution regions, the machine learning algorithm was able to smoothly interpolate across the whole solution landscape. This flexibility comes at a cost, lack of numerical precision, but allows a complementarity approach to traditional numerical methods. It is in that spirit that we have demonstrated in Section~\ref{sec:z1landscape} that one can utilize the smoothness of the machine learning loss landscape to perform gradient descent and find ambiguous solutions with low $\sin \mu$. There the inflexible, but powerful, classical algorithms allowed further refinement in order to obtain the lowest possible solution precisely. Extensions to other amplitude parameterizations are immediate with our machine learning framework whereas developing the corresponding classical iterative schemes (if possible) would require considerable amounts of effort.

Although here we focused on the narrow problem of the relationship between the modulus of an amplitude and its phase in the elastic scattering regime, a similar methodology can be used for much broader questions. The $S$-matrix bootstrap approach attempts to apply a set of general constraints such as unitarity, analyticity, and crossing to constrain the form of amplitudes. Implementing these constraints directly\footnote{The so-called primal approach, see e.g. the discussion in \cite{Guerrieri:2021tak}.} is a nontrivial task and the subject of many ongoing works, see e.g. \cite{Paulos2019,Chen2022,EliasMiro:2022xaa,Tourkine:2023xtu}. Machine learning offers the potential to search through a broad class of functions and perform the gradient descent efficiently, as we have seen here for phase-ambiguities in the elastic regime. In addition, a similar methodology could help with the analytic $S$-matrix bootstrap which applies constraints such as collinear limits or possible locations of singularities to perturbative scattering amplitudes. The classical approach has already been very successful, bootstrapping the 6-point amplitude in ${\cal N}=4$ super-Yang-Mills theory to 6 loops this way~\cite{Caron-Huot:2019vjl}. Additional constraints are known, such as those on sequential discontinuities~\cite{Hannesdottir:2022xki}, but have not been incorporated. Machine learning could make it easier to apply additional constraints and
it would be very interesting to explore the potential of machine learning for the $S$-matrix bootstrap further. This paper represents just a small first step into a field with enormous possibilities. 

\section*{Acknowledgements}
We would like to thank Filip Niewinski for their collaboration in the early stages of this work. We also thank Zohar Komargodski, Piotr Tourkine, and Jiaxin Qiao for useful discussions. AD and MDS are supported in part by the National Science Foundation under Cooperative Agreement PHY-2019786 (The NSF AI Institute for Artificial Intelligence and Fundamental Interactions).
This project has received funding from the European Research Council (ERC) under the European Union’s Horizon 2020 research and innovation programme (grant agreement number 949077).

\appendix
\section{Finite partial wave decomposition} \label{sec:appendixfinite}
In our study of polynomial amplitudes, we were mainly concerned with recovering solutions that admitted an infinite partial wave decomposition. In order to verify this point we characterize here the space of unitary amplitude solutions that admit a finite partial wave decomposition and a corresponding polynomial $B(z)$. We start our analysis by looking for solutions that could have a corresponding linear $B(z) = a z+b$. Since $B(z)^2$ is a polynomial of order 2, our finite partial wave solution must be of order 1 and parameterized as 
\begin{equation}
    F_1(z) = e^{i \delta_0} \sin \delta_0 + 3 e^{i \delta_1} \sin \delta_1 z
\end{equation}
where we used the explicit representation for the first two Legendre polynomials.  Equating $|F_1(z)|^2 = (az+b)^2$ we find the system 
\begin{equation}
    \left\{\begin{array}{ccc}
         a^2 &=& 9 \sin^2 \delta_1 \\
         b^2 &=&  \sin^2 \delta_0 \\
         2 ab &=& 6 \sin \delta_0  \sin \delta_1  \cos(\delta_0 - \delta_1)
    \end{array}\right.
\end{equation}
which can only be realized for $a=3b$ with $\delta_0=\delta_1$. Then we have $ F_1(z) = \pm e^{i\delta_0} \left(az+\frac{a}{3}\right)$ that satisfies the unitarity constraint and $|F_1(z)|^2 = |a|^2 |z+\frac{1}{3}|^2$.  The corresponding modulus is $B(z) = |az + \frac{a}{3}|$ where the absolute value is necessary to ensure positivity for $-1<z<1$.  Thus we do not have a simple finite partial wave amplitude that is associated with $B(z) = az+b$ where $b> |a|$ as was considered in Section~\ref{sec:linearamplitudes}.

For the quadratic modulus $B(z) = az^2 + c$ we can proceed in a similar fashion, parameterizing 
\begin{equation}
     F_2(z) = e^{i \delta_0} \sin \delta_0 + 3 e^{i \delta_1} \sin \delta_1 z + \frac{5}{2} e^{i \delta_2} \sin \delta_2 (3z^2-1)
\end{equation}
and deriving a similar system of equations
\begin{equation}
    \left\{\begin{array}{ccc}
         a^2 &=& \frac{225}{4} \sin^2 \delta_2 \\
         0 &=& \sin \delta_1  \sin \delta_2  \cos(\delta_1 - \delta_2) \\
         2 ac &=& 9 \sin^2 \delta_1 - \frac{2}{3} a^2 + 15 \sin \delta_0  \sin \delta_2  \cos(\delta_0 - \delta_2) \\
          0 &=& \sin \delta_1  \sin \delta_0  \cos(\delta_0 - \delta_1) \\
           c^2 &=& \sin^2 \delta_0 +\frac{a^2}{9} - 5 \sin \delta_2  \sin \delta_0  \cos(\delta_0 - \delta_2)
    \end{array}\right.
\end{equation}

We have 3 different solution sets. The first one with $\delta_1=0$ does not lead to a valid solution with both $a>0$ and $c>0$. The second solution set has $\delta_0 = 0$ and $\delta_1  = \delta_2 \pm \frac{\pi}{2}$. For $\delta_1  = \delta_2 - \frac{\pi}{2}$ we can have $a = \frac{15}{4} \sqrt{\frac{3}{7}}$ with $a = 3c$ that leads to a valid quadratic differential cross section. The third solution set has $\delta_1 = \delta_2 \pm \frac{\pi}{2}$ and $\delta_0 = \delta_1 \pm \frac{\pi}{2}$. For $\delta_1 = \delta_2 - \frac{\pi}{2}$ and $\delta_0 = \delta_1 - \frac{\pi}{2}$ we can have $a = \frac{5}{2}\sqrt{\frac{3}{2}}$ with $a = 5c$ that leads to a valid solution. Summarizing, for $B(z) = a z^2 + c$ we have two valid solutions with $a>0$ and $c>0$:

\begin{align}
    F_2^a(z) &= -\frac{15 z}{2 \sqrt{7}} e^{-i \sin ^{-1}\left(\frac{5}{2 \sqrt{7}}\right)} + \frac{5}{4} \sqrt{\frac{3}{7}} \left(3 z^2-1\right) e^{i \sin ^{-1}\left(\frac{1}{2}\sqrt{\frac{3}{7}}\right)}\\
        F_2^b(z) &= \frac{1}{\sqrt{6}} e^{i \sin ^{-1}\left(\frac{1}{\sqrt{6}}\right)}-\sqrt{\frac{15}{2}} z e^{-i \sin ^{-1}\left(\sqrt{\frac{5}{6}}\right)} + \frac{5 }{2 \sqrt{6}} \left(3 z^2-1\right) e^{i \sin ^{-1}\left(\frac{1}{\sqrt{6}}\right)}
\end{align}

which are respectively associated with the moduli
\begin{align} \label{eq:solquadb1}
B_2^a(z) &= \frac{5}{4}  \sqrt{\frac{3}{7}} \left(3 z^2+1\right) \\  \label{eq:solquadb2}
B_2^b(z) &= \sqrt{\frac{3}{8}} \left(5 z^2+1\right)
\end{align}

\begin{figure}[hbt!]
     \centering
     \begin{subfigure}[b]{0.45\textwidth}
         \centering
         \includegraphics[width=\textwidth]{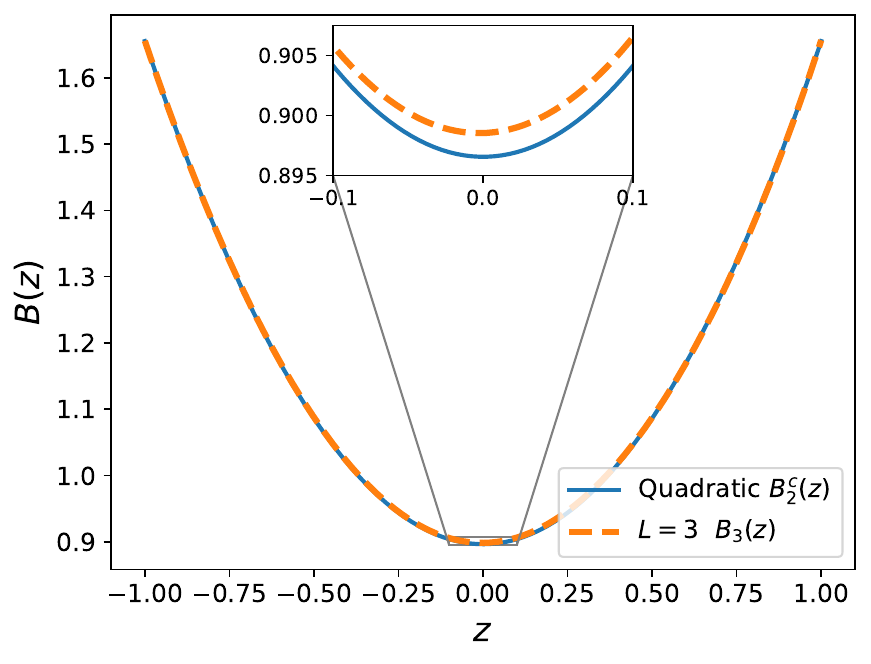}
         \caption{$B(z)$ moduli}
         \label{fig:comparisonmod}
     \end{subfigure}
          \hfill
          \begin{subfigure}[b]{0.45\textwidth}
         \centering
         \includegraphics[width=\textwidth]{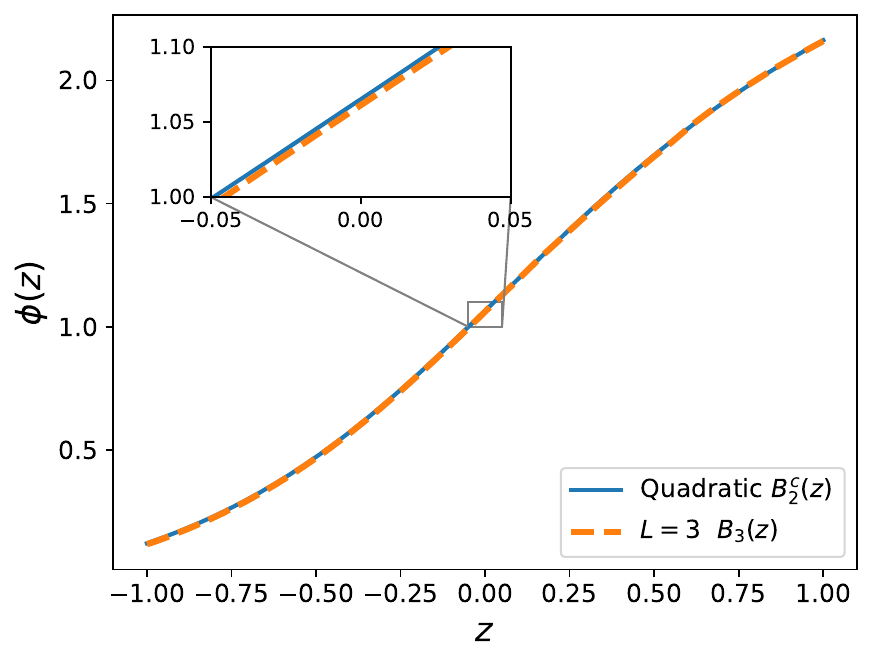}
         \caption{$\phi(z)$ phase}
         \label{fig:comparisonphases}
     \end{subfigure}
     \caption{Comparison between the exact $L=3$ finite partial wave solution and the machine-learned one associated with the quadratic input modulus $B_2^c(z)$. On the left panel, we display the numerical closeness of the two moduli and on the right panel we compare the exact $L=3$ phase and the machine learned one associated with the input $B_2^c(z)$.}
      \label{fig:comparisonmodphases}
\end{figure}

As discussed in Section~\ref{sec:scansquad}, scans over quadratic moduli revealed a 1D curve of low loss, whose corresponding $B(z)$ do not match any of the ones of Eq.~(\ref{eq:solquadb1}-\ref{eq:solquadb2}). Upon closer inspection, the low loss values are explained by the numerical closeness of the input moduli with ones corresponding to finite partial wave solutions with $L \neq 2$. For instance the modulus $B_2^c(z)=\frac{22}{29} z^2 + \frac{26}{29}$ has a loss $\mathcal{L}_S^E < 10 ^{-6}$ and is numerically within $0.3 \%$ of another modulus corresponding to the $L=3$ solution 
\begin{equation}
    F_3(z) = \sin \delta_0 e^{i \delta_0} + 3 z \sin \delta_1 e^{i \delta_1} + \frac{5}{2} (3z^2 - 1) \sin \delta_2 e^{i \delta_2} + \frac{7}{2} (5 z^ 3 -3 z) \sin \delta_3 e^{i \delta_3} \,.
\end{equation}
For the best fit values of $\delta_0=2.051, \delta_1 = 0.4578, \delta_2=-3.131, \delta_3 = -3.128$, we have an associated $B_3(z)$ that is numerically close to $B_2^c(z)$ which we display on the Fig.~\ref{fig:comparisonmod}. We refine the phase learned during our quadratic scans by letting the network run for 5000 epochs with $B_2^c(z)$ as input. We then compare the resulting phase to the exact $L=3$ solution as shown in Fig.~\ref{fig:comparisonphases}. The agreement between the learned phase and the finite $L=3$ solution explains why the associated loss value is so low for this spurious solution. Along the 1D curve of low loss, all of the factious solutions found have the same property in that their quadratic modulus is well approximated by a finite partial wave solution with $L \neq 2$. It is to be noted that the finite partial wave solutions of Fig.~\ref{fig:comparisonphases} look qualitatively different from the infinite partial wave solutions found in the $\sin \mu < 1$ region, with one example displayed in the bottom right panel of Fig.~\ref{fig:kernelslinquad}.

\section{Phase shift ambiguities with finite partial waves}
\label{sec:appendixfiniteatkinson}
\begin{figure}[hbt!]
     \centering
     \begin{subfigure}[b]{0.45\textwidth}
         \centering
         \includegraphics[width=\textwidth]{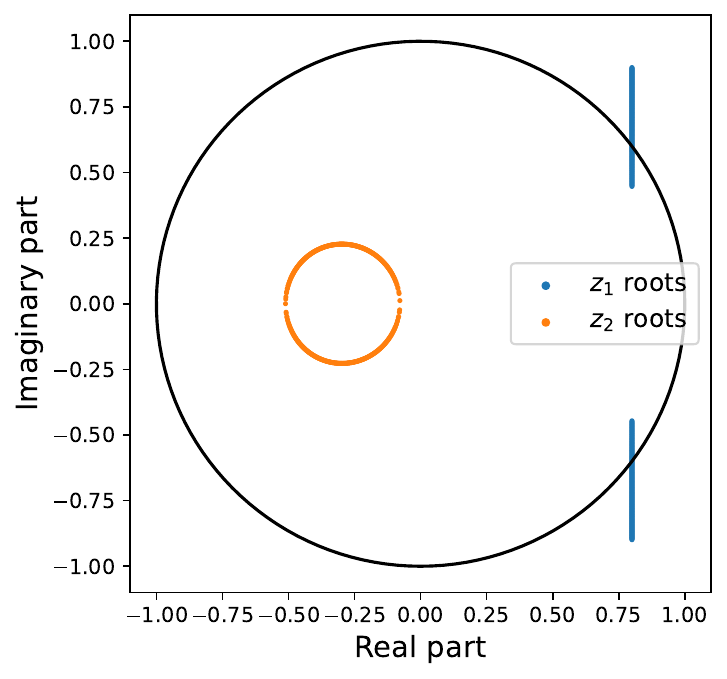}
         \caption{$L=2$}
         \label{fig:l2roots}
     \end{subfigure}
          \hfill
          \begin{subfigure}[b]{0.45\textwidth}
         \centering
         \includegraphics[width=\textwidth]{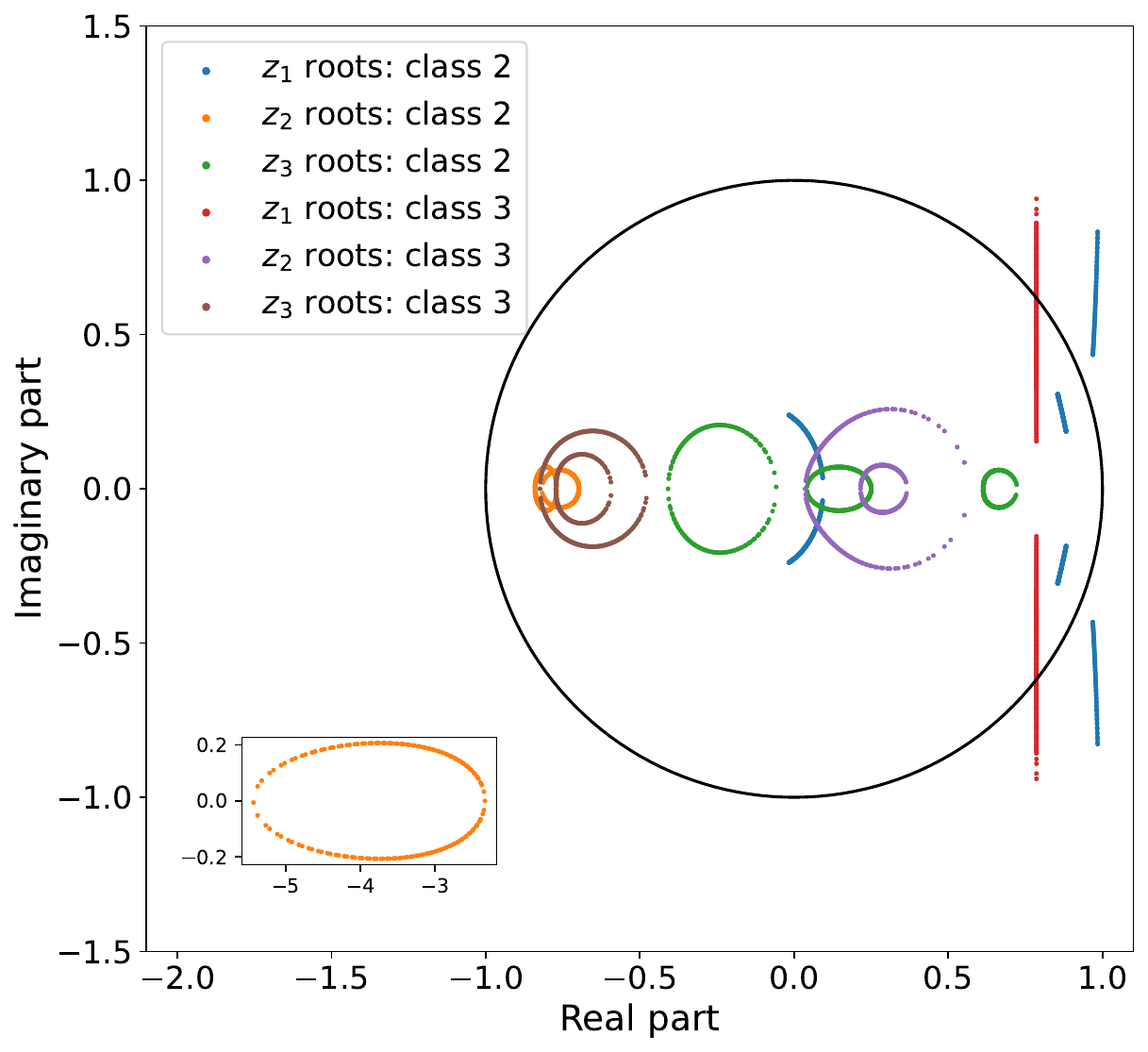}
         \caption{$L=3$}
         \label{fig:l3roots}
     \end{subfigure}
     \caption{Roots of the ambiguous polynomial ambiguities at $L=2,3$. We only represent solutions where the two ambiguous amplitudes have a single distinct root, respectively $z_1$ and $z_1^\star$. At $L=2$ the solutions are all of Ref's.~\cite{Atkinsonambiguity} class 2, as described in Section~\ref{sec:infiniteLclassical}. At $L=3$ the solutions can be class 2 or class 3.}
\end{figure}

 Whereas the main convergence region of the algorithm by Atkinson et al. is limited to $|z_1|>1$ for solutions with a large number of non-zero partial waves, it can also resolve a select few solutions within the unit circle, corresponding to true finite partial wave amplitudes. Finite partial wave ambiguities can be resolved exactly when the number of partial waves is small, as has been described in the literature for $L=2,3,4$~\cite{ATKINSON1973125, BERENDS1973507, Cornille1974}. In particular, the $L=2$ ambiguous solutions are all of the type described by Ref.~\cite{Atkinsonambiguity}. At $L=2$ the two ambiguous amplitudes $F(z)$ and $\tilde{F}(z)$ are polynomials of order 2 and possess one common root $z_2$. The other root is distinct and is respectively $z_1$ and $z_1^\star$. Since the difference $F(z)- (-\tilde{F}^\star(z))$ is a linear polynomial in $z$, the amplitudes correspond to genuine class 2 solutions. We represent in Fig.~\ref{fig:l2roots} the roots associated with these ambiguous amplitudes, noticing that we have the real part of $z_1$ that precisely equates $\frac{4}{5}$. As expected we have recovered solutions both outside the $z_1$ unit circle but also inside of it. All of these solutions can be recovered by both the classical descending algorithm and our machine learning implementation.

 At $L=3$ we have two different families of ambiguous amplitude solutions which share two roots $z_2, z_3$ and differ only by a single root, $z_1$ and $z_1^\star$. One family has $\text{Im}(z_2+z_3)=0$ and the other family has $\text{Im}(z_2+z_3)\neq 0$. In the first case the difference $F(z)- (-\tilde{F}^\star(z))$ is a linear polynomial in $z$, a class 2 solution, while in the second case, the same difference is a quadratic polynomial, hence a class 3 solution. The first family is resolved by the original implementation of the descending algorithm while the second family would require a shooting method that aims at finding $\gamma_2(C)=0$ for $\gamma_2$ appearing in Eq.~\eqref{onezero}. Both family and their respective roots are displayed in Fig.~\ref{fig:l3roots}, where notably we have a plethora of points lying within $|z_1|<1$.

\section{Scaled and non-scaled losses}
The difference between using the non-scaled or scaled losses of respectively Eq.~(\ref{eq:lossfunc}) and Eq.~(\ref{eq:lossfunc_scaled}) becomes apparent when studying edge cases. One example of interest concerns differential cross sections that are almost vanishing at a particular $z$ value, making $\sin \mu$ blow up. The $B(z)$ term in the denominator of Eq.~(\ref{eq:lossfunc_scaled}) makes the whole loss expression close to singular in that case. A simple example to probe this is to take $B(z) = z^2/2 + \epsilon$. The differential cross section is positive but almost vanishes at $z=0$, such that $\sin \mu = (60 \epsilon ^2+20 \epsilon +1)/(60 \epsilon)$.  For $\epsilon < (10 - \sqrt{85})/30$ we have $\sin \mu > 1$ and the existence of a solution is not guaranteed. In particular, the classical fixed point iterative scheme of \cite{atkinson1970} does not converge. 

\begin{figure}[hbt!]
     \centering
  \begin{subfigure}[b]{0.45\textwidth}
         \centering
         \includegraphics[width=\textwidth]{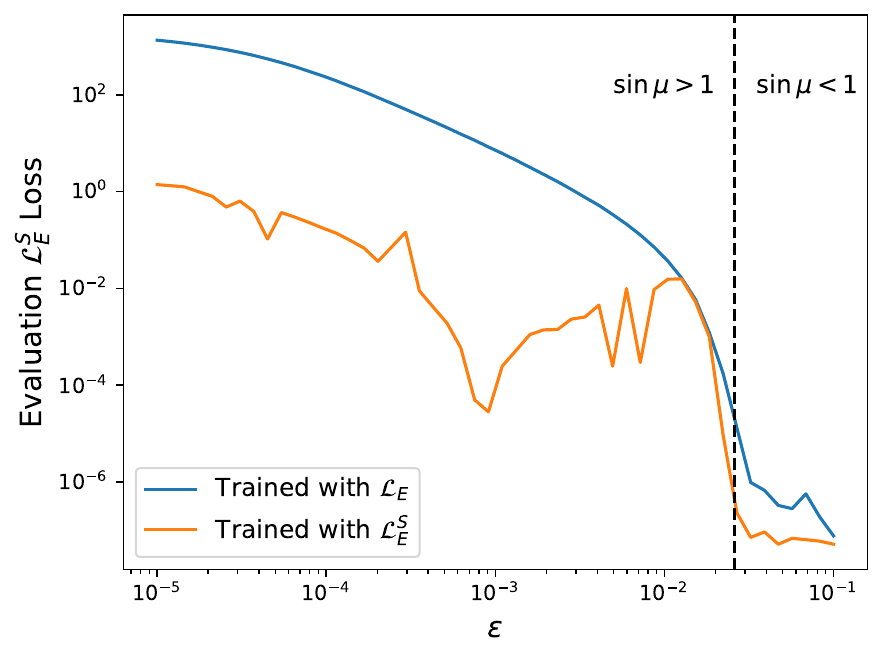}
         \caption{Scaled loss of Eq.~(\ref{eq:lossfunc_scaled}) at evaluation}
         \label{fig:lossesscaleddiff}
     \end{subfigure}
          \hfill
               \begin{subfigure}[b]{0.45\textwidth}
         \centering
         \includegraphics[width=\textwidth]{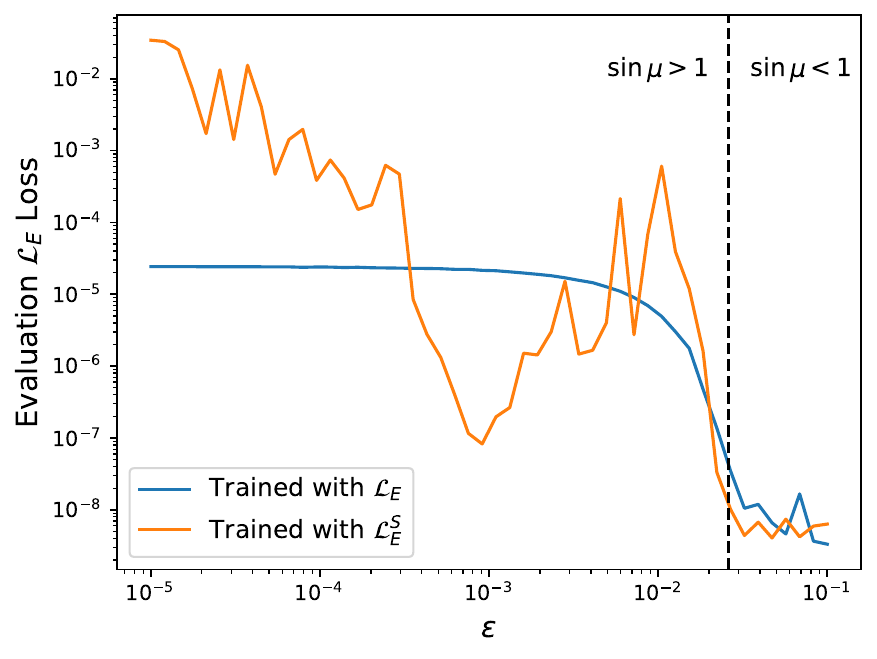}
         \caption{Non-scaled loss of Eq.~(\ref{eq:lossfunc}) at evaluation}
         \label{fig:lossesbasediff}
     \end{subfigure}
     \caption{Training on the input modulus $B(z)=az^2 + \epsilon$ using either a scaled (orange) or a non-scaled (blue) loss function. The panels compare the different loss metrics at evaluation time. The dashed black line indicates the transition between moduli with $\sin \mu < 1$ and moduli with $\sin \mu > 1$.}
\end{figure}

We can compare how the choice of the loss function plays a role in this edge case scenario, by training different neural networks using either Eq.~(\ref{eq:lossfunc}) or  Eq.~(\ref{eq:lossfunc_scaled}) as a loss function. We create a series of different $\epsilon$ values, distributed in $\epsilon \in [10^{-5}, 10^{-1}]$, and train a neural network for 2000 epochs at each point, using either loss function. We show in Fig.~\ref{fig:lossesscaleddiff} the scaled loss $\mathcal{L}_E^S$ at evaluation and in Fig.~\ref{fig:lossesbasediff} the non-scaled loss $\mathcal{L}_E$ at evaluation. In the $\sin \mu < 1$ region the networks trained using the scaled loss (orange curves) perform better on both evaluation metrics and accurately recover the phase solutions. In the $\sin \mu > 1$ region the networks trained with a scaled loss also perform better overall. In particular, we observe a dip in the evaluation losses around $\epsilon \sim 10^{-3}$, where $\mathcal{L}_E^S < 10^{-4}$ and $\mathcal{L}_E < 10^{-7}$. For the networks trained with the non-scaled loss (blue curves), no such dip is observed and instead, the scaled loss at evaluation blows up when $\epsilon < 10^{-2}$, as shown in Fig.~\ref{fig:lossesscaleddiff}.

\begin{figure}[hbt!]
     \centering
     \begin{subfigure}[b]{0.45\textwidth}
         \centering
         \includegraphics[width=\textwidth]{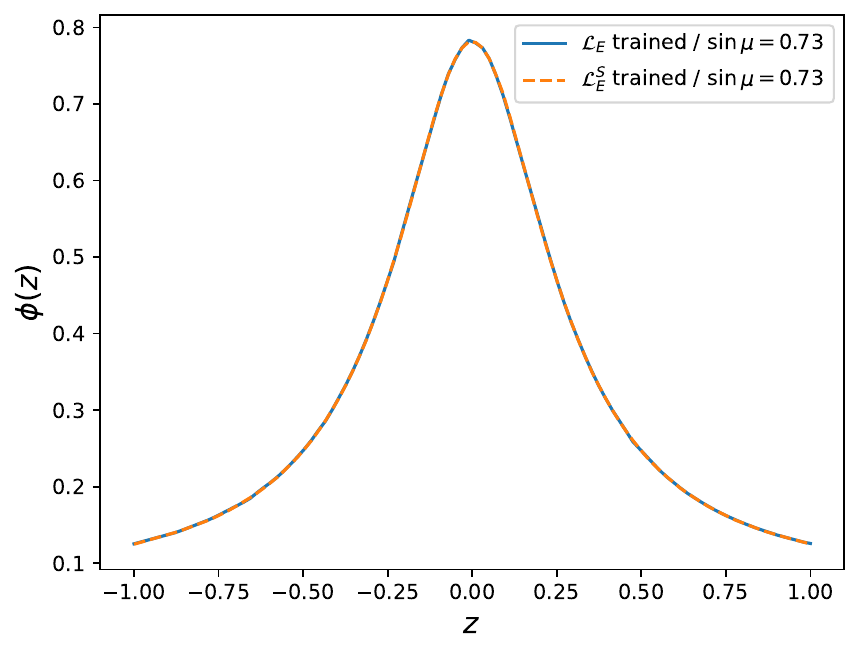}
         \caption{Low $\sin \mu$ value}
         \label{fig:phaseslowsinmu}
     \end{subfigure}
          \hfill
          \begin{subfigure}[b]{0.45\textwidth}
         \centering
         \includegraphics[width=\textwidth]{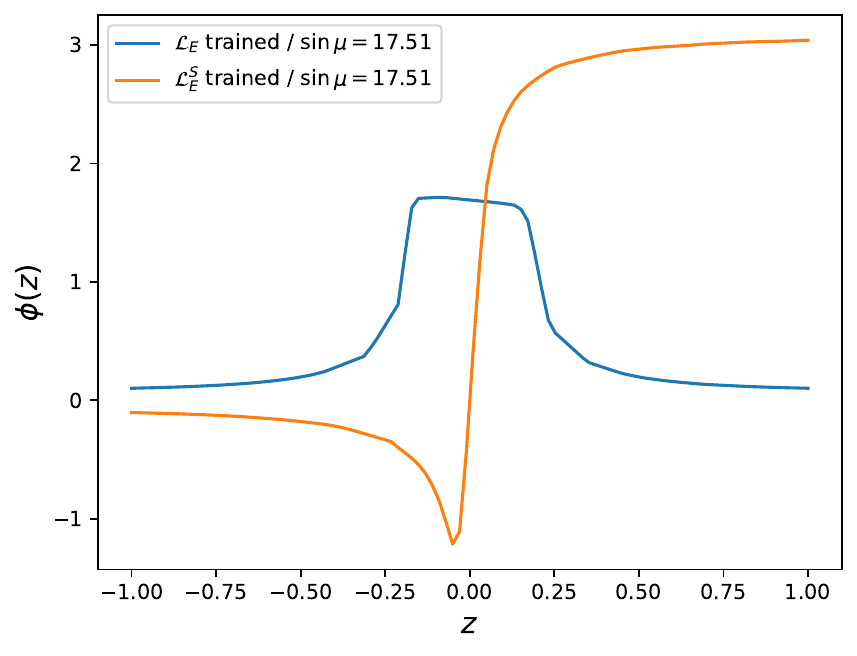}
         \caption{High $\sin \mu$ value}
         \label{fig:phaseshighsinmu}
     \end{subfigure}
     \caption{Learned phases for the input modulus $B(z)=z^2/2 + \epsilon$  where networks are trained using either a scaled (orange) or non-scaled (blue) loss function. On the left panel we use $\epsilon \sim 0.047$ where $\sin \mu < 1$ and on the right panel we use $\epsilon \sim 9.1 \times10^{-4}$ where $\sin \mu > 1$. On the right panel, the network trained with the non-scaled loss does not lead to a physical phase.}
\end{figure}

To understand this point better we can plot the learned phases at a few specific values of $\epsilon$. In Fig.~\ref{fig:phaseslowsinmu} and Fig.~\ref{fig:phaseshighsinmu} we plot the phases learned at respectively $\epsilon=0.04715$ and $\epsilon=0.00091$ where the former corresponds to a point in the $\sin \mu <1$ region and the latter to the dip in the $\sin \mu > 1$ region. For $\sin \mu < 1$, both phases are identical and both networks properly resolve the phase solution. However, as $\sin \mu > 1$, the phases learned by the two networks become drastically different. The networks trained on the base loss $\mathcal{L}_E$ will learn a simple deformation of the phase solution while the networks trained on $\mathcal{L}_E^S$ will learn a brand new phase shape. The dip in the evaluation losses that we observed can be explained by the numerical proximity of the phase to a genuine finite partial wave solution, as discussed in Appendix~\ref{sec:appendixfinite}. This feature will only be able to be captured by the networks trained using the scaled loss. It is important to mention however that in most circumstances (including for physical differential cross sections) we do not expect $B(z)$ to be close to vanishing and thus both training losses will perform similarly.

\section{Simple dual bounds}\label{sec:dual_bound}

Let us consider the following problem: can a given function $B(z)$ be an elastic differential cross-section?
One obvious requirement is that $B(z)\geq 0$ but there are more constraints.

Let us show that not any $B(z)$ can arise as a differential cross-section. We consider the following integral
\begin{equation}
\int_{-1}^1 d z F(z) F^*(z) = \int_{-1}^1 d z B(z)^2 . 
\end{equation}
By plugging the partial wave expansion for $F(z)$ into the integral we get
\begin{equation}
\int_{-1}^1 d z F(z) F^*(z) = 2 \sum_{\ell=0}^\infty (2 \ell + 1) | f_\ell |^2 = 2 \sum_{\ell=0}^\infty  (2 \ell + 1) {\rm Im} f_\ell =2 {\rm Im} F(1)  ,   
\end{equation}
where we used elastic unitarity ${\rm Im} f_\ell = | f_\ell |^2$. Using the fact that $B(1) = |F(1)| \geq  {\rm Im} F(1)$ we thus get the simplest constraint
\begin{equation}
\label{eq:basicbound}
2 B(1) \geq   \int_{-1}^1 d z B(z)^2,  
\end{equation}
which also immediately follows from considering elastic unitarity equation at $z=1$.

Consider next the next to simplest integral with spin one Legendre polynomial
\begin{equation}
\int_{-1}^1 d z P_1(z) F(z) F^*(z) = \int_{-1}^1 d z P_1(z) B(z)^2 . 
\end{equation}
This time we get a product of three Legendre polynomials which produces the Wigner 3j-symbol. It is only non-zero when $\ell - \ell' = \pm 1$.

Plugging the explicit expression we get
\begin{equation}
\int_{-1}^1 d z P_1(z) F(z) F^*(z) =\sum_{\ell=0}^\infty 2 (\ell+1) \left( f_{\ell+1} f^*_\ell + f_{\ell+1}^* f_\ell \right) = 2 \sum_{\ell=0}^\infty 2 (\ell+1) \left( {\rm Im}f_{\ell+1} {\rm Im} f_\ell + {\rm Re}f_{\ell+1} {\rm Re} f_\ell \right) .
\end{equation}
Consider next the following sum
\begin{equation}
\label{eq:dualineqleg1}
\sum_{\ell=0}^\infty 2(\ell+1) \left(|f_{\ell+1}|^2+|f_{\ell}|^2 - 2 ( {\rm Im}f_{\ell+1} {\rm Im} f_\ell + {\rm Re}f_{\ell+1} {\rm Re} f_\ell) \right) \geq 0 \ . 
\end{equation}
We then notice that using elastic unitarity
\begin{equation}
\sum_{\ell=0}^\infty (\ell+1) \left(|f_{\ell+1}|^2+|f_{\ell}|^2 \right) = \sum_{\ell=0}^\infty (2 \ell+1) {\rm Im} f_{\ell} = {\rm Im}F(1) .     
\end{equation}
As before, using the fact that $B(1) = |F(1)| \geq  {\rm Im} F(1)$, we thus get from \eqref{eq:dualineqleg1} (and its analog where we flip sign in front of the second term)
\begin{equation}
 2 B(1) \geq \int_{-1}^1 d z P_1(z) B(z)^2 \geq - 2 B(1) .   
\end{equation}
This bound is correct but it is trivially satisfied given \eqref{eq:basicbound}.

To get better bounds we need to put more constraints. Imagine that we know that $B(z)^2$ is a polynomial of a maximal degree $N$. We can then consider zero projections
\begin{equation}
\int_{-1}^1 d z P_{\ell>N}(z) B(z)^2 = 0 .     
\end{equation}
We can try to add these zero projections to the argument above. Consider for example the spin three zero projection
\begin{eqnarray}
 \int_{-1}^1 d z P_{3}(z) B(z)^2 &=&0 \\
 \sum_{\ell=0}^\infty d_{1,\ell} \left( f_{\ell+1} f^*_\ell + f_{\ell+1}^* f_\ell \right) &+& \sum_{\ell=0}^\infty d_{3,\ell} \left( f_{\ell+3} f^*_\ell + f_{\ell+3}^* f_\ell \right)=0,
\end{eqnarray}
where in the second line we rewrote it in terms of partial waves.

We can now derive bounds by considering the following positive semi-definite problem, see e.g. \cite{Caron-Huot:2020cmc} for a similar analysis in the case of dispersion relations,
\begin{equation}
c_{\pm} \ \mathbb{I} \pm T + \sum_i n_i N_i \succcurlyeq 0 ,
\end{equation}
where $N_i$ are zero projection matrices $f^* N_i f = 0$, $T$ is the target quantity that we want to bound $t = f^* T f$, and $\mathbb{I}$ is the diagonal matrix with elements being $2 \ell+1$.

If we have found $c_{\pm}$ and $n_i$ such that the matrix above is positive semi-definite, we get the bound
\begin{equation}
- c_- B(1) \leq t \leq c_+ B(1) \ . 
\end{equation}
Numerically, this can be done first by truncating in spin, and then extrapolating the cut-off to infinity.

Implementing the spin-3 zero projection constraint we get that 
\begin{equation}
\text{Spin 3:}~~~ 1.56 B(1) \geq \int_{-1}^1 d z P_1(z) B(z)^2 \geq - 1.56 B(1) ,
\end{equation}
which is an improvement of the previous bound. 
Similarly, we can consider the spin-four zero projection
\begin{equation}
\int_{-1}^1 d z P_{4}(z) B(z)^2 =0 ,
\end{equation}
and repeat the derivation above. This time we get the following bound
\begin{equation}
\text{Spin 4:}~~~    1.24 B(1) \geq \int_{-1}^1 d z P_2(z) B(z)^2 \geq - 0.67 B(1) .
\end{equation}

For polynomial cross-sections we have infinitely many null constraints which we could try to use to derive the dual bounds. These bounds must be satisfied and they do not depend on the existence of the actual solution. 

\printbibliography

@Article{Bessis1967,
author={Bessis, D.
and Martin, A.},
title={A theorem of uniqueness},
journal={Il Nuovo Cimento A (1965-1970)},
year={1967},
month={12},
day={01},
volume={52},
number={3},
pages={719-726},
issn={1826-9869},
doi={10.1007/BF02738839},
url={https://doi.org/10.1007/BF02738839}
}

@article{ALVAREZESTRADA1971196,
title = {On the construction of scattering amplitudes from experimental data and analyticity},
journal = {Annals of Physics},
volume = {68},
number = {1},
pages = {196-243},
year = {1971},
issn = {0003-4916},
doi = {https://doi.org/10.1016/0003-4916(71)90247-8},
url = {https://www.sciencedirect.com/science/article/pii/0003491671902478},
author = {R.F Alvarez-Estrada}
}

@article{Martin1973,
     author = {Martin, A.},
     title = {Reconstruction of {Scattering} {Amplitudes} {From} {Differential} {Cross-Section}},
     journal = {Les rencontres physiciens-math\'ematiciens de Strasbourg -RCP25},
     note = {talk:2},
     publisher = {Institut de Recherche Math\'ematique Avanc\'ee - Universit\'e Louis Pasteur},
     volume = {20},
     year = {1974},
     language = {en},
     url = {http://www.numdam.org/item/RCP25_1974__20__A2_0/}
}

@Article{Martin1969,
author={Martin, A.},
title={Construction of the scattering amplitude from the differential cross-sections},
journal={Il Nuovo Cimento A (1965-1970)},
year={1969},
month={1},
day={01},
volume={59},
number={1},
pages={131-152},
issn={1826-9869},
doi={10.1007/BF02756351},
url={https://doi.org/10.1007/BF02756351}
}

@book{Chadan_1989,
	doi = {10.1007/978-3-642-83317-5},
	url = {https://doi.org/10.1007%2F978-3-642-83317-5},
	year = 1989,
	publisher = {Springer Berlin Heidelberg},
	author = {K. Chadan and P. C. Sabatier and R. G. Newton},
	title = {Inverse Problems in Quantum Scattering Theory}
}

@InProceedings{atkinson1970,
author={Atkinson, D.},
editor={Urban, Paul},
title={Introduction to the Use of Non-Linear Techniques in S-Matrix Theory},
booktitle={Developments in High Energy Physics},
year={1970},
publisher={Springer Vienna},
address={Vienna},
pages={32-70},
isbn={978-3-7091-5835-7}
}

@article{RAISSI2019686,
title = {Physics-informed neural networks: A deep learning framework for solving forward and inverse problems involving nonlinear partial differential equations},
journal = {Journal of Computational Physics},
volume = {378},
pages = {686-707},
year = {2019},
issn = {0021-9991},
doi = {https://doi.org/10.1016/j.jcp.2018.10.045},
url = {https://www.sciencedirect.com/science/article/pii/S0021999118307125},
author = {M. Raissi and P. Perdikaris and G.E. Karniadakis}
}

@Article{Butler2019,
	title={{The Machine Learning landscape of top taggers}},
	author={Gregor Kasieczka and Tilman Plehn and Anja Butter and Kyle Cranmer and Dipsikha Debnath and Barry M. Dillon and Malcolm Fairbairn and Darius A. Faroughy and Wojtek Fedorko and Christophe Gay and Loukas Gouskos and Jernej F. Kamenik and Patrick T. Komiske and Simon Leiss and Alison Lister and Sebastian Macaluso and Eric M. Metodiev and Liam Moore and Ben Nachman and Karl Nordström and Jannicke Pearkes and Huilin Qu and Yannik Rath and Marcel Rieger and David Shih and Jennifer M. Thompson and Sreedevi Varma},
	journal={SciPost Phys.},
	volume={7},
	pages={014},
	year={2019},
	publisher={SciPost},
	doi={10.21468/SciPostPhys.7.1.014},
	url={https://scipost.org/10.21468/SciPostPhys.7.1.014},
}

@article{calogan,
  title = {CaloGAN: Simulating 3D high energy particle showers in multilayer electromagnetic calorimeters with generative adversarial networks},
  author = {Paganini, Michela and de Oliveira, Luke and Nachman, Benjamin},
  journal = {Phys. Rev. D},
  volume = {97},
  issue = {1},
  pages = {014021},
  numpages = {12},
  year = {2018},
  month = {1},
  publisher = {American Physical Society},
  doi = {10.1103/PhysRevD.97.014021},
  url = {https://link.aps.org/doi/10.1103/PhysRevD.97.014021}
}

@article{Tegmark2020,
author = {Silviu-Marian Udrescu  and Max Tegmark },
title = {AI Feynman: A physics-inspired method for symbolic regression},
journal = {Science Advances},
volume = {6},
number = {16},
pages = {eaay2631},
year = {2020},
doi = {10.1126/sciadv.aay2631},
URL = {https://www.science.org/doi/abs/10.1126/sciadv.aay2631},
eprint = {https://www.science.org/doi/pdf/10.1126/sciadv.aay2631}}

@misc{kamienny2023deep,
      title={Deep Generative Symbolic Regression with Monte-Carlo-Tree-Search}, 
      author={Pierre-Alexandre Kamienny and Guillaume Lample and Sylvain Lamprier and Marco Virgolin},
      year={2023},
      eprint={2302.11223},
      archivePrefix={arXiv},
      primaryClass={cs.LG}
}

@article{HORNIK1989359,
title = {Multilayer feedforward networks are universal approximators},
journal = {Neural Networks},
volume = {2},
number = {5},
pages = {359-366},
year = {1989},
issn = {0893-6080},
doi = {https://doi.org/10.1016/0893-6080(89)90020-8},
url = {https://www.sciencedirect.com/science/article/pii/0893608089900208},
author = {Kurt Hornik and Maxwell Stinchcombe and Halbert White}
}

@misc{neuralpde,
  doi = {10.48550/ARXIV.2107.09443},
  url = {https://arxiv.org/abs/2107.09443},
  author = {Zubov, Kirill and McCarthy, Zoe and Ma, Yingbo and Calisto, Francesco and Pagliarino, Valerio and Azeglio, Simone and Bottero, Luca and Luján, Emmanuel and Sulzer, Valentin and Bharambe, Ashutosh and Vinchhi, Nand and Balakrishnan, Kaushik and Upadhyay, Devesh and Rackauckas, Chris},
  title = {NeuralPDE: Automating Physics-Informed Neural Networks (PINNs) with Error Approximations},
  publisher = {arXiv},
  year = {2021},
  copyright = {Creative Commons Attribution Non Commercial Share Alike 4.0 International}
}

@article{DeepXDE,
author = {Lu, Lu and Meng, Xuhui and Mao, Zhiping and Karniadakis, George Em},
title = {DeepXDE: A Deep Learning Library for Solving Differential Equations},
journal = {SIAM Review},
volume = {63},
number = {1},
pages = {208-228},
year = {2021},
doi = {10.1137/19M1274067},
URL = {https://doi.org/10.1137/19M1274067}
}

@article{YUAN2022111260,
title = {A-PINN: Auxiliary physics informed neural networks for forward and inverse problems of nonlinear integro-differential equations},
journal = {Journal of Computational Physics},
volume = {462},
pages = {111260},
year = {2022},
issn = {0021-9991},
doi = {https://doi.org/10.1016/j.jcp.2022.111260},
url = {https://www.sciencedirect.com/science/article/pii/S0021999122003229},
author = {Lei Yuan and Yi-Qing Ni and Xiang-Yun Deng and Shuo Hao}
}

@article{PANG2020109760,
title = {nPINNs: Nonlocal physics-informed neural networks for a parametrized nonlocal universal Laplacian operator. Algorithms and applications},
journal = {Journal of Computational Physics},
volume = {422},
pages = {109760},
year = {2020},
issn = {0021-9991},
doi = {https://doi.org/10.1016/j.jcp.2020.109760},
url = {https://www.sciencedirect.com/science/article/pii/S0021999120305349},
author = {G. Pang and M. D'Elia and M. Parks and G.E. Karniadakis}
}

@article{JEBowcock_1975,
doi = {10.1088/0034-4885/38/9/002},
url = {https://dx.doi.org/10.1088/0034-4885/38/9/002},
year = {1975},
month = {9},
publisher = {},
volume = {38},
number = {9},
pages = {1099},
author = {J E Bowcock and  H Burkhardt},
title = {Principles and problems of phase-shift analysis},
journal = {Reports on Progress in Physics}
}

@Article{Gangal1984,
author={Gangal, A. D.
and Kupsch, J.},
title={Determination of the scattering amplitude},
journal={Communications in Mathematical Physics},
year={1984},
month={9},
day={01},
volume={93},
number={3},
pages={333-339},
issn={1432-0916},
doi={10.1007/BF01258532},
url={https://doi.org/10.1007/BF01258532}
}

@Article{Itzykson1973,
author={Itzykson, C.
and Martin, A.},
title={Phase-shift ambiguities for analytic amplitudes},
journal={Il Nuovo Cimento A (1971-1996)},
year={1973},
month={9},
day={01},
volume={17},
number={2},
pages={245-287},
issn={1826-9869},
doi={10.1007/BF02777935},
url={https://doi.org/10.1007/BF02777935}
}

@Article{Paulos2019,
author={Paulos, Miguel F.
and Penedones, Joao
and Toledo, Jonathan
and van Rees, Balt C.
and Vieira, Pedro},
title={The S-matrix bootstrap. Part III: higher dimensional amplitudes},
journal={Journal of High Energy Physics},
year={2019},
month={12},
day={04},
volume={2019},
number={12},
pages={40},
issn={1029-8479},
doi={10.1007/JHEP12(2019)040},
url={https://doi.org/10.1007/JHEP12(2019)040}
}

@Article{Chen2022,
author={Chen, Hongbin
and Fitzpatrick, A. Liam
and Karateev, Denis},
title={Nonperturbative bounds on scattering of massive scalar particles in d $\geq$ 2},
journal={Journal of High Energy Physics},
year={2022},
month={12},
day={16},
volume={2022},
number={12},
pages={92},
issn={1029-8479},
doi={10.1007/JHEP12(2022)092},
url={https://doi.org/10.1007/JHEP12(2022)092}
}

@article{Atkinsonambiguity,
  title = {Crichton ambiguities with infinitely many partial waves},
  author = {Atkinson, D. and Kok, L. P. and de Roo, M.},
  journal = {Phys. Rev. D},
  volume = {17},
  issue = {9},
  pages = {2492--2502},
  numpages = {0},
  year = {1978},
  month = {5},
  publisher = {American Physical Society},
  doi = {10.1103/PhysRevD.17.2492},
  url = {https://link.aps.org/doi/10.1103/PhysRevD.17.2492}
}

@article{ATKINSON1973125,
title = {Crichton's phase-shift ambiguity},
journal = {Nuclear Physics B},
volume = {55},
number = {1},
pages = {125-131},
year = {1973},
issn = {0550-3213},
doi = {https://doi.org/10.1016/0550-3213(73)90413-6},
url = {https://www.sciencedirect.com/science/article/pii/0550321373904136},
author = {D. Atkinson and P.W. Johnson and N. Mehta and M. {de Roo}},
}

@Article{Crichton1966,
author={Crichton, J. H.},
title={Phase-shift ambiguities for spin-independent scattering},
journal={Il Nuovo Cimento A (1965-1970)},
year={1966},
month={9},
day={01},
volume={45},
number={1},
pages={256-258},
issn={1826-9869},
doi={10.1007/BF02738098},
url={https://doi.org/10.1007/BF02738098}
}

@inproceedings{DiGiovanni2020FindingMS,
  title={Finding Multiple Solutions of ODEs with Neural Networks},
  author={Marco Di Giovanni and David Sondak and Pavlos Protopapas and Marco Brambilla},
  booktitle={AAAI Spring Symposium: MLPS},
  year={2020}
}

@article{BERENDS1973507,
title = {Examples of phase-shift ambiguities for spinless elastic scattering},
journal = {Nuclear Physics B},
volume = {56},
number = {2},
pages = {507-524},
year = {1973},
issn = {0550-3213},
doi = {https://doi.org/10.1016/0550-3213(73)90044-8},
url = {https://www.sciencedirect.com/science/article/pii/0550321373900448},
author = {F.A. Berends and S.N.M. Ruijsenaars}
}

@article{GERSTEN1969537,
title = {Ambiguities of complex phase-shift analysis},
journal = {Nuclear Physics B},
volume = {12},
number = {3},
pages = {537-548},
year = {1969},
issn = {0550-3213},
doi = {https://doi.org/10.1016/0550-3213(69)90072-8},
url = {https://www.sciencedirect.com/science/article/pii/0550321369900728},
author = {A. Gersten},
}

@Article{Cornille1974,
author={Cornille, H.
and Drouffe, J. M.},
title={Phase-shift ambiguities for spinless and $L_\text{max} \leq$4 elastic scattering},
journal={Il Nuovo Cimento A (1965-1970)},
year={1974},
month={4},
day={01},
volume={20},
number={3},
pages={401-436},
issn={1826-9869},
doi={10.1007/BF02821973},
url={https://doi.org/10.1007/BF02821973}
}

@article{newton1968,
    author = {Newton, Roger G.},
    title = "{Determination of the Amplitude from the Differential Cross Section by Unitarity}",
    journal = {Journal of Mathematical Physics},
    volume = {9},
    number = {12},
    pages = {2050-2055},
    year = {2003},
    month = {10},
    issn = {0022-2488},
    doi = {10.1063/1.1664543},
    url = {https://doi.org/10.1063/1.1664543},
    eprint = {https://pubs.aip.org/aip/jmp/article-pdf/9/12/2050/8182953/2050\_1\_online.pdf},
}

@inproceedings{Adam2014,
    author = "Kingma, Diederik P. and Ba, Jimmy",
    title = "{Adam: A Method for Stochastic Optimization}",
    eprint = "1412.6980",
    archivePrefix = "arXiv",
    primaryClass = "cs.LG",
    month = "12",
    year = "2014"
}

@article{Bart:1973cf,
    author = "Bart, G. R. and Johnson, P. W. and Warnock, R. L.",
    title = "{Continuum ambiguity in the construction of unitary analytic amplitudes from fixed-energy-scattering data}",
    doi = "10.1063/1.1666226",
    journal = "J. Math. Phys.",
    volume = "14",
    pages = "1558--1565",
    year = "1973"
}

@article{Atkinson:1972hr,
    author = "Atkinson, D. and Johnson, P. W. and Warnock, R. L.",
    title = "{Determination of the scattering amplitude from the differential cross-section and unitarity}",
    doi = "10.1007/BF01645512",
    journal = "Commun. Math. Phys.",
    volume = "28",
    pages = "133--158",
    year = "1972"
}

@article{EliasMiro:2022xaa,
    author = "Elias Miro, Joan and Guerrieri, Andrea and Gumus, Mehmet Asim",
    title = "{Bridging positivity and S-matrix bootstrap bounds}",
    eprint = "2210.01502",
    archivePrefix = "arXiv",
    primaryClass = "hep-th",
    doi = "10.1007/JHEP05(2023)001",
    journal = "JHEP",
    volume = "05",
    pages = "001",
    year = "2023"
}

@article{Guerrieri:2021tak,
    author = "Guerrieri, Andrea and Sever, Amit",
    title = "{Rigorous Bounds on the Analytic S Matrix}",
    eprint = "2106.10257",
    archivePrefix = "arXiv",
    primaryClass = "hep-th",
    doi = "10.1103/PhysRevLett.127.251601",
    journal = "Phys. Rev. Lett.",
    volume = "127",
    number = "25",
    pages = "251601",
    year = "2021"
}

@article{Tourkine:2023xtu,
    author = "Tourkine, Piotr and Zhiboedov, Alexander",
    title = "{Scattering amplitudes from dispersive iterations of unitarity}",
    eprint = "2303.08839",
    archivePrefix = "arXiv",
    primaryClass = "hep-th",
    reportNumber = "CERN-TH-2023-025",
    month = "3",
    year = "2023"
}

@incollection{pytorch,
title = {PyTorch: An Imperative Style, High-Performance Deep Learning Library},
author = {Paszke, Adam and Gross, Sam and Massa, Francisco and Lerer, Adam and Bradbury, James and Chanan, Gregory and Killeen, Trevor and Lin, Zeming and Gimelshein, Natalia and Antiga, Luca and Desmaison, Alban and Kopf, Andreas and Yang, Edward and DeVito, Zachary and Raison, Martin and Tejani, Alykhan and Chilamkurthy, Sasank and Steiner, Benoit and Fang, Lu and Bai, Junjie and Chintala, Soumith},
booktitle = {Advances in Neural Information Processing Systems 32},
pages = {8024--8035},
year = {2019},
publisher = {Curran Associates, Inc.},
url = {http://papers.neurips.cc/paper/9015-pytorch-an-imperative-style-high-performance-deep-learning-library.pdf}
}

@article{Atkinson:1973wt,
    author = "Atkinson, D. and Mahoux, G. and Yndurain, F. J.",
    title = "{Construction of a unitary analytic scattering amplitude (i). scalar particles}",
    doi = "10.1016/0550-3213(73)90078-3",
    journal = "Nucl. Phys. B",
    volume = "54",
    pages = "263--284",
    year = "1973"
}

@article{Martin:2020jlu,
    author = "Martin, Andr\'e and Richard, Jean-Marc",
    title = "{New result on phase shift analysis}",
    eprint = "2004.11156",
    archivePrefix = "arXiv",
    primaryClass = "math-ph",
    reportNumber = "CERN-TH-2020-066",
    doi = "10.1103/PhysRevD.101.094014",
    journal = "Phys. Rev. D",
    volume = "101",
    number = "9",
    pages = "094014",
    year = "2020"
}

@article{Kruczenski:2022lot,
    author = "Kruczenski, Martin and Penedones, Joao and van Rees, Balt C.",
    title = "{Snowmass White Paper: S-matrix Bootstrap}",
    eprint = "2203.02421",
    archivePrefix = "arXiv",
    primaryClass = "hep-th",
    month = "3",
    year = "2022"
}

@article{Caron-Huot:2019vjl,
    author = "Caron-Huot, Simon and Dixon, Lance J. and Dulat, Falko and von Hippel, Matt and McLeod, Andrew J. and Papathanasiou, Georgios",
    title = "{Six-Gluon amplitudes in planar $ \mathcal{N} $ = 4 super-Yang-Mills theory at six and seven loops}",
    eprint = "1903.10890",
    archivePrefix = "arXiv",
    primaryClass = "hep-th",
    reportNumber = "DESY 19-042, HU-EP-19/04, DESY-19-042, HU-EP-19-04, SLAC-PUB-17413",
    doi = "10.1007/JHEP08(2019)016",
    journal = "JHEP",
    volume = "08",
    pages = "016",
    year = "2019"
}

@article{Hannesdottir:2022xki,
    author = "Hannesdottir, Holmfridur S. and McLeod, Andrew J. and Schwartz, Matthew D. and Vergu, Cristian",
    title = "{Constraints on sequential discontinuities from the geometry of on-shell spaces}",
    eprint = "2211.07633",
    archivePrefix = "arXiv",
    primaryClass = "hep-th",
    reportNumber = "CERN-TH-2022-189",
    doi = "10.1007/JHEP07(2023)236",
    journal = "JHEP",
    volume = "07",
    pages = "236",
    year = "2023"
}

@article{Caron-Huot:2020cmc,
    author = "Caron-Huot, Simon and Van Duong, Vincent",
    title = "{Extremal Effective Field Theories}",
    eprint = "2011.02957",
    archivePrefix = "arXiv",
    primaryClass = "hep-th",
    doi = "10.1007/JHEP05(2021)280",
    journal = "JHEP",
    volume = "05",
    pages = "280",
    year = "2021"
}

@article{Correia:2020xtr,
    author = "Correia, Miguel and Sever, Amit and Zhiboedov, Alexander",
    title = "{An analytical toolkit for the S-matrix bootstrap}",
    eprint = "2006.08221",
    archivePrefix = "arXiv",
    primaryClass = "hep-th",
    reportNumber = "CERN-TH-2020-095",
    doi = "10.1007/JHEP03(2021)013",
    journal = "JHEP",
    volume = "03",
    pages = "013",
    year = "2021"
}

@book{Martin:1969ina,
    author = "Martin, Andr\'e",
    title = "{Scattering Theory: Unitarity, Analyticity and Crossing}",
    doi = "10.1007/BFb0101043",
    volume = "3",
    year = "1969"
}

@article{Ananthanarayan:2000ht,
    author = "Ananthanarayan, B. and Colangelo, G. and Gasser, J. and Leutwyler, H.",
    title = "{Roy equation analysis of pi pi scattering}",
    eprint = "hep-ph/0005297",
    archivePrefix = "arXiv",
    reportNumber = "IISC-CTS-12-99, ZU-TH-10-00, BUTP-99-33",
    doi = "10.1016/S0370-1573(01)00009-6",
    journal = "Phys. Rept.",
    volume = "353",
    pages = "207--279",
    year = "2001"
}

@article{Atkinson:1968hza,
    author = "Atkinson, D.",
    title = "{A proof of the existence of functions that satisfy exactly both crossing and unitarity}: {I. Neutral pion-pion scattering. No subtractions.}",
    doi = "10.1016/0550-3213(70)90120-3",
    journal = "Nucl. Phys. B",
    volume = "7",
    pages = "375--408",
    year = "1968",
    note = "[Erratum: Nucl.Phys.B 15, 331--331 (1970)]"
}

@article{Atkinson:1968exe,
    author = "Atkinson, D.",
    title = "{A proof of the existence of functions that satisfy exactly both crossing and unitarity (ii) charged pions. no subtractions}",
    doi = "10.1016/0550-3213(70)90121-5",
    journal = "Nucl. Phys. B",
    volume = "8",
    pages = "377--390",
    year = "1968",
    note = "[Erratum: Nucl.Phys.B 15, 331--331 (1970)]"
}

@article{Atkinson:1969eh,
    author = "Atkinson, D.",
    title = "{A proof of the existence of functions that satisfy exactly both crossing and unitarity (iii). subtractions}",
    doi = "10.1016/0550-3213(69)90245-4",
    journal = "Nucl. Phys. B",
    volume = "13",
    pages = "415--436",
    year = "1969"
}

@article{Atkinson:1970pe,
    author = "Atkinson, D.",
    title = "{A proof of the existence of functions that satisfy exactly both crossing and unitarity. iv. nearly constant asymptotic cross-sections}",
    doi = "10.1016/0550-3213(70)90157-4",
    journal = "Nucl. Phys. B",
    volume = "23",
    pages = "397--412",
    year = "1970"
}

\end{document}